\documentclass[longauth]{aa}

\usepackage{graphicx}
\usepackage{natbib}
\usepackage{scalerel}

\usepackage[table]{xcolor}
\usepackage{multirow}
\usepackage{txfonts}
\usepackage[pdfencoding=auto,psdextra]{hyperref}
\hypersetup{
    colorlinks=true,
    linkcolor=blue,
    filecolor=magenta,      
    urlcolor=blue,
    citecolor=blue
}
\makeatletter
\renewcommand*\aa@pageof{, page \thepage{} of \pageref*{LastPage}}

\newcommand{\mass}{\Sigma_{\star}}

\makeatother

\usepackage[utf8]{inputenc}

\usepackage[switch, modulo]{lineno}

\usepackage{euclid}

\begin{document}

\title{\Euclid\/ preparation}
 \subtitle{LXVIII. Extracting physical parameters from galaxies with machine learning}    

   
\newcommand{\orcid}[1]{} 
\author{Euclid Collaboration: I.~Kova{\v{c}}i{\'{c}}\orcid{0000-0001-6751-3263}\thanks{\email{inja.kovacic@ugent.be}}\inst{\ref{aff1}}
\and M.~Baes\orcid{0000-0002-3930-2757}\inst{\ref{aff1}}
\and A.~Nersesian\orcid{0000-0001-6843-409X}\inst{\ref{aff2},\ref{aff1}}
\and N.~Andreadis\orcid{0009-0001-9915-6325}\inst{\ref{aff1}}
\and L.~Nemani\inst{\ref{aff3}}
\and Abdurro'uf\orcid{0000-0002-5258-8761}\inst{\ref{aff4}}
\and L.~Bisigello\orcid{0000-0003-0492-4924}\inst{\ref{aff5},\ref{aff6}}
\and M.~Bolzonella\orcid{0000-0003-3278-4607}\inst{\ref{aff7}}
\and C.~Tortora\orcid{0000-0001-7958-6531}\inst{\ref{aff8}}
\and A.~van~der~Wel\inst{\ref{aff1}}
\and S.~Cavuoti\orcid{0000-0002-3787-4196}\inst{\ref{aff8},\ref{aff9}}
\and C.~J.~Conselice\orcid{0000-0003-1949-7638}\inst{\ref{aff10}}
\and A.~Enia\orcid{0000-0002-0200-2857}\inst{\ref{aff11},\ref{aff7}}
\and L.~K.~Hunt\orcid{0000-0001-9162-2371}\inst{\ref{aff12}}
\and P.~Iglesias-Navarro\orcid{0009-0009-8959-2404}\inst{\ref{aff13},\ref{aff14}}
\and E.~Iodice\orcid{0000-0003-4291-0005}\inst{\ref{aff8}}
\and J.~H.~Knapen\orcid{0000-0003-1643-0024}\inst{\ref{aff13},\ref{aff14}}
\and F.~R.~Marleau\orcid{0000-0002-1442-2947}\inst{\ref{aff15}}
\and O.~M\"uller\orcid{0000-0003-4552-9808}\inst{\ref{aff16}}
\and R.~F.~Peletier\orcid{0000-0001-7621-947X}\inst{\ref{aff17}}
\and J.~Rom\'an\orcid{0000-0002-3849-3467}\inst{\ref{aff14},\ref{aff13}}
\and R.~Ragusa\inst{\ref{aff8}}
\and P.~Salucci\inst{\ref{aff18},\ref{aff19}}
\and T.~Saifollahi\orcid{0000-0002-9554-7660}\inst{\ref{aff20}}
\and M.~Scodeggio\inst{\ref{aff21}}
\and M.~Siudek\orcid{0000-0002-2949-2155}\inst{\ref{aff22},\ref{aff23}}
\and T.~De~Waele\orcid{0000-0001-6518-9834}\inst{\ref{aff1}}
\and A.~Amara\inst{\ref{aff24}}
\and S.~Andreon\orcid{0000-0002-2041-8784}\inst{\ref{aff25}}
\and N.~Auricchio\orcid{0000-0003-4444-8651}\inst{\ref{aff7}}
\and C.~Baccigalupi\orcid{0000-0002-8211-1630}\inst{\ref{aff26},\ref{aff27},\ref{aff19},\ref{aff18}}
\and M.~Baldi\orcid{0000-0003-4145-1943}\inst{\ref{aff11},\ref{aff7},\ref{aff28}}
\and S.~Bardelli\orcid{0000-0002-8900-0298}\inst{\ref{aff7}}
\and P.~Battaglia\orcid{0000-0002-7337-5909}\inst{\ref{aff7}}
\and R.~Bender\orcid{0000-0001-7179-0626}\inst{\ref{aff29},\ref{aff30}}
\and C.~Bodendorf\inst{\ref{aff29}}
\and D.~Bonino\orcid{0000-0002-3336-9977}\inst{\ref{aff31}}
\and W.~Bon\inst{\ref{aff32}}
\and E.~Branchini\orcid{0000-0002-0808-6908}\inst{\ref{aff33},\ref{aff34},\ref{aff25}}
\and M.~Brescia\orcid{0000-0001-9506-5680}\inst{\ref{aff35},\ref{aff8},\ref{aff9}}
\and J.~Brinchmann\orcid{0000-0003-4359-8797}\inst{\ref{aff36},\ref{aff37}}
\and S.~Camera\orcid{0000-0003-3399-3574}\inst{\ref{aff38},\ref{aff39},\ref{aff31}}
\and V.~Capobianco\orcid{0000-0002-3309-7692}\inst{\ref{aff31}}
\and C.~Carbone\orcid{0000-0003-0125-3563}\inst{\ref{aff21}}
\and J.~Carretero\orcid{0000-0002-3130-0204}\inst{\ref{aff40},\ref{aff41}}
\and S.~Casas\orcid{0000-0002-4751-5138}\inst{\ref{aff42}}
\and F.~J.~Castander\orcid{0000-0001-7316-4573}\inst{\ref{aff23},\ref{aff43}}
\and M.~Castellano\orcid{0000-0001-9875-8263}\inst{\ref{aff3}}
\and G.~Castignani\orcid{0000-0001-6831-0687}\inst{\ref{aff7}}
\and A.~Cimatti\inst{\ref{aff44}}
\and C.~Colodro-Conde\inst{\ref{aff13}}
\and G.~Congedo\orcid{0000-0003-2508-0046}\inst{\ref{aff45}}
\and L.~Conversi\orcid{0000-0002-6710-8476}\inst{\ref{aff46},\ref{aff47}}
\and Y.~Copin\orcid{0000-0002-5317-7518}\inst{\ref{aff48}}
\and F.~Courbin\orcid{0000-0003-0758-6510}\inst{\ref{aff16},\ref{aff49},\ref{aff50}}
\and H.~M.~Courtois\orcid{0000-0003-0509-1776}\inst{\ref{aff51}}
\and A.~Da~Silva\orcid{0000-0002-6385-1609}\inst{\ref{aff52},\ref{aff53}}
\and H.~Degaudenzi\orcid{0000-0002-5887-6799}\inst{\ref{aff54}}
\and G.~De~Lucia\orcid{0000-0002-6220-9104}\inst{\ref{aff27}}
\and A.~M.~Di~Giorgio\orcid{0000-0002-4767-2360}\inst{\ref{aff55}}
\and J.~Dinis\orcid{0000-0001-5075-1601}\inst{\ref{aff52},\ref{aff53}}
\and M.~Douspis\orcid{0000-0003-4203-3954}\inst{\ref{aff56}}
\and F.~Dubath\orcid{0000-0002-6533-2810}\inst{\ref{aff54}}
\and X.~Dupac\inst{\ref{aff47}}
\and S.~Dusini\orcid{0000-0002-1128-0664}\inst{\ref{aff57}}
\and A.~Ealet\orcid{0000-0003-3070-014X}\inst{\ref{aff48}}
\and M.~Farina\orcid{0000-0002-3089-7846}\inst{\ref{aff55}}
\and S.~Farrens\orcid{0000-0002-9594-9387}\inst{\ref{aff58}}
\and F.~Faustini\orcid{0000-0001-6274-5145}\inst{\ref{aff59},\ref{aff3}}
\and S.~Ferriol\inst{\ref{aff48}}
\and P.~Fosalba\orcid{0000-0002-1510-5214}\inst{\ref{aff43},\ref{aff23}}
\and M.~Frailis\orcid{0000-0002-7400-2135}\inst{\ref{aff27}}
\and E.~Franceschi\orcid{0000-0002-0585-6591}\inst{\ref{aff7}}
\and S.~Galeotta\orcid{0000-0002-3748-5115}\inst{\ref{aff27}}
\and B.~Gillis\orcid{0000-0002-4478-1270}\inst{\ref{aff45}}
\and C.~Giocoli\orcid{0000-0002-9590-7961}\inst{\ref{aff7},\ref{aff60}}
\and A.~Grazian\orcid{0000-0002-5688-0663}\inst{\ref{aff61}}
\and F.~Grupp\inst{\ref{aff29},\ref{aff30}}
\and L.~Guzzo\orcid{0000-0001-8264-5192}\inst{\ref{aff62},\ref{aff25}}
\and S.~V.~H.~Haugan\orcid{0000-0001-9648-7260}\inst{\ref{aff63}}
\and W.~Holmes\inst{\ref{aff64}}
\and I.~Hook\orcid{0000-0002-2960-978X}\inst{\ref{aff65}}
\and F.~Hormuth\inst{\ref{aff66}}
\and A.~Hornstrup\orcid{0000-0002-3363-0936}\inst{\ref{aff67},\ref{aff68}}
\and K.~Jahnke\orcid{0000-0003-3804-2137}\inst{\ref{aff69}}
\and M.~Jhabvala\inst{\ref{aff70}}
\and B.~Joachimi\orcid{0000-0001-7494-1303}\inst{\ref{aff71}}
\and E.~Keih\"anen\orcid{0000-0003-1804-7715}\inst{\ref{aff72}}
\and S.~Kermiche\orcid{0000-0002-0302-5735}\inst{\ref{aff73}}
\and A.~Kiessling\orcid{0000-0002-2590-1273}\inst{\ref{aff64}}
\and M.~Kilbinger\orcid{0000-0001-9513-7138}\inst{\ref{aff58}}
\and B.~Kubik\orcid{0009-0006-5823-4880}\inst{\ref{aff48}}
\and K.~Kuijken\orcid{0000-0002-3827-0175}\inst{\ref{aff74}}
\and M.~K\"ummel\orcid{0000-0003-2791-2117}\inst{\ref{aff30}}
\and M.~Kunz\orcid{0000-0002-3052-7394}\inst{\ref{aff75}}
\and H.~Kurki-Suonio\orcid{0000-0002-4618-3063}\inst{\ref{aff76},\ref{aff77}}
\and S.~Ligori\orcid{0000-0003-4172-4606}\inst{\ref{aff31}}
\and P.~B.~Lilje\orcid{0000-0003-4324-7794}\inst{\ref{aff63}}
\and V.~Lindholm\orcid{0000-0003-2317-5471}\inst{\ref{aff76},\ref{aff77}}
\and I.~Lloro\inst{\ref{aff78}}
\and D.~Maino\inst{\ref{aff62},\ref{aff21},\ref{aff79}}
\and E.~Maiorano\orcid{0000-0003-2593-4355}\inst{\ref{aff7}}
\and O.~Mansutti\orcid{0000-0001-5758-4658}\inst{\ref{aff27}}
\and S.~Marcin\inst{\ref{aff80}}
\and O.~Marggraf\orcid{0000-0001-7242-3852}\inst{\ref{aff81}}
\and K.~Markovic\orcid{0000-0001-6764-073X}\inst{\ref{aff64}}
\and M.~Martinelli\orcid{0000-0002-6943-7732}\inst{\ref{aff3},\ref{aff82}}
\and N.~Martinet\orcid{0000-0003-2786-7790}\inst{\ref{aff32}}
\and F.~Marulli\orcid{0000-0002-8850-0303}\inst{\ref{aff83},\ref{aff7},\ref{aff28}}
\and R.~Massey\orcid{0000-0002-6085-3780}\inst{\ref{aff84}}
\and E.~Medinaceli\orcid{0000-0002-4040-7783}\inst{\ref{aff7}}
\and S.~Mei\orcid{0000-0002-2849-559X}\inst{\ref{aff85}}
\and M.~Melchior\inst{\ref{aff80}}
\and Y.~Mellier\inst{\ref{aff86},\ref{aff87}}
\and M.~Meneghetti\orcid{0000-0003-1225-7084}\inst{\ref{aff7},\ref{aff28}}
\and E.~Merlin\orcid{0000-0001-6870-8900}\inst{\ref{aff3}}
\and G.~Meylan\inst{\ref{aff16}}
\and M.~Moresco\orcid{0000-0002-7616-7136}\inst{\ref{aff83},\ref{aff7}}
\and L.~Moscardini\orcid{0000-0002-3473-6716}\inst{\ref{aff83},\ref{aff7},\ref{aff28}}
\and S.-M.~Niemi\inst{\ref{aff88}}
\and J.~W.~Nightingale\orcid{0000-0002-8987-7401}\inst{\ref{aff89}}
\and C.~Padilla\orcid{0000-0001-7951-0166}\inst{\ref{aff90}}
\and S.~Paltani\orcid{0000-0002-8108-9179}\inst{\ref{aff54}}
\and F.~Pasian\orcid{0000-0002-4869-3227}\inst{\ref{aff27}}
\and K.~Pedersen\inst{\ref{aff91}}
\and V.~Pettorino\inst{\ref{aff88}}
\and S.~Pires\orcid{0000-0002-0249-2104}\inst{\ref{aff58}}
\and G.~Polenta\orcid{0000-0003-4067-9196}\inst{\ref{aff59}}
\and M.~Poncet\inst{\ref{aff92}}
\and L.~A.~Popa\inst{\ref{aff93}}
\and L.~Pozzetti\orcid{0000-0001-7085-0412}\inst{\ref{aff7}}
\and F.~Raison\orcid{0000-0002-7819-6918}\inst{\ref{aff29}}
\and R.~Rebolo\inst{\ref{aff13},\ref{aff14},\ref{aff94}}
\and A.~Renzi\orcid{0000-0001-9856-1970}\inst{\ref{aff6},\ref{aff57}}
\and J.~Rhodes\orcid{0000-0002-4485-8549}\inst{\ref{aff64}}
\and G.~Riccio\inst{\ref{aff8}}
\and E.~Romelli\orcid{0000-0003-3069-9222}\inst{\ref{aff27}}
\and M.~Roncarelli\orcid{0000-0001-9587-7822}\inst{\ref{aff7}}
\and E.~Rossetti\orcid{0000-0003-0238-4047}\inst{\ref{aff11}}
\and R.~Saglia\orcid{0000-0003-0378-7032}\inst{\ref{aff30},\ref{aff29}}
\and Z.~Sakr\orcid{0000-0002-4823-3757}\inst{\ref{aff95},\ref{aff96},\ref{aff97}}
\and A.~G.~S\'anchez\orcid{0000-0003-1198-831X}\inst{\ref{aff29}}
\and D.~Sapone\orcid{0000-0001-7089-4503}\inst{\ref{aff98}}
\and B.~Sartoris\orcid{0000-0003-1337-5269}\inst{\ref{aff30},\ref{aff27}}
\and M.~Schirmer\orcid{0000-0003-2568-9994}\inst{\ref{aff69}}
\and P.~Schneider\orcid{0000-0001-8561-2679}\inst{\ref{aff81}}
\and T.~Schrabback\orcid{0000-0002-6987-7834}\inst{\ref{aff15}}
\and A.~Secroun\orcid{0000-0003-0505-3710}\inst{\ref{aff73}}
\and G.~Seidel\orcid{0000-0003-2907-353X}\inst{\ref{aff69}}
\and S.~Serrano\orcid{0000-0002-0211-2861}\inst{\ref{aff43},\ref{aff99},\ref{aff23}}
\and C.~Sirignano\orcid{0000-0002-0995-7146}\inst{\ref{aff6},\ref{aff57}}
\and G.~Sirri\orcid{0000-0003-2626-2853}\inst{\ref{aff28}}
\and L.~Stanco\orcid{0000-0002-9706-5104}\inst{\ref{aff57}}
\and J.~Steinwagner\orcid{0000-0001-7443-1047}\inst{\ref{aff29}}
\and P.~Tallada-Cresp\'{i}\orcid{0000-0002-1336-8328}\inst{\ref{aff40},\ref{aff41}}
\and D.~Tavagnacco\orcid{0000-0001-7475-9894}\inst{\ref{aff27}}
\and A.~N.~Taylor\inst{\ref{aff45}}
\and H.~I.~Teplitz\orcid{0000-0002-7064-5424}\inst{\ref{aff100}}
\and I.~Tereno\inst{\ref{aff52},\ref{aff101}}
\and R.~Toledo-Moreo\orcid{0000-0002-2997-4859}\inst{\ref{aff102}}
\and F.~Torradeflot\orcid{0000-0003-1160-1517}\inst{\ref{aff41},\ref{aff40}}
\and I.~Tutusaus\orcid{0000-0002-3199-0399}\inst{\ref{aff96}}
\and L.~Valenziano\orcid{0000-0002-1170-0104}\inst{\ref{aff7},\ref{aff103}}
\and T.~Vassallo\orcid{0000-0001-6512-6358}\inst{\ref{aff30},\ref{aff27}}
\and G.~Verdoes~Kleijn\orcid{0000-0001-5803-2580}\inst{\ref{aff17}}
\and A.~Veropalumbo\orcid{0000-0003-2387-1194}\inst{\ref{aff25},\ref{aff34},\ref{aff104}}
\and Y.~Wang\orcid{0000-0002-4749-2984}\inst{\ref{aff100}}
\and J.~Weller\orcid{0000-0002-8282-2010}\inst{\ref{aff30},\ref{aff29}}
\and A.~Zacchei\orcid{0000-0003-0396-1192}\inst{\ref{aff27},\ref{aff26}}
\and G.~Zamorani\orcid{0000-0002-2318-301X}\inst{\ref{aff7}}
\and E.~Zucca\orcid{0000-0002-5845-8132}\inst{\ref{aff7}}
\and A.~Biviano\orcid{0000-0002-0857-0732}\inst{\ref{aff27},\ref{aff26}}
\and E.~Bozzo\orcid{0000-0002-8201-1525}\inst{\ref{aff54}}
\and C.~Burigana\orcid{0000-0002-3005-5796}\inst{\ref{aff5},\ref{aff103}}
\and M.~Calabrese\orcid{0000-0002-2637-2422}\inst{\ref{aff105},\ref{aff21}}
\and D.~Di~Ferdinando\inst{\ref{aff28}}
\and J.~A.~Escartin~Vigo\inst{\ref{aff29}}
\and F.~Finelli\orcid{0000-0002-6694-3269}\inst{\ref{aff7},\ref{aff103}}
\and J.~Gracia-Carpio\inst{\ref{aff29}}
\and S.~Matthew\orcid{0000-0001-8448-1697}\inst{\ref{aff45}}
\and N.~Mauri\orcid{0000-0001-8196-1548}\inst{\ref{aff44},\ref{aff28}}
\and M.~P\"ontinen\orcid{0000-0001-5442-2530}\inst{\ref{aff76}}
\and V.~Scottez\inst{\ref{aff86},\ref{aff106}}
\and M.~Tenti\orcid{0000-0002-4254-5901}\inst{\ref{aff28}}
\and M.~Viel\orcid{0000-0002-2642-5707}\inst{\ref{aff26},\ref{aff27},\ref{aff18},\ref{aff19},\ref{aff107}}
\and M.~Wiesmann\orcid{0009-0000-8199-5860}\inst{\ref{aff63}}
\and Y.~Akrami\orcid{0000-0002-2407-7956}\inst{\ref{aff108},\ref{aff109}}
\and V.~Allevato\orcid{0000-0001-7232-5152}\inst{\ref{aff8}}
\and S.~Alvi\orcid{0000-0001-5779-8568}\inst{\ref{aff110}}
\and S.~Anselmi\orcid{0000-0002-3579-9583}\inst{\ref{aff57},\ref{aff6},\ref{aff111}}
\and M.~Archidiacono\orcid{0000-0003-4952-9012}\inst{\ref{aff62},\ref{aff79}}
\and F.~Atrio-Barandela\orcid{0000-0002-2130-2513}\inst{\ref{aff112}}
\and M.~Ballardini\orcid{0000-0003-4481-3559}\inst{\ref{aff110},\ref{aff7},\ref{aff113}}
\and M.~Bethermin\orcid{0000-0002-3915-2015}\inst{\ref{aff20},\ref{aff32}}
\and L.~Blot\orcid{0000-0002-9622-7167}\inst{\ref{aff114},\ref{aff111}}
\and S.~Borgani\orcid{0000-0001-6151-6439}\inst{\ref{aff115},\ref{aff26},\ref{aff27},\ref{aff19}}
\and S.~Bruton\orcid{0000-0002-6503-5218}\inst{\ref{aff116}}
\and R.~Cabanac\orcid{0000-0001-6679-2600}\inst{\ref{aff96}}
\and A.~Calabro\orcid{0000-0003-2536-1614}\inst{\ref{aff3}}
\and B.~Camacho~Quevedo\orcid{0000-0002-8789-4232}\inst{\ref{aff43},\ref{aff23}}
\and G.~Ca\~nas-Herrera\orcid{0000-0003-2796-2149}\inst{\ref{aff88},\ref{aff117}}
\and A.~Cappi\inst{\ref{aff7},\ref{aff118}}
\and F.~Caro\inst{\ref{aff3}}
\and C.~S.~Carvalho\inst{\ref{aff101}}
\and T.~Castro\orcid{0000-0002-6292-3228}\inst{\ref{aff27},\ref{aff19},\ref{aff26},\ref{aff107}}
\and K.~C.~Chambers\orcid{0000-0001-6965-7789}\inst{\ref{aff119}}
\and T.~Contini\orcid{0000-0003-0275-938X}\inst{\ref{aff96}}
\and A.~R.~Cooray\orcid{0000-0002-3892-0190}\inst{\ref{aff120}}
\and O.~Cucciati\orcid{0000-0002-9336-7551}\inst{\ref{aff7}}
\and G.~Desprez\orcid{0000-0001-8325-1742}\inst{\ref{aff121}}
\and A.~D\'iaz-S\'anchez\orcid{0000-0003-0748-4768}\inst{\ref{aff122}}
\and J.~J.~Diaz\inst{\ref{aff22}}
\and S.~Di~Domizio\orcid{0000-0003-2863-5895}\inst{\ref{aff33},\ref{aff34}}
\and H.~Dole\orcid{0000-0002-9767-3839}\inst{\ref{aff56}}
\and S.~Escoffier\orcid{0000-0002-2847-7498}\inst{\ref{aff73}}
\and A.~G.~Ferrari\orcid{0009-0005-5266-4110}\inst{\ref{aff44},\ref{aff28}}
\and P.~G.~Ferreira\orcid{0000-0002-3021-2851}\inst{\ref{aff123}}
\and I.~Ferrero\orcid{0000-0002-1295-1132}\inst{\ref{aff63}}
\and A.~Finoguenov\orcid{0000-0002-4606-5403}\inst{\ref{aff76}}
\and A.~Fontana\orcid{0000-0003-3820-2823}\inst{\ref{aff3}}
\and F.~Fornari\orcid{0000-0003-2979-6738}\inst{\ref{aff103}}
\and L.~Gabarra\orcid{0000-0002-8486-8856}\inst{\ref{aff123}}
\and K.~Ganga\orcid{0000-0001-8159-8208}\inst{\ref{aff85}}
\and J.~Garc\'ia-Bellido\orcid{0000-0002-9370-8360}\inst{\ref{aff108}}
\and T.~Gasparetto\orcid{0000-0002-7913-4866}\inst{\ref{aff27}}
\and V.~Gautard\inst{\ref{aff124}}
\and E.~Gaztanaga\orcid{0000-0001-9632-0815}\inst{\ref{aff23},\ref{aff43},\ref{aff125}}
\and F.~Giacomini\orcid{0000-0002-3129-2814}\inst{\ref{aff28}}
\and F.~Gianotti\orcid{0000-0003-4666-119X}\inst{\ref{aff7}}
\and G.~Gozaliasl\orcid{0000-0002-0236-919X}\inst{\ref{aff126},\ref{aff76}}
\and C.~M.~Gutierrez\orcid{0000-0001-7854-783X}\inst{\ref{aff127}}
\and A.~Hall\orcid{0000-0002-3139-8651}\inst{\ref{aff45}}
\and S.~Hemmati\orcid{0000-0003-2226-5395}\inst{\ref{aff128}}
\and H.~Hildebrandt\orcid{0000-0002-9814-3338}\inst{\ref{aff129}}
\and J.~Hjorth\orcid{0000-0002-4571-2306}\inst{\ref{aff91}}
\and A.~Jimenez~Mu\~noz\orcid{0009-0004-5252-185X}\inst{\ref{aff130}}
\and J.~J.~E.~Kajava\orcid{0000-0002-3010-8333}\inst{\ref{aff131},\ref{aff132}}
\and V.~Kansal\orcid{0000-0002-4008-6078}\inst{\ref{aff133},\ref{aff134}}
\and D.~Karagiannis\orcid{0000-0002-4927-0816}\inst{\ref{aff135},\ref{aff136}}
\and C.~C.~Kirkpatrick\inst{\ref{aff72}}
\and A.~M.~C.~Le~Brun\orcid{0000-0002-0936-4594}\inst{\ref{aff111}}
\and J.~Le~Graet\orcid{0000-0001-6523-7971}\inst{\ref{aff73}}
\and J.~Lesgourgues\orcid{0000-0001-7627-353X}\inst{\ref{aff42}}
\and T.~I.~Liaudat\orcid{0000-0002-9104-314X}\inst{\ref{aff137}}
\and A.~Loureiro\orcid{0000-0002-4371-0876}\inst{\ref{aff138},\ref{aff139}}
\and J.~Macias-Perez\orcid{0000-0002-5385-2763}\inst{\ref{aff130}}
\and G.~Maggio\orcid{0000-0003-4020-4836}\inst{\ref{aff27}}
\and M.~Magliocchetti\orcid{0000-0001-9158-4838}\inst{\ref{aff55}}
\and F.~Mannucci\orcid{0000-0002-4803-2381}\inst{\ref{aff12}}
\and R.~Maoli\orcid{0000-0002-6065-3025}\inst{\ref{aff140},\ref{aff3}}
\and J.~Mart\'{i}n-Fleitas\orcid{0000-0002-8594-569X}\inst{\ref{aff141}}
\and C.~J.~A.~P.~Martins\orcid{0000-0002-4886-9261}\inst{\ref{aff142},\ref{aff36}}
\and L.~Maurin\orcid{0000-0002-8406-0857}\inst{\ref{aff56}}
\and R.~B.~Metcalf\orcid{0000-0003-3167-2574}\inst{\ref{aff83},\ref{aff7}}
\and M.~Miluzio\inst{\ref{aff47},\ref{aff143}}
\and P.~Monaco\orcid{0000-0003-2083-7564}\inst{\ref{aff115},\ref{aff27},\ref{aff19},\ref{aff26}}
\and A.~Montoro\orcid{0000-0003-4730-8590}\inst{\ref{aff23},\ref{aff43}}
\and A.~Mora\orcid{0000-0002-1922-8529}\inst{\ref{aff141}}
\and C.~Moretti\orcid{0000-0003-3314-8936}\inst{\ref{aff18},\ref{aff107},\ref{aff27},\ref{aff26},\ref{aff19}}
\and G.~Morgante\inst{\ref{aff7}}
\and Nicholas~A.~Walton\orcid{0000-0003-3983-8778}\inst{\ref{aff144}}
\and L.~Patrizii\inst{\ref{aff28}}
\and V.~Popa\orcid{0000-0002-9118-8330}\inst{\ref{aff93}}
\and D.~Potter\orcid{0000-0002-0757-5195}\inst{\ref{aff145}}
\and I.~Risso\orcid{0000-0003-2525-7761}\inst{\ref{aff146}}
\and P.-F.~Rocci\inst{\ref{aff56}}
\and M.~Sahl\'en\orcid{0000-0003-0973-4804}\inst{\ref{aff147}}
\and E.~Sarpa\orcid{0000-0002-1256-655X}\inst{\ref{aff18},\ref{aff107},\ref{aff19}}
\and C.~Scarlata\orcid{0000-0002-9136-8876}\inst{\ref{aff116}}
\and A.~Schneider\orcid{0000-0001-7055-8104}\inst{\ref{aff145}}
\and M.~Sereno\orcid{0000-0003-0302-0325}\inst{\ref{aff7},\ref{aff28}}
\and F.~Shankar\orcid{0000-0001-8973-5051}\inst{\ref{aff148}}
\and P.~Simon\inst{\ref{aff81}}
\and A.~Spurio~Mancini\orcid{0000-0001-5698-0990}\inst{\ref{aff149},\ref{aff150}}
\and J.~Stadel\orcid{0000-0001-7565-8622}\inst{\ref{aff145}}
\and S.~A.~Stanford\orcid{0000-0003-0122-0841}\inst{\ref{aff151}}
\and K.~Tanidis\inst{\ref{aff123}}
\and C.~Tao\orcid{0000-0001-7961-8177}\inst{\ref{aff73}}
\and G.~Testera\inst{\ref{aff34}}
\and R.~Teyssier\orcid{0000-0001-7689-0933}\inst{\ref{aff152}}
\and S.~Toft\orcid{0000-0003-3631-7176}\inst{\ref{aff153},\ref{aff154}}
\and S.~Tosi\orcid{0000-0002-7275-9193}\inst{\ref{aff33},\ref{aff34}}
\and A.~Troja\orcid{0000-0003-0239-4595}\inst{\ref{aff6},\ref{aff57}}
\and M.~Tucci\inst{\ref{aff54}}
\and C.~Valieri\inst{\ref{aff28}}
\and J.~Valiviita\orcid{0000-0001-6225-3693}\inst{\ref{aff76},\ref{aff77}}
\and D.~Vergani\orcid{0000-0003-0898-2216}\inst{\ref{aff7}}
\and G.~Verza\orcid{0000-0002-1886-8348}\inst{\ref{aff155},\ref{aff156}}
\and P.~Vielzeuf\orcid{0000-0003-2035-9339}\inst{\ref{aff73}}}
                                                                                   
\institute{Sterrenkundig Observatorium, Universiteit Gent, Krijgslaan 281 S9, 9000 Gent, Belgium\label{aff1}
\and
STAR Institute, Quartier Agora - All\'ee du six Ao\^ut, 19c B-4000 Li\`ege, Belgium\label{aff2}
\and
INAF-Osservatorio Astronomico di Roma, Via Frascati 33, 00078 Monteporzio Catone, Italy\label{aff3}
\and
Johns Hopkins University 3400 North Charles Street Baltimore, MD 21218, USA\label{aff4}
\and
INAF, Istituto di Radioastronomia, Via Piero Gobetti 101, 40129 Bologna, Italy\label{aff5}
\and
Dipartimento di Fisica e Astronomia "G. Galilei", Universit\`a di Padova, Via Marzolo 8, 35131 Padova, Italy\label{aff6}
\and
INAF-Osservatorio di Astrofisica e Scienza dello Spazio di Bologna, Via Piero Gobetti 93/3, 40129 Bologna, Italy\label{aff7}
\and
INAF-Osservatorio Astronomico di Capodimonte, Via Moiariello 16, 80131 Napoli, Italy\label{aff8}
\and
INFN section of Naples, Via Cinthia 6, 80126, Napoli, Italy\label{aff9}
\and
Jodrell Bank Centre for Astrophysics, Department of Physics and Astronomy, University of Manchester, Oxford Road, Manchester M13 9PL, UK\label{aff10}
\and
Dipartimento di Fisica e Astronomia, Universit\`a di Bologna, Via Gobetti 93/2, 40129 Bologna, Italy\label{aff11}
\and
INAF-Osservatorio Astrofisico di Arcetri, Largo E. Fermi 5, 50125, Firenze, Italy\label{aff12}
\and
Instituto de Astrof\'isica de Canarias, Calle V\'ia L\'actea s/n, 38204, San Crist\'obal de La Laguna, Tenerife, Spain\label{aff13}
\and
Departamento de Astrof\'isica, Universidad de La Laguna, 38206, La Laguna, Tenerife, Spain\label{aff14}
\and
Universit\"at Innsbruck, Institut f\"ur Astro- und Teilchenphysik, Technikerstr. 25/8, 6020 Innsbruck, Austria\label{aff15}
\and
Institute of Physics, Laboratory of Astrophysics, Ecole Polytechnique F\'ed\'erale de Lausanne (EPFL), Observatoire de Sauverny, 1290 Versoix, Switzerland\label{aff16}
\and
Kapteyn Astronomical Institute, University of Groningen, PO Box 800, 9700 AV Groningen, The Netherlands\label{aff17}
\and
SISSA, International School for Advanced Studies, Via Bonomea 265, 34136 Trieste TS, Italy\label{aff18}
\and
INFN, Sezione di Trieste, Via Valerio 2, 34127 Trieste TS, Italy\label{aff19}
\and
Universit\'e de Strasbourg, CNRS, Observatoire astronomique de Strasbourg, UMR 7550, 67000 Strasbourg, France\label{aff20}
\and
INAF-IASF Milano, Via Alfonso Corti 12, 20133 Milano, Italy\label{aff21}
\and
Instituto de Astrof\'isica de Canarias (IAC); Departamento de Astrof\'isica, Universidad de La Laguna (ULL), 38200, La Laguna, Tenerife, Spain\label{aff22}
\and
Institute of Space Sciences (ICE, CSIC), Campus UAB, Carrer de Can Magrans, s/n, 08193 Barcelona, Spain\label{aff23}
\and
School of Mathematics and Physics, University of Surrey, Guildford, Surrey, GU2 7XH, UK\label{aff24}
\and
INAF-Osservatorio Astronomico di Brera, Via Brera 28, 20122 Milano, Italy\label{aff25}
\and
IFPU, Institute for Fundamental Physics of the Universe, via Beirut 2, 34151 Trieste, Italy\label{aff26}
\and
INAF-Osservatorio Astronomico di Trieste, Via G. B. Tiepolo 11, 34143 Trieste, Italy\label{aff27}
\and
INFN-Sezione di Bologna, Viale Berti Pichat 6/2, 40127 Bologna, Italy\label{aff28}
\and
Max Planck Institute for Extraterrestrial Physics, Giessenbachstr. 1, 85748 Garching, Germany\label{aff29}
\and
Universit\"ats-Sternwarte M\"unchen, Fakult\"at f\"ur Physik, Ludwig-Maximilians-Universit\"at M\"unchen, Scheinerstrasse 1, 81679 M\"unchen, Germany\label{aff30}
\and
INAF-Osservatorio Astrofisico di Torino, Via Osservatorio 20, 10025 Pino Torinese (TO), Italy\label{aff31}
\and
Aix-Marseille Universit\'e, CNRS, CNES, LAM, Marseille, France\label{aff32}
\and
Dipartimento di Fisica, Universit\`a di Genova, Via Dodecaneso 33, 16146, Genova, Italy\label{aff33}
\and
INFN-Sezione di Genova, Via Dodecaneso 33, 16146, Genova, Italy\label{aff34}
\and
Department of Physics "E. Pancini", University Federico II, Via Cinthia 6, 80126, Napoli, Italy\label{aff35}
\and
Instituto de Astrof\'isica e Ci\^encias do Espa\c{c}o, Universidade do Porto, CAUP, Rua das Estrelas, PT4150-762 Porto, Portugal\label{aff36}
\and
Faculdade de Ci\^encias da Universidade do Porto, Rua do Campo de Alegre, 4150-007 Porto, Portugal\label{aff37}
\and
Dipartimento di Fisica, Universit\`a degli Studi di Torino, Via P. Giuria 1, 10125 Torino, Italy\label{aff38}
\and
INFN-Sezione di Torino, Via P. Giuria 1, 10125 Torino, Italy\label{aff39}
\and
Centro de Investigaciones Energ\'eticas, Medioambientales y Tecnol\'ogicas (CIEMAT), Avenida Complutense 40, 28040 Madrid, Spain\label{aff40}
\and
Port d'Informaci\'{o} Cient\'{i}fica, Campus UAB, C. Albareda s/n, 08193 Bellaterra (Barcelona), Spain\label{aff41}
\and
Institute for Theoretical Particle Physics and Cosmology (TTK), RWTH Aachen University, 52056 Aachen, Germany\label{aff42}
\and
Institut d'Estudis Espacials de Catalunya (IEEC),  Edifici RDIT, Campus UPC, 08860 Castelldefels, Barcelona, Spain\label{aff43}
\and
Dipartimento di Fisica e Astronomia "Augusto Righi" - Alma Mater Studiorum Universit\`a di Bologna, Viale Berti Pichat 6/2, 40127 Bologna, Italy\label{aff44}
\and
Institute for Astronomy, University of Edinburgh, Royal Observatory, Blackford Hill, Edinburgh EH9 3HJ, UK\label{aff45}
\and
European Space Agency/ESRIN, Largo Galileo Galilei 1, 00044 Frascati, Roma, Italy\label{aff46}
\and
ESAC/ESA, Camino Bajo del Castillo, s/n., Urb. Villafranca del Castillo, 28692 Villanueva de la Ca\~nada, Madrid, Spain\label{aff47}
\and
Universit\'e Claude Bernard Lyon 1, CNRS/IN2P3, IP2I Lyon, UMR 5822, Villeurbanne, F-69100, France\label{aff48}
\and
Institut de Ci\`{e}ncies del Cosmos (ICCUB), Universitat de Barcelona (IEEC-UB), Mart\'{i} i Franqu\`{e}s 1, 08028 Barcelona, Spain\label{aff49}
\and
Instituci\'o Catalana de Recerca i Estudis Avan\c{c}ats (ICREA), Passeig de Llu\'{\i}s Companys 23, 08010 Barcelona, Spain\label{aff50}
\and
UCB Lyon 1, CNRS/IN2P3, IUF, IP2I Lyon, 4 rue Enrico Fermi, 69622 Villeurbanne, France\label{aff51}
\and
Departamento de F\'isica, Faculdade de Ci\^encias, Universidade de Lisboa, Edif\'icio C8, Campo Grande, PT1749-016 Lisboa, Portugal\label{aff52}
\and
Instituto de Astrof\'isica e Ci\^encias do Espa\c{c}o, Faculdade de Ci\^encias, Universidade de Lisboa, Campo Grande, 1749-016 Lisboa, Portugal\label{aff53}
\and
Department of Astronomy, University of Geneva, ch. d'Ecogia 16, 1290 Versoix, Switzerland\label{aff54}
\and
INAF-Istituto di Astrofisica e Planetologia Spaziali, via del Fosso del Cavaliere, 100, 00100 Roma, Italy\label{aff55}
\and
Universit\'e Paris-Saclay, CNRS, Institut d'astrophysique spatiale, 91405, Orsay, France\label{aff56}
\and
INFN-Padova, Via Marzolo 8, 35131 Padova, Italy\label{aff57}
\and
Universit\'e Paris-Saclay, Universit\'e Paris Cit\'e, CEA, CNRS, AIM, 91191, Gif-sur-Yvette, France\label{aff58}
\and
Space Science Data Center, Italian Space Agency, via del Politecnico snc, 00133 Roma, Italy\label{aff59}
\and
Istituto Nazionale di Fisica Nucleare, Sezione di Bologna, Via Irnerio 46, 40126 Bologna, Italy\label{aff60}
\and
INAF-Osservatorio Astronomico di Padova, Via dell'Osservatorio 5, 35122 Padova, Italy\label{aff61}
\and
Dipartimento di Fisica "Aldo Pontremoli", Universit\`a degli Studi di Milano, Via Celoria 16, 20133 Milano, Italy\label{aff62}
\and
Institute of Theoretical Astrophysics, University of Oslo, P.O. Box 1029 Blindern, 0315 Oslo, Norway\label{aff63}
\and
Jet Propulsion Laboratory, California Institute of Technology, 4800 Oak Grove Drive, Pasadena, CA, 91109, USA\label{aff64}
\and
Department of Physics, Lancaster University, Lancaster, LA1 4YB, UK\label{aff65}
\and
Felix Hormuth Engineering, Goethestr. 17, 69181 Leimen, Germany\label{aff66}
\and
Technical University of Denmark, Elektrovej 327, 2800 Kgs. Lyngby, Denmark\label{aff67}
\and
Cosmic Dawn Center (DAWN), Denmark\label{aff68}
\and
Max-Planck-Institut f\"ur Astronomie, K\"onigstuhl 17, 69117 Heidelberg, Germany\label{aff69}
\and
NASA Goddard Space Flight Center, Greenbelt, MD 20771, USA\label{aff70}
\and
Department of Physics and Astronomy, University College London, Gower Street, London WC1E 6BT, UK\label{aff71}
\and
Department of Physics and Helsinki Institute of Physics, Gustaf H\"allstr\"omin katu 2, 00014 University of Helsinki, Finland\label{aff72}
\and
Aix-Marseille Universit\'e, CNRS/IN2P3, CPPM, Marseille, France\label{aff73}
\and
Leiden Observatory, Leiden University, Einsteinweg 55, 2333 CC Leiden, The Netherlands\label{aff74}
\and
Universit\'e de Gen\`eve, D\'epartement de Physique Th\'eorique and Centre for Astroparticle Physics, 24 quai Ernest-Ansermet, CH-1211 Gen\`eve 4, Switzerland\label{aff75}
\and
Department of Physics, P.O. Box 64, 00014 University of Helsinki, Finland\label{aff76}
\and
Helsinki Institute of Physics, Gustaf H{\"a}llstr{\"o}min katu 2, University of Helsinki, Helsinki, Finland\label{aff77}
\and
NOVA optical infrared instrumentation group at ASTRON, Oude Hoogeveensedijk 4, 7991PD, Dwingeloo, The Netherlands\label{aff78}
\and
INFN-Sezione di Milano, Via Celoria 16, 20133 Milano, Italy\label{aff79}
\and
University of Applied Sciences and Arts of Northwestern Switzerland, School of Engineering, 5210 Windisch, Switzerland\label{aff80}
\and
Universit\"at Bonn, Argelander-Institut f\"ur Astronomie, Auf dem H\"ugel 71, 53121 Bonn, Germany\label{aff81}
\and
INFN-Sezione di Roma, Piazzale Aldo Moro, 2 - c/o Dipartimento di Fisica, Edificio G. Marconi, 00185 Roma, Italy\label{aff82}
\and
Dipartimento di Fisica e Astronomia "Augusto Righi" - Alma Mater Studiorum Universit\`a di Bologna, via Piero Gobetti 93/2, 40129 Bologna, Italy\label{aff83}
\and
Department of Physics, Centre for Extragalactic Astronomy, Durham University, South Road, Durham, DH1 3LE, UK\label{aff84}
\and
Universit\'e Paris Cit\'e, CNRS, Astroparticule et Cosmologie, 75013 Paris, France\label{aff85}
\and
Institut d'Astrophysique de Paris, 98bis Boulevard Arago, 75014, Paris, France\label{aff86}
\and
Institut d'Astrophysique de Paris, UMR 7095, CNRS, and Sorbonne Universit\'e, 98 bis boulevard Arago, 75014 Paris, France\label{aff87}
\and
European Space Agency/ESTEC, Keplerlaan 1, 2201 AZ Noordwijk, The Netherlands\label{aff88}
\and
School of Mathematics, Statistics and Physics, Newcastle University, Herschel Building, Newcastle-upon-Tyne, NE1 7RU, UK\label{aff89}
\and
Institut de F\'{i}sica d'Altes Energies (IFAE), The Barcelona Institute of Science and Technology, Campus UAB, 08193 Bellaterra (Barcelona), Spain\label{aff90}
\and
DARK, Niels Bohr Institute, University of Copenhagen, Jagtvej 155, 2200 Copenhagen, Denmark\label{aff91}
\and
Centre National d'Etudes Spatiales -- Centre spatial de Toulouse, 18 avenue Edouard Belin, 31401 Toulouse Cedex 9, France\label{aff92}
\and
Institute of Space Science, Str. Atomistilor, nr. 409 M\u{a}gurele, Ilfov, 077125, Romania\label{aff93}
\and
Consejo Superior de Investigaciones Cientificas, Calle Serrano 117, 28006 Madrid, Spain\label{aff94}
\and
Institut f\"ur Theoretische Physik, University of Heidelberg, Philosophenweg 16, 69120 Heidelberg, Germany\label{aff95}
\and
Institut de Recherche en Astrophysique et Plan\'etologie (IRAP), Universit\'e de Toulouse, CNRS, UPS, CNES, 14 Av. Edouard Belin, 31400 Toulouse, France\label{aff96}
\and
Universit\'e St Joseph; Faculty of Sciences, Beirut, Lebanon\label{aff97}
\and
Departamento de F\'isica, FCFM, Universidad de Chile, Blanco Encalada 2008, Santiago, Chile\label{aff98}
\and
Satlantis, University Science Park, Sede Bld 48940, Leioa-Bilbao, Spain\label{aff99}
\and
Infrared Processing and Analysis Center, California Institute of Technology, Pasadena, CA 91125, USA\label{aff100}
\and
Instituto de Astrof\'isica e Ci\^encias do Espa\c{c}o, Faculdade de Ci\^encias, Universidade de Lisboa, Tapada da Ajuda, 1349-018 Lisboa, Portugal\label{aff101}
\and
Universidad Polit\'ecnica de Cartagena, Departamento de Electr\'onica y Tecnolog\'ia de Computadoras,  Plaza del Hospital 1, 30202 Cartagena, Spain\label{aff102}
\and
INFN-Bologna, Via Irnerio 46, 40126 Bologna, Italy\label{aff103}
\and
Dipartimento di Fisica, Universit\`a degli studi di Genova, and INFN-Sezione di Genova, via Dodecaneso 33, 16146, Genova, Italy\label{aff104}
\and
Astronomical Observatory of the Autonomous Region of the Aosta Valley (OAVdA), Loc. Lignan 39, I-11020, Nus (Aosta Valley), Italy\label{aff105}
\and
ICL, Junia, Universit\'e Catholique de Lille, LITL, 59000 Lille, France\label{aff106}
\and
ICSC - Centro Nazionale di Ricerca in High Performance Computing, Big Data e Quantum Computing, Via Magnanelli 2, Bologna, Italy\label{aff107}
\and
Instituto de F\'isica Te\'orica UAM-CSIC, Campus de Cantoblanco, 28049 Madrid, Spain\label{aff108}
\and
CERCA/ISO, Department of Physics, Case Western Reserve University, 10900 Euclid Avenue, Cleveland, OH 44106, USA\label{aff109}
\and
Dipartimento di Fisica e Scienze della Terra, Universit\`a degli Studi di Ferrara, Via Giuseppe Saragat 1, 44122 Ferrara, Italy\label{aff110}
\and
Laboratoire Univers et Th\'eorie, Observatoire de Paris, Universit\'e PSL, Universit\'e Paris Cit\'e, CNRS, 92190 Meudon, France\label{aff111}
\and
Departamento de F{\'\i}sica Fundamental. Universidad de Salamanca. Plaza de la Merced s/n. 37008 Salamanca, Spain\label{aff112}
\and
Istituto Nazionale di Fisica Nucleare, Sezione di Ferrara, Via Giuseppe Saragat 1, 44122 Ferrara, Italy\label{aff113}
\and
Center for Data-Driven Discovery, Kavli IPMU (WPI), UTIAS, The University of Tokyo, Kashiwa, Chiba 277-8583, Japan\label{aff114}
\and
Dipartimento di Fisica - Sezione di Astronomia, Universit\`a di Trieste, Via Tiepolo 11, 34131 Trieste, Italy\label{aff115}
\and
Minnesota Institute for Astrophysics, University of Minnesota, 116 Church St SE, Minneapolis, MN 55455, USA\label{aff116}
\and
Institute Lorentz, Leiden University, Niels Bohrweg 2, 2333 CA Leiden, The Netherlands\label{aff117}
\and
Universit\'e C\^{o}te d'Azur, Observatoire de la C\^{o}te d'Azur, CNRS, Laboratoire Lagrange, Bd de l'Observatoire, CS 34229, 06304 Nice cedex 4, France\label{aff118}
\and
Institute for Astronomy, University of Hawaii, 2680 Woodlawn Drive, Honolulu, HI 96822, USA\label{aff119}
\and
Department of Physics \& Astronomy, University of California Irvine, Irvine CA 92697, USA\label{aff120}
\and
Department of Astronomy \& Physics and Institute for Computational Astrophysics, Saint Mary's University, 923 Robie Street, Halifax, Nova Scotia, B3H 3C3, Canada\label{aff121}
\and
Departamento F\'isica Aplicada, Universidad Polit\'ecnica de Cartagena, Campus Muralla del Mar, 30202 Cartagena, Murcia, Spain\label{aff122}
\and
Department of Physics, Oxford University, Keble Road, Oxford OX1 3RH, UK\label{aff123}
\and
CEA Saclay, DFR/IRFU, Service d'Astrophysique, Bat. 709, 91191 Gif-sur-Yvette, France\label{aff124}
\and
Institute of Cosmology and Gravitation, University of Portsmouth, Portsmouth PO1 3FX, UK\label{aff125}
\and
Department of Computer Science, Aalto University, PO Box 15400, Espoo, FI-00 076, Finland\label{aff126}
\and
Instituto de Astrof\'\i sica de Canarias, c/ Via Lactea s/n, La Laguna E-38200, Spain. Departamento de Astrof\'\i sica de la Universidad de La Laguna, Avda. Francisco Sanchez, La Laguna, E-38200, Spain\label{aff127}
\and
Caltech/IPAC, 1200 E. California Blvd., Pasadena, CA 91125, USA\label{aff128}
\and
Ruhr University Bochum, Faculty of Physics and Astronomy, Astronomical Institute (AIRUB), German Centre for Cosmological Lensing (GCCL), 44780 Bochum, Germany\label{aff129}
\and
Univ. Grenoble Alpes, CNRS, Grenoble INP, LPSC-IN2P3, 53, Avenue des Martyrs, 38000, Grenoble, France\label{aff130}
\and
Department of Physics and Astronomy, Vesilinnantie 5, 20014 University of Turku, Finland\label{aff131}
\and
Serco for European Space Agency (ESA), Camino bajo del Castillo, s/n, Urbanizacion Villafranca del Castillo, Villanueva de la Ca\~nada, 28692 Madrid, Spain\label{aff132}
\and
ARC Centre of Excellence for Dark Matter Particle Physics, Melbourne, Australia\label{aff133}
\and
Centre for Astrophysics \& Supercomputing, Swinburne University of Technology,  Hawthorn, Victoria 3122, Australia\label{aff134}
\and
School of Physics and Astronomy, Queen Mary University of London, Mile End Road, London E1 4NS, UK\label{aff135}
\and
Department of Physics and Astronomy, University of the Western Cape, Bellville, Cape Town, 7535, South Africa\label{aff136}
\and
IRFU, CEA, Universit\'e Paris-Saclay 91191 Gif-sur-Yvette Cedex, France\label{aff137}
\and
Oskar Klein Centre for Cosmoparticle Physics, Department of Physics, Stockholm University, Stockholm, SE-106 91, Sweden\label{aff138}
\and
Astrophysics Group, Blackett Laboratory, Imperial College London, London SW7 2AZ, UK\label{aff139}
\and
Dipartimento di Fisica, Sapienza Universit\`a di Roma, Piazzale Aldo Moro 2, 00185 Roma, Italy\label{aff140}
\and
Aurora Technology for European Space Agency (ESA), Camino bajo del Castillo, s/n, Urbanizacion Villafranca del Castillo, Villanueva de la Ca\~nada, 28692 Madrid, Spain\label{aff141}
\and
Centro de Astrof\'{\i}sica da Universidade do Porto, Rua das Estrelas, 4150-762 Porto, Portugal\label{aff142}
\and
HE Space for European Space Agency (ESA), Camino bajo del Castillo, s/n, Urbanizacion Villafranca del Castillo, Villanueva de la Ca\~nada, 28692 Madrid, Spain\label{aff143}
\and
Institute of Astronomy, University of Cambridge, Madingley Road, Cambridge CB3 0HA, UK\label{aff144}
\and
Department of Astrophysics, University of Zurich, Winterthurerstrasse 190, 8057 Zurich, Switzerland\label{aff145}
\and
INAF-Osservatorio Astronomico di Brera, Via Brera 28, 20122 Milano, Italy, and INFN-Sezione di Genova, Via Dodecaneso 33, 16146, Genova, Italy\label{aff146}
\and
Theoretical astrophysics, Department of Physics and Astronomy, Uppsala University, Box 515, 751 20 Uppsala, Sweden\label{aff147}
\and
School of Physics \& Astronomy, University of Southampton, Highfield Campus, Southampton SO17 1BJ, UK\label{aff148}
\and
Department of Physics, Royal Holloway, University of London, TW20 0EX, UK\label{aff149}
\and
Mullard Space Science Laboratory, University College London, Holmbury St Mary, Dorking, Surrey RH5 6NT, UK\label{aff150}
\and
Department of Physics and Astronomy, University of California, Davis, CA 95616, USA\label{aff151}
\and
Department of Astrophysical Sciences, Peyton Hall, Princeton University, Princeton, NJ 08544, USA\label{aff152}
\and
Cosmic Dawn Center (DAWN)\label{aff153}
\and
Niels Bohr Institute, University of Copenhagen, Jagtvej 128, 2200 Copenhagen, Denmark\label{aff154}
\and
Center for Cosmology and Particle Physics, Department of Physics, New York University, New York, NY 10003, USA\label{aff155}
\and
Center for Computational Astrophysics, Flatiron Institute, 162 5th Avenue, 10010, New York, NY, USA\label{aff156}}

\titlerunning{MCRT with explicit absorption}

\date{\today}

\abstract{The \Euclid\/\ mission is generating a vast amount of imaging data in four broadband filters at a high angular resolution. This data will allow for the detailed study of mass, metallicity, and stellar populations across galaxies that will constrain their formation and evolutionary pathways. Transforming the \Euclid\/\ imaging for large samples of galaxies into maps of physical parameters in an efficient and reliable manner is an outstanding challenge. Here, we investigate the power and reliability of machine learning techniques to extract the distribution of physical parameters within well-resolved galaxies. We focus on estimating stellar mass surface density, mass-averaged stellar metallicity, and age. We generated noise-free synthetic high-resolution ($100{\rm \, pc} \times100 {\rm \, pc}$) imaging data in the \Euclid\/\ photometric bands for a set of 1154 galaxies from the TNG50 cosmological simulation. The images were generated with the SKIRT radiative transfer code, taking into account the complex 3D distribution of stellar populations and interstellar dust attenuation. We used a machine learning framework to map the idealised mock observational data to the physical parameters on a pixel-by-pixel basis. We find that stellar mass surface density can be accurately recovered with a $\leq 0.130 {\rm \,dex}$ scatter. 
Conversely, stellar metallicity and age estimates are, as expected, less robust, but they still contain significant information that originates from underlying correlations at a sub-kiloparsec scales between stellar mass surface density and stellar population properties. As a corollary, we show that TNG50 follows a spatially resolved mass-metallicity relation that is consistent with observations. Due to its relatively low computational and time requirements, which has a time-frame of minutes without dedicated high performance computing infrastructure once it has been trained, our method allows for fast and robust estimates of the stellar mass surface density distributions of nearby galaxies from four-filter \Euclid\/ imaging data. Equivalent estimates of stellar population properties (stellar metallicity and age) are less robust but still hold value as first-order approximations across large samples.} 


   \titlerunning{\Euclid\/: Distribution of physical parameters with machine learning}
   \authorrunning{I. Kova{\v{c}}i{\'{c}} et al.}
   
   \maketitle

\section{Introduction}

Understanding the underlying processes that drive galactic evolution through cosmic time has been one of the burning issues in astrophysics for the past decades. An important class of constraints for galaxy evolution theories is scaling relations between the physical properties of galaxies. For example, galaxy stellar mass, considered one of the fundamental parameters of galaxies, has been found to be tightly correlated with star-formation rate \citep[SFR;][]{Brinchmann_2004, Whitaker_2012, Speagle_2014, Watkins_2022, Chamba_2022, Popesso_2023} as well as with gas phase \citep{Lequeux_1979, Tremonti_2004, Ly_2016, Zahid_2017} and stellar metallicity \citep{Kauffmann_2003, Gallazzi_2005}. These correlations are well reproduced by observations of large data sets of galaxies \citep[for a detailed review, see][and references within]{Maiolino_2019}.

Due to observational constraints, early scaling relations were formulated by treating each galaxy as a single observational point \citep[e.g.][]{Schmidt_1959,Faber_1973,Lequeux_1979}. However, galaxies themselves are complex systems, and taking into account the powerful interplay between stars, gas, and dust in a galactic disc, the existence of scaling relations on sub-kiloparsec scales \citep{Rosales-Ortega_2012, Sanchez_2013,Gao_2018} becomes an even more powerful diagnostic of galaxy evolution.

A number of large imaging and integral-field spectroscopic surveys have become available in the recent years, and others will soon become operational, including Mapping Nearby Galaxies at Apache Point Observatory \citep[MaNGA;][]{Bundy_2015,Abdurrouf_2022}, the \textit{Vera C.\ Rubin} Observatory's Legacy Survey of Space and Time \citep[LSST;][]{Ivezic_2019}, and the Euclid Wide Survey \citep[EWS;][]{laureijs2011euclid, Scaramella_2022}. These surveys will provide detailed insight into thousands of nearby galaxies. Therefore, developing methods to accurately determine spatially resolved galaxy properties, such as stellar mass and surface density, is of great importance. Recent studies have shown that the structural properties of the stellar components in galaxies are also related to those of the surrounding dark matter halos \citep[e.g.][]{Salucci_2019}. This entanglement is thought to be due to tuned baryonic feedback or due to specific properties of the dark particles. This link between the luminous and the dark components of galaxies could make the availability of these observational properties a primary task in observational cosmology.

Traditionally, the physical properties of galaxies are estimated by fitting template spectral energy distributions (SEDs) to the observations \citep{Walcher_2010, Conroy_2013, Pacifici_2023}. SED fitting is usually applied to the global integrated SEDs of galaxies, and the result is an estimate of the main global physical properties, such as stellar mass, SFR, stellar metallicity, star-formation history (SFH), and level of dust attenuation \citep[e.g.][]{Leja_2017, Leja_2019, Nersesian_2019}. More recently, SED fitting has also been applied at resolved scales, that is, to individual pixels, yielding distributions of these physical properties at kiloparsec or sub-kiloparsec scales \citep[e.g.][]{Zibetti_2009, Viaene_2014, Sorba_2015, Abdurro'uf_2021, Abdurro'uf_2022, Abdurro'uf_2023}. 

Several issues arise due to the assumptions that have to be made when deriving the physical properties of galaxies. For example, as the main component of light produced within the galaxy, evolutionary tracks of stars can be used to fit the SEDs of stellar populations \citep[so-called isochrone synthesis, e.g.][]{Charlot_1991, Bruzual_1993, Charlot_2003}. However, complications occur from, among other issues, the incomplete libraries of stellar spectra (e.g. the lack of rest-frame UV libraries), which could be used to obtain accurate galactic parameters for high-redshift galaxies \citep{Pellerin_2009}. Another common issue arises from the limitations of our understanding of the complex history in the interaction between the interstellar medium (ISM) and stars \citep[e.g.][]{Matteucci_2008} as well as modelling of the short and complex evolutionary tracks of the short-lived and difficult to observe high-mass stars.

{Another complicating factor is the effect of interstellar dust. The contribution of interstellar dust to galaxy SEDs consists of two effects: attenuation of the intrinsic UV to near-IR (NIR) emission and thermal emission in the mid-IR (MIR) to milimetre regime. In SED fitting, the former is typically accounted for using a parameterised attenuation curve \citep[e.g.][]{Calzetti_2000, Charlot_2000, LoFaro_2017, Salim_2018, Decleir_2019}, whereas the latter effect is taken into account using empirical or radiative transfer-based template SEDs \citep[e.g.][]{Dale_2002, Lagache_2004, Siebenmorgen_2007, Law_2018}. However, our understanding of the complex dust contribution in the ISM is very incomplete, leaving a number of open questions, such as the details of the shape, composition, size distribution, fluffiness, and formation and destruction processes of interstellar dust \citep[e.g.][]{Draine_2003, Galliano_2021}. The attenuation curve and dust emission libraries are limited by the parameter space they were derived from, and linking dust attenuation and dust emission is complex due to non-linear radiative transfer effects \citep{Baes_2001a, Steinacker_2013}. Even with the number of large-scale panchromatic and spectroscopic surveys increasing the understanding of the complex dust grain physics in recent years, there is still a great diversity both in extinction and attenuation curves that contributes to the uncertainty of the properties derived via SED fitting \citep{Salim_2020}.}

On top of that, SED fitting is usually computationally demanding, especially when investigating a large parameter space. Scaling up these methods becomes unfeasible if one is to effectively analyse the vast amount of data that will become available with the upcoming surveys. This points to the necessity of exploring alternative methods, and one such method that has been employed in recent years is machine learning (ML). In this work, we focus on supervised ML, which learns the relationship between a set of input parameters and a target via a set of examples \citep[for a review on different ML algorithms and their applications in astronomy, we recommend][and references within]{Baron_2019}.

One of the early issues in astronomy addressed by ML has been measuring the photometric redshift of galaxies due to the difficulty of gaining true values of other galactic parameters from observations \citep{Tagliaferri_2003,Collister_2004,Dabrusco_2007,Carliles_2010,Hildebrandt_2010,Abdalla_2011,Cavuoti_2012,Brescia_2013,Sadeh_2016,Disanto_2018, Leistedt_2023,Alsing_2023, Alsing_2024}, with some attempts to infer SFRs from photometric data \citep[e.g.][]{Stensbo_2017,Delli_2019}. In recent years, a number of authors have focused on data from cosmological simulations \citep[][]{Lovell_2019, Gilda_2021, Simet_2021} and observational data \citep[][]{Acquaviva_2015,Surana_2020,Iglesias_Navarro_2024, Alsing_2024}, demonstrating that ML techniques can accurately recover a number of galactic parameters even on limited data sets and with much smaller computational and time requirements. One interesting application of ML is to infer hard to measure observations \citep[for example far-IR][]{Luhman_FIR}, thus bridging the gap in the galactic spectral libraries on both global \citep{Dobbels_2020} and resolved \citep{Dobbels_2021} scales. Interestingly, in \citet{Dobbels_2021} the authors find no significant dependence of the prediction by the ML algorithm on the pixel scale of the observations in their work.

These recent successful developments in deploying ML to solve astronomical problems have a potential to be very significant for the \Euclid\/\ mission \citep{2024arXiv240513491E}. Successfully launched on 1 July 2023, it has started to generate a vast amount of imaging data in four broadband filters. In particular, the Euclid Wide Survey \citep{laureijs2011euclid} is expected to detect $\sim 2 \times 10^9$ galaxies up to $z \sim$\,3, with the majority of galaxies at $z \sim$\,1 \citep{Scaramella_2022}. Mapping the distribution of the physical properties of these galaxies as a function of cosmic time, and the correlation between them, offers an additional constraint to the scaling relations for theories of galaxy formation and evolution. At the same time, the \Euclid\/\ surveys will spatially resolve tens of thousands of nearby galaxies to kiloparsec or sub-kiloparsec scales. Generating physically resolved maps at kiloparsec scales of the most important physical properties of galaxies offers valuable input to galaxy evolution theories. The sheer amount of data makes ML techniques a necessary tool to efficiently generate physical maps at high spatial resolution.

There have been a number of attempts to test the feasibility of ML methods on simulated \Euclid\/ data, for example, modelling photometric redshift \citep{Desprez_2020}, galaxy classification \citep{Humphrey_2023}, identification of galaxy-galaxy strong lensing events \citep{Leuzzi_2023}, deriving galaxy properties \citep{Bretonniere_2022,Bretonniere_2023,Bisigello_2023,Aussel_2024}, and inferring global physical parameters of galaxies \citep{EP-Enia}. The studies demonstrate that ML could be a viable alternative to traditional methods; however, caution should be taken into account due to the high impact of the parameter space on the algorithm's predictive power and the difficulty of interpreting the results, both of which are common pitfalls with ML. 

\begin{figure*}
    \centering
    \includegraphics[width=0.98\textwidth]{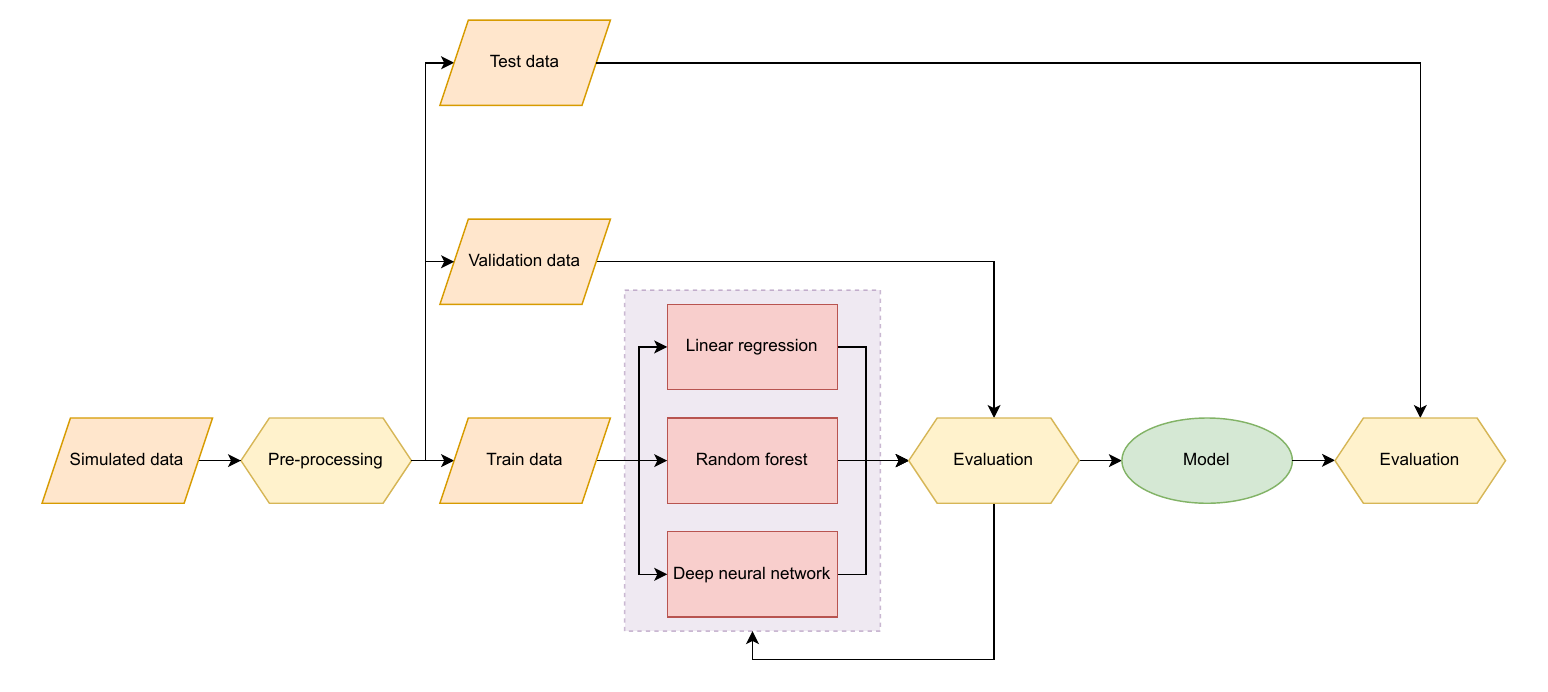}
    \caption{Workflow of the entire ML process described in this work.}
    \label{fig:workflow}
\end{figure*}

In this work, we expand on the approach explored by \citet{Bisigello_2023}, who attempted to infer galaxy physical properties with deep ML from mock \Euclid\/\ magnitudes and $H$-band images. We focus on the spatially resolved distribution of physical properties within nearby galaxies. More specifically, we investigate to which degree we can infer physical property maps from sets of synthetic \Euclid\/\ images of galaxies extracted from the TNG50 cosmological hydrodynamical simulation \citep{Pillepich_2019}. We employ two ML methods, random forest (RF) and a deep neural network (DNN), and we compare them to an ordinary least squares linear regression (LR). We focus on stellar mass surface density ($\mass$), stellar-mass-averaged stellar metallicity ($Z_{\star}$), and stellar-mass-averaged stellar age ($t_\star$). The workflow of the entire ML process described in this work can be seen in Fig.\,\ref{fig:workflow}. In this work, we focus on noise-free data; however, we plan to include noise and test this relationship on more realistic simulated data as well as on actual \Euclid\/\ observations.

This paper is organised as follows: In Sect.\,\ref{sec:data}, we introduce the synthetic images and describe the steps undertaken to prepare them for the ML method. In Sect.\,\ref{sec:methods}, we describe the ML algorithms considered in this work and the metrics used to analyse the viability of this method. Section~\ref{sec:results_general} contains the estimates of galaxy properties, while Sect.\,\ref{sec:discussion} focuses on the underlying processes that drive these results, including a focused discussion on the effects of the underlying resolved mass-metallicity relation. The main findings and the future outlook are summarised in Sect.\,\ref{sec:summary}.


\section{Data}\label{sec:data}

\subsection{The TNG50 simulation}

IllustrisTNG \citep[][hereafter TNG]{Springel_2017, Pillepich_2017, Pillepich_2018, Naiman_2018, Marinacci_2018} is a publicly available state-of-the-art large-volume magnetohydrodynamics cosmological simulation that uses the moving mesh code \textsc{AREPO} \citep{Springel_2010} to model the formation and evolution of galaxies in a $\Lambda$CDM universe. As a successor of the Illustris project \citep{Vogelsberger_2014, Genel_2014}, it offers an upgrade and refinement in its treatment of cosmic magnetism, supermassive black holes, galactic winds, and stellar evolution \citep{Weinberger_2016}. The TNG project consists of three runs: TNG50, TNG100, and TNG300. They are defined by the size of their physical simulation box with side lengths of 50, 100, and 300\,Mpc, respectively \citep{Nelson_2018}. Out of the three, the highest-resolution TNG50 \citep{Pillepich_2019, Nelson_2019} is best suited for the purpose of our work. It follows the evolution of a cubic volume of 51.7 comoving Mpc on the side, with cosmological parameters based on the {\it{Planck}} mission \citep{Planck_2016}: $\Omega_{\text{m}} = 0.3089$, $\Omega_{\text{b}} = 0.0486$, $\Omega_{\Lambda} = 0.6911$, $H_0=100\,h\,{\rm km\,s^{-1}\,Mpc^{-1}}$ with $h = 0.6774$, and a \citet{Chabrier_2003} initial mass function (IMF). It reaches a baryonic mass resolution of $8.5\times10^{4} \,{\rm M_{\odot}}$ and the average cell size of 70--140\,pc in the star-forming regions of the galaxies. For more details on this simulation, we refer to \citet{Pillepich_2019}.

\subsection{Synthetic images}

Our data set consists of 1154 galaxies at $z = 0$ with a total mass range between $10^{9.8}\,{\rm M_{\odot}}$, and $10^{12}\,{\rm M_{\odot}}$ extracted from the TNG50 simulation. Based on this set of galaxies, \citet{baes2024tng50skirt} recently released an image atlas, the TNG50-SKIRT Atlas, which contains synthetic high-resolution images in 18 broadband filters from UV to NIR wavelengths. The images are generated with the SKIRT radiative transfer code \citep{Baes_2011, Camps_2015, Camps_Baes_2020} and take into account different stellar populations, absorption and scattering by ISM dust, and localised dust attenuation in star-forming regions. The atlas contains images for each galaxy at five different projections, where the angle of the first projection is arbitrarily determined, and the other four are spread over a unit sphere in a way to maximise the angular separation between them \citep{tammes1930origin}. Beside the dust-obscured images, the atlas also contains dust-free images in all 18 bands, as well as matching maps of the stellar mass surface density, dust mass surface density, and mass-weighted stellar age and metallicity. These physical parameter maps were generated by projecting the 3D physical fields such as the stellar mass density on the observer's plane of the sky, using the `probe' functionality in the SKIRT code \citep{Camps_Baes_2020}. The sample contains 869 star-forming galaxies and 285 quiescent galaxies \citep[for a characterisation of the sample, see Fig.~1 in][]{baes2024tng50skirt}. 

{We extend the TNG50-SKIRT Atlas image database presented in \citet{baes2024tng50skirt} with additional synthetic images in the four \Euclid\/\ broadband filters, that is, the optical $\IE$ band of the VIS instrument \citep{Cropper_2018, 2024arXiv240513491E} and the $\YE$, $\JE$, and $\HE$ bands of the NIR NISP instrument \citep{Schirmer_2022, 2024arXiv240513493E}. The images are generated in a similar way as for the original atlas: we use SKIRT to generate both dust-obscured and dust-free images, but now with the \Euclid\/\ transmission curves. Apart from that, all other settings, such as the observer positions, the spatial resolution, and the field of view, remain identical as for the original images \citep[see][Table~2]{baes2024tng50skirt}. The additional SKIRT simulations were run on the supercomputer facilities of the Vlaams Supercomputer Centre (VSC)\footnote{\url{https://www.ugent.be/hpc/en/infrastructure/overview.htm}}.}

All maps and images have a pixel scale of $100\,\mathrm{pc}$ and a field of view of $160\,\mathrm{kpc}$ on the side, corresponding to $1600 \times 1600$ pixels in size, thus ensuring the outer regions of even the most extended galaxies are covered. Except for the presence of Monte Carlo noise in dust-obscured images, the images are noise-free.

For this work, we opt only for galaxies belonging to one randomly chosen projection (O5), since the first projection angle is arbitrary, thus ensuring a similar randomness in the inclination of the observations. Of course, we can assume that the projection angle affects the accuracy of the result, but our choice is motivated by the real world scenario of observing a large number of galaxies. Additionally, we decide to infer the galaxy parameters on a pixel-by-pixel basis which, due to the high resolution of the images, gives us a high number of data points. Due to this wealth of data, the inclusion of other angles should not offer any additional information for the ML method explored here, but it might prove its usefulness in, for example, training a convolutional neural network since in that case each data point would have to be a tuple consisting of a number of neighbouring pixels, thus decreasing the number of available data points from this data set significantly; however that is beyond the scope of this work. Regardless, we test whether the inclusion of other projections would affect the result on a subset of 200 randomly selected galaxies using a DNN (for the description, see Sect.\,\ref{sec:methods_ML}). At five angles each, the total data set consists of 1000 “unique” observations. The results are shown in Appendix\,\ref{app:angles}. For all three target parameters, the differences in accuracy of inferred galaxy properties are negligible.

\subsection{Data pre-processing}\label{sec:preprocessing}

The SKIRT generated mock images contain Monte Carlo noise, which increases towards fainter surface brightness levels. To assess the reliability of the surface brightness levels, we select a sample of ten galaxies for which we re-ran the SKIRT simulations with the reliability statistics module switched on. The result of these additional simulations are diagnostic statistics on the reliability of the Monte Carlo simulation in each pixel \citep[for detailed information on this procedure, we refer to][]{Camps_2018, Camps_Baes_2020}. Based on these statistics, we created a mask in the $\HE$-band images that removes all pixels where the Monte Carlo results are deemed unreliable. The motivation behind the choice of the $\HE$ band was twofold. Traditionally, stellar mass is reconstructed as the stellar mass-to-light ratio (that is, the ratio between stellar mass and light that reaches the observer) from the optical-near-IR colour \citep{Zibetti_2009,Zibetti_2019}. The variations in this ratio are smaller at longer wavelengths due to NIR being dominated by the light from the low-mass long-lived stars, as well as the lower dust extinction at those wavelengths \citep{Bell_2001}, making the ${\HE}$ band at $1.65 {\rm \, \mu m}$ close to the ideal choice for reconstructing stellar mass \citep{Walcher_2010}. The second reason for this choice, as it will be discussed in detail in Sect.\,\ref{sec:ml_alg_dependence}, comes from the fact that it is the most important feature when modelling both $\mass$ and $Z_{\star}$. The application of this mask translates to the removal of all pixels where $I_\text{\HE} \leq 0.55\, {\rm MJy\,sr^{-1}}$, which removes a large fraction of the pixels in the outskirts of the images. The full data set of pixels surviving this clipping still consists of more than 71 million individual pixels.

We convert both the surface brightness levels and the galactic parameters onto logarithmic scale, thus avoiding issues that could occur due to the large dynamic range. Surface brightness levels are in units of ${\rm MJy\,sr^{-1}}$. The $\mass$ is in units of ${\rm M_{\odot}\, pc^{-2}}$ and spans seven orders of magnitude in total (from $-2.5\text{\,dex}$ to $5.3\text{\,dex}$); however, 99\% of all points span three (from $0\text{\,dex}$ to $3\text{\,dex}$). The $Z_{\star}$ is dimensionless but normalised with $Z_{\odot} = 0.02$ in the analysis, and spans less than two orders of magnitude (from $-2.8\text{\,dex}$ to $-1.1\text{\,dex}$), with 99\% of all points spanning only from $-2.2\text{\,dex}$ to $-1.2\text{\,dex}$. The $t_{\star}$, in units of Gyr, spans two orders of magnitude in total (from $-3.3\text{\,dex}$ to $1.1\text{\,dex}$), and less than one for the inner 99\% of all pixels within the test set (from $0.2\text{\,dex}$ to $1.1\text{\,dex}$). Afterwards, we normalise the data in such a way that all the values lie between zero and one (a standard procedure when preparing a data set for ML problems) using the linear transformation
\begin{equation}
    X_{\text{std}} = \frac{X - X_{\text{min}}}{X_{\text{max}}-X_{\text{min}}} \quad ,
\end{equation}
where $X_{\text{min}}$ and $X_{\text{max}}$ are the minimal and maximal value of the training set range, respectively.

\begin{figure*}
    \centering
    \includegraphics[width=0.9\textwidth]{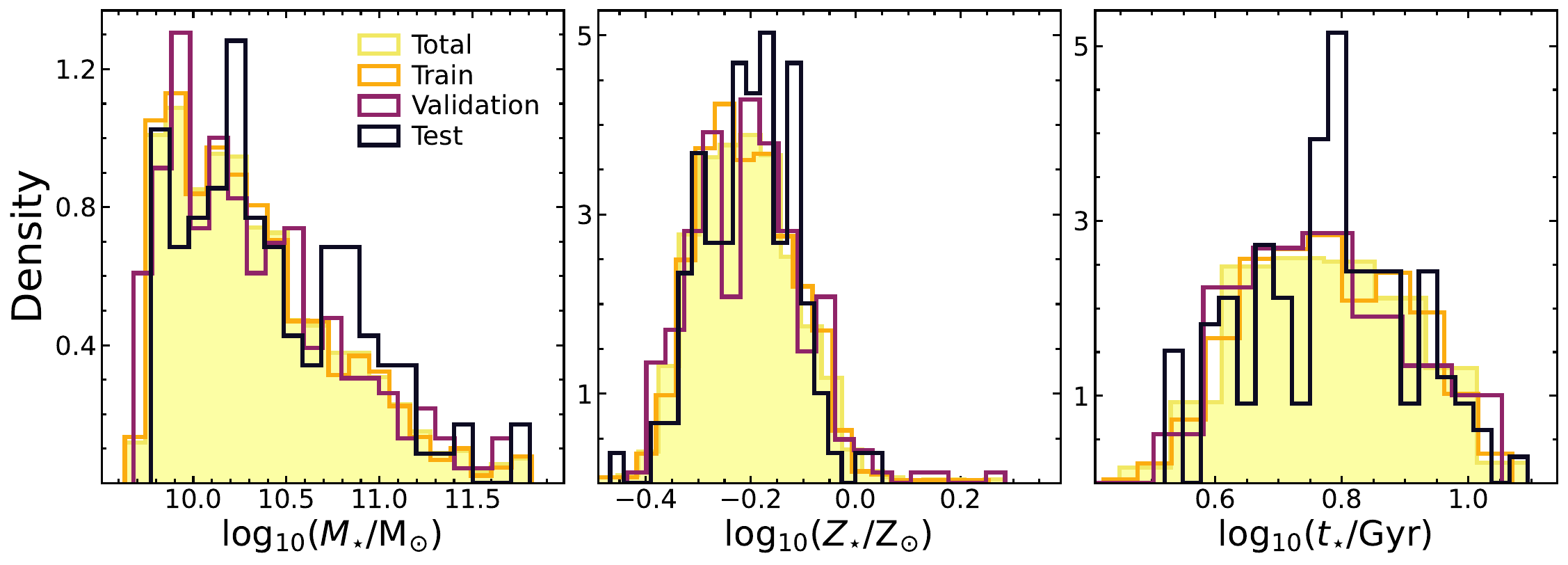}
    \caption{Normalised distribution of $\logten M_{\star}$, $\logten Z_{\star}$, and $\logten t_{\star}$ global values of the galaxy properties for the entire data set, training set, and test set (see legend).}
    \label{fig:hist}
\end{figure*}
Finally, we opt to split our data on a galaxy instead of a pixel basis. This translates to a random fraction of $\sim$\,10\% of galaxies, equalling to a total of 115 entire galaxies (of which 86 are star-forming and 29 are quiescent) being selected into the test set, which is never seen by any of the algorithms during training. The remaining 90\% (totalling 1039 galaxies, of which 783 are star-forming and 256 are quiescent) is used by the algorithm for training. However, 20\% of pixels are randomly set aside for the validation of the algorithm during training. This split is random and changes at each run of the algorithm. The total data distribution can be seen in Fig.\,\ref{fig:hist}.


\section{Methods}\label{sec:methods}

\subsection{Machine learning approach}\label{sec:methods_ML}

We tested two supervised ML algorithms, namely an RF and a DNN, to map \Euclid\/\ mock observations of nearby galaxies to galactic physical parameters on a pixel-by-pixel basis, and to analyse the relationship between them. We compare these two algorithms against a baseline, namely an ordinary least squares linear regression, which is modelled from the Python {\tt{SciPy}} library as implemented in Python {\tt{scikit-learn}} \citep{scikit-learn} and concerns the linear component of this relationship.

\begin{table}[]
    \caption{Hyper-parameter setup for the RF algorithm.}

    \centering
    \begin{tabular}{l r}
    \hline \hline
        Hyper-parameter & Setup \\
        \hline 
         Number of trees in the forest & 100 \\
         Minimum number of samples required in a leaf & 12 \\
         Minimum number of samples to split a node & 20 \\
         Maximum depth of the tree & 80 \\
         Maximum number of samples at base estimator & 2\,000\,000 \\
         Loss function & MSE \\
         \hline \hline
    \end{tabular}
    \label{tab:rf_hyper-parameters}
\end{table}

The two ML algorithms concern the non-linear component of this relationship. For the first one, we selected RF, an ensemble method algorithm \citep{Breiman_2001} from the Python {\tt{scikit-learn}} library. The RF algorithm builds a forest of decision trees, where each decision tree starts at the root node and then recursively splits until the minimum node size is reached. At each node, it selects a random given subset of variables, determines the best variable at this node, and splits it into two daughter nodes, then it averages the output of an ensemble of such trees to make a prediction \citep{Hastie2009elements,ISLP}. RF is one of the most robust methods for ML, as it was designed to avoid over-fitting (the algorithm becoming too good at predicting values on the set it was trained on, but bad at generalising on new data), and it requires minimal user input when tuning the hyper-parameters (for instance, the size of the decision tree, number of samples contained within a leaf, etc). Prior to this choice, we tested a number of different algorithms from the {\texttt{scikit-learn}} Python library on a subset of our data set, and did not notice any major effect on the accuracy of our predictions. The additional benefit of RF when compared to other ML algorithms that show similar speed and accuracy is its ability to determine feature importance, which offers a certain degree of interpretability and helps in the analysis of the results.

When building the RF, we started with a brute force approach at tuning the hyper-parameters using the {\tt{scikit-learn}} module {\tt{GridSearchCV}} on a subset of 20\% randomly selected pixels of the training set, followed by manual tuning on the entire data set until we converge to a solution that minimises computational requirements without sacrifices to the accuracy of the prediction (using metrics described in Sect.\,\ref{sec:ML_metrics}). As a final step for this method, we perform a $k$-fold cross-validation across the entire training set, which consists of 1039 whole galaxies. In the cross-validation procedure, we bin the data into $k$ (in this case $k = 10$) bins that consist of random $\sim$\,10\% ($\sim$\,104) galaxies. We set aside one bin as a validation set, train the algorithm on the remaining nine bins, and repeat the aforementioned process in a loop through all ten bins. This is a standard procedure for hyper-parameter tuning for RF, but it is also a useful additional step to avoid any possibility of over-fitting that might occur with the sample selection itself, as there is always a small risk of the algorithm behaving well on one validation set, but not on the others.

The RF setup used in this work is described in Table\,\ref{tab:rf_hyper-parameters}. We limit the maximum size of a sub-sample required to build one tree to two million, as this quantity is the most responsible for high memory requirements of RF. We decrease this value from the size of the full set until we reach the minimum at which the accuracy of the algorithm is not affected. We test the RF on a desktop machine to determine feasibility of modelling galactic parameters in optimal amount of time without requiring dedicated infrastructure. With the final model, we are able to train the algorithm on the entire data set in less than an hour on an $11^\text{th}$ generation Intel i7 2.30\,GHz 16-core laptop with 62.5\,GiB memory and 128\,GiB swap.

The third and final approach consists of building a DNN using the Google Python library {\tt{TensorFlow}} \citep{abadi2016tensorflow}. An artificial neural network is a type of a ML technique that is inspired by the functioning of neurons in a biological brain. It consists of a series of layers of a certain width, that is the number of neurons within a layer. The architecture of a DNN used in this work is summarised in Table\,\ref{tab:dnn_hyper-parameters} and consists of six layers. The first layer is the input layer, its size is determined by the number of features loaded into the algorithm. This is followed by four hidden layers that consist of 64 neurons each, which are fully connected with all the neurons in the previous and following layers. Each neuron receives a signal in the form of all the features ($x_{i}$) whose number is determined by the first layer, multiplies them by a certain weight ($w_{i})$, which is different for each feature, and is arbitrarily determined at the initialisation of the network. The weighted features are then summed up together with a bias ($b$). A final, single output layer returns the predictions of a single of the three galaxy parameters ($\hat{y}$). Each of these transformations would be linear; however, a non-linear so-called activation function is then applied on this output, returning a non-linear transformation:
\begin{equation}
    \hat{y} = f \left( \sum_{i=1}^n (w_i \times x_i) + b \right) \quad .
\end{equation}

The most common activation function used for regression problems \citep[e.g.][]{6639346,6638312,Krizhevsky_relu}, as well as in this work, is a rectified linear unit activation function \citep[ReLU;][]{Nair2010RectifiedLU, pmlr-v15-glorot11a}.

The network takes a sub-sample of 8192 points from our data set. This batch sub-sample is then propagated through the network, obtaining predicted values of the target parameter, which is then assessed for accuracy against a true value in this batch via a loss function. For the loss function we choose the mean squared error (MSE). MSE calculates the average of a squared distance between the predicted and the real value, it is the most common loss function used for regression problems, and it has proven to work the best in our case as well. The result of this comparison between the predicted and true values in the batch determine the direction in which the weights are updated, and this information is then back-propagated through the network. This process is repeated until the entire data set is used.

Another important hyper-parameter in building a DNN is the learning rate, which determines the size of the step the algorithm takes in the optimisation process \citep{murphy2012machine}. Instead of choosing a constant learning rate, we decide to implement a learning rate scheduler, whose main function is decreasing the network training time. The network starts with a learning rate of $10^{-4}$, a lower end of a fairly common starting point. This learning rate is then multiplied by 0.8 every time the loss function does not improve for more than 300 iterations. A learning rate that is too low requires a large number of iterations before the network converges, and it also increases the risk of the network becoming stuck in a local minimum, and never reaching a global one. On the contrary, a learning rate that is too high can cause the network to overshoot, that is arbitrarily jump between points without ever “settling” at a minimum. A learning rate scheduler is a simple way to circumvent that common downfall while still benefiting from increased performance earlier in the learning. And finally, we also set an early stop, signalling to the algorithm to stop its learning process if the loss function hasn't decreased for 2000 iterations. Even though it is rather large, this value allows the learning rate scheduler at least five updates, allowing it to decrease to around one third of its initial value. From our tests, this is a large enough number of epochs to ensure the best result, but it also still decreases the training time significantly.

For the optimiser, that is the algorithm which dynamically adjusts the learning rate for each parameter, we choose Adam \citep[adaptive moment estimation;][]{kingma2017adam}, which has shown superior performance for large data sets \citep{optimization_2011}. For the details on how DNNs function, we recommend \citet{Goodfellow2016}.

\begin{table}[]
    \caption{Hyper-parameter setup for the DNN.}

    \centering
    \begin{tabular}{l r}
    \hline \hline
        Hyper-parameter & Setup \\
        \hline 
         Number of hidden layers & 4 \\
         Neurons per layer & 64 \\
         Training epoch & 10\,000 \\
         Activation function & ReLU \\
         Batch size & 8192 \\
         Optimiser & Adam \\
         Initial learning rate & $10^{-4}$ \\
         Learning rate scheduler patience & 300 \\
         Learning rate scheduler factor & 0.8 \\
         Loss function & MSE \\
         Early stop & 2\,000\\
         \hline \hline
    \end{tabular}
        \hspace{0.15cm}

    \label{tab:dnn_hyper-parameters}
\end{table}

Unlike RF, a DNN requires a high level of engagement when tuning its hyper-parameters and its training is much more computationally demanding. In this work, it requires a dedicated GPU architecture to be viable, and was run on the supercomputer facilities of the VSC\footnote{Specifically, the \textit{accelgor} GPU cluster with the NVIDIA Ampere A100 GPU and no parallelisation}. In case of all three galaxy parameters, specifically the ones that show strong non-linearity, a DNN might offer a better performance over RF, but at the expense of computational power, though as it will be shown in this work, the difference between the predictive power of an RF and a DNN is negligible. Regardless, we present the results of both non-linear methods.

\subsection{Input data}

The input features from the mock observations consist of four \Euclid\/\ bands. To test our three ML methods, we organised them into the following three different categories:
\begin{itemize}
    \item Each of the single bands, and we selected the best fit (according to metrics presented in Sect.\,\ref{sec:ML_metrics}), which is $\HE$ in all cases.
    \item All four bands: $\IE$, $\YE$, $\JE$, $\HE$.
    \item Four bands, $\IE$, $\YE$, $\JE$, $\HE$, as well as six colours modelled from them: $(\IE - \YE)$, $(\IE - \JE)$, $(\IE - \HE)$, $(\YE - \JE)$, $(\YE - \HE)$, and $(\JE - \HE)$.
\end{itemize}

\subsection{Metrics of model reliability}\label{sec:ML_metrics}

Throughout this work, we use a number of metrics to determine the reliability of ML. The coefficient of determination \citep[$R^2$;][]{Wright_1921} is a commonly used term in regression analysis that can span from $-\infty$ to $+1$, where the latter denotes a model that perfectly matches the ground truth. $R^2$ determines how much the dependent variable, which in the case of ML is the prediction of the model, is determined by the independent variable, or the ground truth. $R^2$ is considered one of the most informative metrics in regression ML \citep{Chicco_2021}. It is calculated as follows:
\begin{equation}
    R^2=1-\frac{\sum_{i = 1}^{n} (\hat{y}_i-y_i)^2}{\sum_{i = 1}^{n} (\overline{y} - y_i)^2} \quad ,
\end{equation}
where $y_i$ is the actual $i$-th value, $\hat{y}_i$ is the predicted $i$-th value, and $\overline{y}$ is the mean of true values, which is defined as
\begin{equation}\label{eq:mean}
    \overline{y} = \frac{1}{n} \sum_{i = 1}^{n} y_i \quad .
\end{equation}
The root-mean-square error (RMSE) of the estimation determines how much the predicted value deviates from the ground truth,
\begin{equation}
    {\rm RMSE} = \sqrt{\frac{1}{n} \sum_{i = 1}^{n} (\hat{y}_i - y_i)^2} \quad ,
\end{equation}
and it can span from zero to $+\infty$, with the former denoting the case where the prediction perfectly matches the ground truth. The RMSE is monotonically anti-correlated to $R^2$.

The normalised median absolute deviation (NMAD) represents a mean absolute deviation that is normalised by a factor of $\sim$\,1.4826, and it is considered the equivalent of standard deviation that is more resilient to outliers \citep{Hohle_2009},
\begin{equation}
    {\rm NMAD} = 1.4826 \times {\rm median}\,|\hat{y}_i - y_i| \quad .
\end{equation}
The bias is represented as
\begin{equation}
    \left< \Delta Y \right> = {\rm median}\,\left(\hat{y}_i - y_i\right) \quad ,
\end{equation}
where $Y$ denotes any of the target galaxy properties (that is, $\mass$, $Z_{\star}$, or $t_{\star}$).

We calculated the fraction of catastrophic outliers ($f_\text{out}$), that is, the number of predicted values that are approximately two times ($0.3 {\rm \, dex}$) above or below the ground truth, matching the uncertainties at recovering stellar mass values via traditional methods \citep{Conroy_2013},
\begin{equation}
    f_\text{out} = \frac{n\left(\left|\hat{y}_i^ -  y_i\right| > 0.3 {\rm \,dex}\right)}{n_\text{tot}}\quad .
    \label{eq:f_out}
\end{equation}

And finally, we calculated the Pearson correlation coefficient \citep[$\rho$;][]{pearson}:
\begin{equation}
    \rho = \frac{\frac{1}{n}\sum_{i=1}^n \left(y_i - \overline{y}\right)\left(\hat{y}_i-\overline{y}_*\right)}{\sigma_y\sigma_{\hat{y}}} \quad,
\end{equation}
where $\overline{y}$ is the mean of true values ($y_i$); $\overline{y}_*$ is the mean of predicted values ($\hat{y}_i$), as defined in Eq.\,(\ref{eq:mean}); and $\sigma_{y}$ and $\sigma_{\hat{y}}$ are the standard deviations of true and predicted values, respectively.

Additionally, as is visible in Fig.\,\ref{fig:hist}, there is a slight bias, especially for $Z_{\star}$ and $t_{\star}$, in the distribution of the test set. To confirm that this bias does not affect the final result, we calculate the metrics of model reliability on both the test and the training sets. The difference between them is minor (within 10\%). The values are shown in Appendix\,\ref{app:samplebias}.

\subsection{Reliability of a machine learning algorithm}\label{sec:ML_errors}

To assess the reliability of a ML algorithm, there are two types of uncertainties: aleatoric and epistemic. Aleatoric (statistical) uncertainty is caused by the presence of random noise in the data set , that is, the irreducible noise from observations, while epistemic (systemic) is caused by the lack of knowledge of the best model \citep{H_llermeier_2021}. To assess the former, we would be required to perform the same ML procedure on idealised data, as well as the data with added noise. Since the goal of this work is to examine the amount of physical information that can be extracted from the \Euclid\/\ observations, and the feasibility of using ML in that process, we are currently only working with idealised data. Introducing noise into the data set would complicate the relationships between the observations and physical parameters with additional variables, so this type of uncertainty cannot be assessed. In the future, we plan to adapt this data set to the actual observations, making this type of error assessment possible. Epistemic uncertainty, on the other hand, informs us of how uncertain the model is about its prediction \citep{kendall2017uncertainties}, and is assessed by running the same algorithm on the same data set a certain number of times. In this work, we opt for ten.

As a deterministic model (that is, an algorithm that creates the same prediction on the same set of data every time), the ordinary least squares linear regression has an epistemic error of zero. RF and DNN, on the other hand, build a model that is conditional on the previous decisions made on the data set, giving them a certain level of randomness that will change the model on every new run during training.

In case of our two ML algorithms, the situation is more complicated due to the algorithms requiring a validation set during training, and the way this validation set is created. While the test data is set to the bin consisting of $\sim\,10\%$ (i.e. 115) whole galaxies, the training data is randomly split into a training and validation set used by the algorithm in a 80\%/20\% ratio.

Additionally, each of the two ML algorithms contribute to the uncertainty due to the stochasticity of the learning process itself. In case of RF, the uncertainty is related to the fact that, for each time the tree splits, a random set of features is selected for training, leading to a different path from the root node to the final leaf. However, in all cases this uncertainty is very small, in the worst case scenario being when the standard deviation ($\pm1\sigma$) is only $\sim\,2\%$ of the mean of $R^2$. In all other cases, it is 20--100 times smaller.

Every time a DNN is initialised, it starts with a different randomly assigned set of weights with which it updates the training data, that is the algorithm starts at a different, random starting point in the optimisation process, and for every epoch, it shuffles the set of samples anew, which affects the update of the weights, and maps a different path through the optimisation. However, these values are still fairly small, with a first standard deviation accounting for 5\% of the mean of $R^2$, a slightly higher value when compared to RF, but still low, and over 20 times higher than the second largest error.

\begin{table*}[]
    \caption{Accuracy scores (as defined in Sect.\,\ref{sec:ML_metrics}) for the three galaxy properties ($\mass$, $Z_{\star}$, and $t_{\star}$) for the three methods (ordinary least squares LR, and two ML algorithms, RF and DNN) for the three distinct cases (single \HE, four \Euclid\/\ bands, and four \Euclid\/\ bands as well as the six colours modelled from them).}
    
    \smallskip
\label{tab:table_all_results}
\smallskip
\centering
\begin{tabular}{l l l c c c c c c }
\hline \hline
  & & & &  & & \\[-0.3cm]

Target & Algorithm & Input features  &  $R^2$  & RMSE  & $f_{\text{out}}$ & NMAD & $\left< \Delta Y \right>$ & $\rho$ \\
 & & & &  & & \\[-0.3cm]
 \hline
 & & & &  & & \\[-0.3cm]
 & &  $I_{\HE}$ & $0.852$& $0.186$ & $0.0757$ & $0.1676$ & $-0.1676$ & 0.923\\
 & LR &  Bands & $0.873$  & $0.147$  &$0.0725$ & $0.1455$& $-0.0055$& 0.935 \\
 & & Bands + colours& $0.908$& $0.150$ &$0.0645$ & $0.1410$&$-0.0059$&0.939 \\
  & & & &  & & & \\[-0.2cm]
& &$I_{\HE}$ & $0.853 $  & $0.185$&$0.0758$ &$0.1685$ &  $-0.0185$  &0.923 \\
 $\mass /{\rm \left(M_{\odot}\, pc^{-2}\right)}$ & RF &       Bands& $0.927$ &  $0.130 $ & $0.0325$ & $0.1102$ & $-0.0021$ & 0.963 \\
& &Bands + colours  &  $0.927 $  & $0.130 $  &$0.0324$ &$0.1102$ & $-0.0022$ & 0.963\\
   & & & &  & & &\\[-0.2cm]

& &  $I_{\HE}$  & $0.854$ & $0.185 $  &$ 0.0736$ & $0.1673$&$-0.0193$ & 0.924\\
   & DNN &     Bands & $0.927$  & $0.130$  & $0.0322$ &  $0.1090$ &$-8.45 \times 10^{-5}$ & 0.963 \\
    & &    Bands + colours & $0.927 $ & $0.130 $  &$0.0322$ & $0.1089$& $-8.03 \times 10^{-5}$  & 0.963\\
      & & & &  & & & \\[-0.3cm]
 \hline
  & & & &  & & & \\[-0.3cm]

& &    $I_{\HE}$ & $0.358$ & $0.096$ & $0.0047$ & $0.0937$ & $-0.0142$ & 0.598 \\
& LR &    Bands  & $0.417$  & $0.091$  & $0.0030$ & $0.0863$ & $-0.0063$& 0.655 \\
& &  Bands + colours &$ 0.435 $ & $0.095$  & $0.0024$&$0.0857$ & $-0.0075$ &0.666 \\

& & & & &  & & \\[-0.2cm]

& &  $I_{\HE}$&$0.356$ &$0.096 $ & $0.0050$ &$0.0947$ & $-0.0141$ & 0.590\\
$Z_{\star}$& RF &Bands &$0.592$&$0.077$&$0.0007$& $0.0725$ & $0.0016$& 0.769 \\
 & &Bands + colours&  $0.595 $ & $0.077 $& $0.0007$& $0.0724$ & $0.0020$&0.770 \\
 
& & & & &  & & \\[-0.2cm]

 & &$I_{\HE}$&$0.385$&$ 0.096$&$0.0048$& $0.0938$&$-0.0145$& 0.599\\
  & DNN&Bands & $0.595 $  & $0.076$ & $0.0007$&$0.0720$&$0.0029$&  0.773 \\
 & &  Bands + colours & $0.595$ & $0.076$&$0.0007$&$0.0719$ &$0.0028$ & 0.773\\
 
& & & &  & & & \\[-0.3cm]   
\hline

& & & &  & & & \\[-0.3cm]
& &      $I_{\HE}$ & $0.010$ & $0.163 $& $0.0509 $& $0.1586 $& $-0.0435$&0.104 \\
& LR &     Bands & $0.199$ & $0.146$ & $0.0368$& $0.1335$ & $-0.0206$& 0.467 \\
& &     Bands + colours & $0.162$ & $0.150 $& $0.0354$&$0.1328$&$- 0.0216$&0.476  \\
& & & &  & & & \\[-0.2cm]

& &$I_{\HE}$&$0.014 $ & $0.162$ &$0.0509$&$0.1576$&$-0.0424$& 0.096\\
$t_{\star}/ {\rm Gyr}$ & RF & Bands & $0.382$ &$0.128 $ &$0.0285$ & $0.1008$&$-0.0061$& 0.618\\
& &Bands + colours &  $0.381$ &$ 0.129$&$0.0286$ & $0.1007$&$-0.0051$& 0.618\\
  & & & &  & & & \\[-0.2cm]

& &$I_{\HE}$&$ 0.130$ &$ 0.162 $&$0.0498$ &$0.1584$& $-0.0430$ & 0.130\\
& DNN& Bands & $0.382 $ & $0.128 $ & $0.0281$ & $0.0997$ &$-0.0024$ & 0.622 \\
& & Bands + colours  & $0.382  $& $0.128$&$0.0280$ & $0.0994$ &$-0.0025$& 0.623\\
      & & & &  & & & \\[-0.3cm]
        \hline \hline
    \end{tabular}
    \tablefoot{ Left to right: Target; ML algorithm; input features; coefficient of determination ($R^2$); root-mean-square error (${\rm RMSE}$); number of catastrophic outliers ($f_\text{out}$); normalised median absolute deviation (${\rm NMAD}$); and bias ($\left< \Delta Y \right>$, where $Y$ represents the target galaxy property); Pearson $\rho$ coefficient. All values are defined in Sect.\,\ref{sec:ML_metrics}.}
\end{table*}


\section{Results}\label{sec:results_general}

In this section, we describe the results for each of our three target values: stellar mass surface density ($\mass$), stellar-mass-averaged stellar metallicity ($Z_{\star}$), and stellar-mass-averaged stellar age ($t_{\star}$). These values were inferred using three methods: ordinary least squares LR and two ML algorithms, namely, RF and DNN. Each algorithm was trained independently for three different selections of
input values: single $\HE$ band, all four \Euclid\/ bands, and when using four \Euclid\/ bands and six colours modelled from them. Altogether, this totals to 27 separate cases, as shown in Table\,\ref{tab:table_all_results}.

\begin{figure*}
    \centering
    \includegraphics[width=\textwidth]{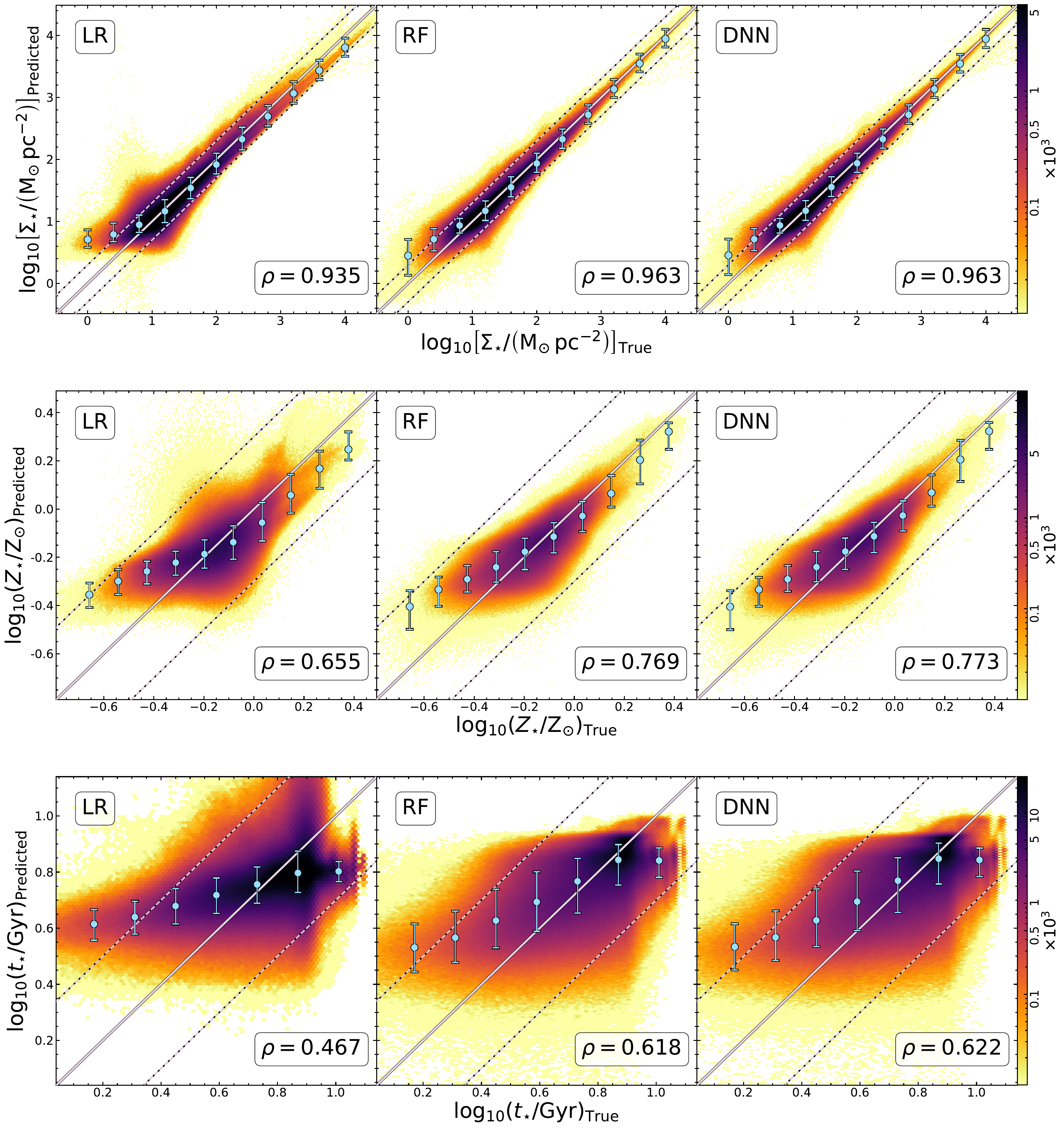}
    \caption{Predictions for $\mass$ (top), $Z_{\star}$ (middle), and $t_{\star}$ (bottom) using the four \Euclid\/\ bands as the input features for the three models: linear regression (left), RF (middle), and DNN (right) as a function of their true value. A diagonal line marks the identity, and the additional two dashed lines mark the limit for the catastrophic outliers (instances where the predicted value is two times larger or smaller than the true value). Running medians, with their 16$^\text{th}$ and 84$^\text{th}$ percentiles are marked with large blue points with error bars.}
    \label{fig:models_all}
\end{figure*}

\subsection{Stellar mass}

Most of the information on stellar mass can be gained from the NIR $1.65 {\rm \, \mu m}$ $\HE$ band \citep{Zibetti_2009,Walcher_2010}.
In our case, when modelling $\mass$ using just one \Euclid\/\ band, the strongest relationship shown is between $\mass$\ and the $\HE$-band surface brightness ($R^2\approx 0.85$), regardless of the algorithm, pointing to a very strong linear relationship between $\HE$ and $\mass$. When we model $\mass$ using all four \Euclid\/\ bands, the increase in accuracy is lower for LR ($R^2 \approx 0.873$), and higher for RF and DNN ($R^2 \approx 0.927$). This indicates that there is a certain non-linear component to the relationship between the other three bands and $\mass$ that is uncovered similarly well by both non-linear algorithms, with the RMSE for DNN with four \Euclid\/\ bands being $0.130 {\rm \,dex}$. The addition of six colours modelled from the four \Euclid\/\ bands does seem to offer a minor benefit ($R^2 \approx 0.908$) for LR, but not for RF or DNN.

We show the relationship between the true $\mass$ and the predicted $\mass$ in the case of using only the four \Euclid\/\ bands for all three methods in Fig.\,\ref{fig:models_all}. The diagonal line marks the identity, with two additional dashed lines on each side that mark $f_\text{out}$ (Eq.\,\ref{eq:f_out}). For higher values of $\mass$, the scatter is fairly low, especially in case of RF and DNN. However, in all three methods, the scatter increases as $\mass$ decreases, especially at $\mass \leq 30 {\rm \,M_{\odot}\,pc^{-2}}$. This is expected, as all our observations were clipped for values $I_\text{\HE} \leq 0.55 {\rm \, MJy\,sr^{-1}}$. With the strong linear relationship between $\mass$\ and \HE, the effect of this clipping is fairly prominent as a lower boundary for inferred $\mass \approx 4 {\rm \, M_{\odot}\,pc^{-2}}$ for LR. Simultaneously, we can expect a stronger effect of Monte Carlo noise at lower surface brightnesses, increasing the uncertainty of ML predictions. For RF and DNN the $f_\text{out} \approx 3\%$, but over double that value ($f_\text{out} \approx 7\%$) for LR (Table\,\ref{tab:table_all_results}).

In conclusion, $\mass$\ can be inferred with a high accuracy from \Euclid\/\ observations using ML (with the Pearson $\rho$ coefficient spanning from 0.923--0.963 in all cases). While the most information is extracted from the \Euclid\/\ \HE band that scales with $\mass$ in a linear fashion, the other three bands offer additional non-linear information that justifies the usage of non-linear ML algorithms such as RF or DNN.

\subsection{Stellar metallicity}

Stellar metallicity can be extracted from the spectral absorption features \citep[line-strength indices, e.g.][]{Maiolino_2019} in the optical range of the spectrum. However many of these optical absorption lines are narrow and shallow, making them difficult to measure \citep{ditrani2023stellar}. For \Euclid\/\, only the $\IE$ band covers the optical range of the spectrum, but due to its broadness, it would make inferring $Z_{\star}$ with \Euclid\/\ imaging very difficult.

The process of predicting $Z_{\star}$ in this work follows similar steps to $\mass$. Out of the single band fits, the most important feature for predicting metallicity is the $\HE$ band ($R^2 \approx 0.36$ for LR and RF, and $R^2 \approx 0.39$ for DNN), pointing to a mild but mostly linear relationship between $Z_{\star}$ and $\HE$. Similarly to $\mass$, the accuracy increases when all four \Euclid\/\ bands are used, and the gain is much higher for RF and DNN ($R^2 \approx 0.595$) than LR ($R^2 \approx 0.417$), pointing to the relationship between $Z_{\star}$ and the other three \Euclid\/\ bands to fall in a more non-linear regime, similarly to $\mass$. The inclusion of colours offers a small benefit to LR ($R^2 \approx 0.435$), but extremely minor to RF, and none to DNN. We recover $Z_{\star}$ with a ${\rm RMSE} < 0.1 {\rm \,dex}$ in all cases, with the best case scenario being with a DNN (${\rm RMSE}= 0.076 {\rm \, dex}$), in case when four \Euclid\/\ bands are used.

We plot the relationship between the true and predicted $Z_{\star}$ in Fig.\,\ref{fig:models_all}. The relationship is somewhat similar to $\mass$, where highest values show the least amount of scatter, which again increases steadily until $Z_{\star} \approx 0.5 {\rm \, Z_{\odot}}$. Compared to the prediction of $\mass$, the inferred value of $Z_{\star}$ shows a much larger scatter. RF and DNN perform significantly better than LR, which seems to struggle with inferring values for $Z_{\star} < 0.4{\rm \, Z_{\odot}}$. Even for higher ranges (where $Z_{\star} \geq 1{\rm \, Z_{\odot}}$), RF and DNN capture the distribution of $Z_{\star}$ much more accurately and with a lot less scatter. When only $\HE$ is used, $f_\text{out}\approx 0.5\%$ (Table\,\ref{tab:table_all_results}), a value similar to all three methods. This value halves with the inclusion of additional data for LR ($f_\text{out}\approx 0.25\%$), and drops seven-fold for RF and DNN ($f_\text{out}\approx 0.07\%$). It is important to note that the range for $Z_{\star}$ is much smaller than $\mass$, making these results seem better in comparison, when in fact $\mass$ is inferred better in all cases. For details on the different ranges of our target values, we refer to Sect.\,\ref{sec:preprocessing}.

We conclude that $Z_{\star}$ can be inferred from \Euclid\/\ observations with a fairly high accuracy, but the prediction is less robust than in case of $\mass$ (Pearson correlation coefficient $\rho = 0.77$ for $Z_{\star}$, compared to $\rho = 0.96$ for $\mass$). Similarly to $\mass$, the information contained in the \Euclid\/\ \HE band is related to $Z_{\star}$ in a linear fashion, while the other three bands contain the information in a more non-linear regime.

\subsection{Stellar age}

The procedure for modelling stellar ages is tightly entangled with $Z_{\star}$ and requires the information located in the 4000\,\r{A} Balmer break \citep{Poggianti_1997} which is outside of the \Euclid observational window for local Universe galaxies. As was the case with $Z_{\star}$, \Euclid\/\ observations do not carry enough information to directly extract stellar ages for our data set either.

When using a single \Euclid\/\ band to model $t_{\star}$, the predictions are equivalent to random chance ($R^2\approx 0$), as shown in Table\,\ref{tab:table_all_results}. The other two attempts yield similar results, with RF and DNN both offering a slightly higher accuracy than LR; however, these values are still low ($R^2 < 0.4$), giving credence to the conclusion that $t_{\star}$ cannot be accurately modelled from \Euclid\/\ observational data alone. As with $Z_{\star}$, RF and DNN can mimic the distribution of $t_\star$ values somewhat, but the accuracy is too low for them to be a feasible method of modelling $t_\star$ with \Euclid\/\ observations.

While the scatter is lower for the higher range of ages, it shows an offset where regardless of the algorithm, $t_{\star}$ is systematically under-predicted, worsening with increasing age (Fig.\,\ref{fig:models_all}). On the lower side of the range, the scatter quickly increases, making predictions for regions with $t_\star \leq 9 {\rm\,Gyr}$ extremely difficult, and for regions with $t_\star \leq 2.5 {\rm \,Gyr}$ predictions are almost impossible. Neither algorithm seems to be able to recover these ages accurately at all. When only $\HE$ band is used to model $t_{\star}$, $f_\text{out} \approx 5 \%$, regardless of the algorithm. Additional features decrease this number somewhat for LR ($f_\text{out} \approx 3.5\%$), and more for RF and DNN  ($f_\text{out} \approx 2.8\%$). However, the range for $t_{\star}$ is very small (spanning from $\sim \,$0 -- 12.7\,Gyr in our entire data set, as discussed in Sect.\,\ref{sec:preprocessing}), making the definition of $f_\text{out}$ very broad in the case of this galaxy property. If we take a narrower limit (for example, $0.2{\rm \,dex}$, i.e. $\sim$\,1.5 times higher or lower than the true values), for the best case scenario (DNN with all input features), $f_\text{out}$ increases to over 10\%.

In conclusion, while some information can be inferred ($\rho \approx 0.62$ for the best case scenario, a DNN using the four \Euclid\/\ bands), $t_{\star}$ cannot be accurately predicted from the \Euclid\/\ observation with ML.

\subsection{Example galaxies}

\begin{figure*}
   \centering
    \includegraphics[width=\textwidth]{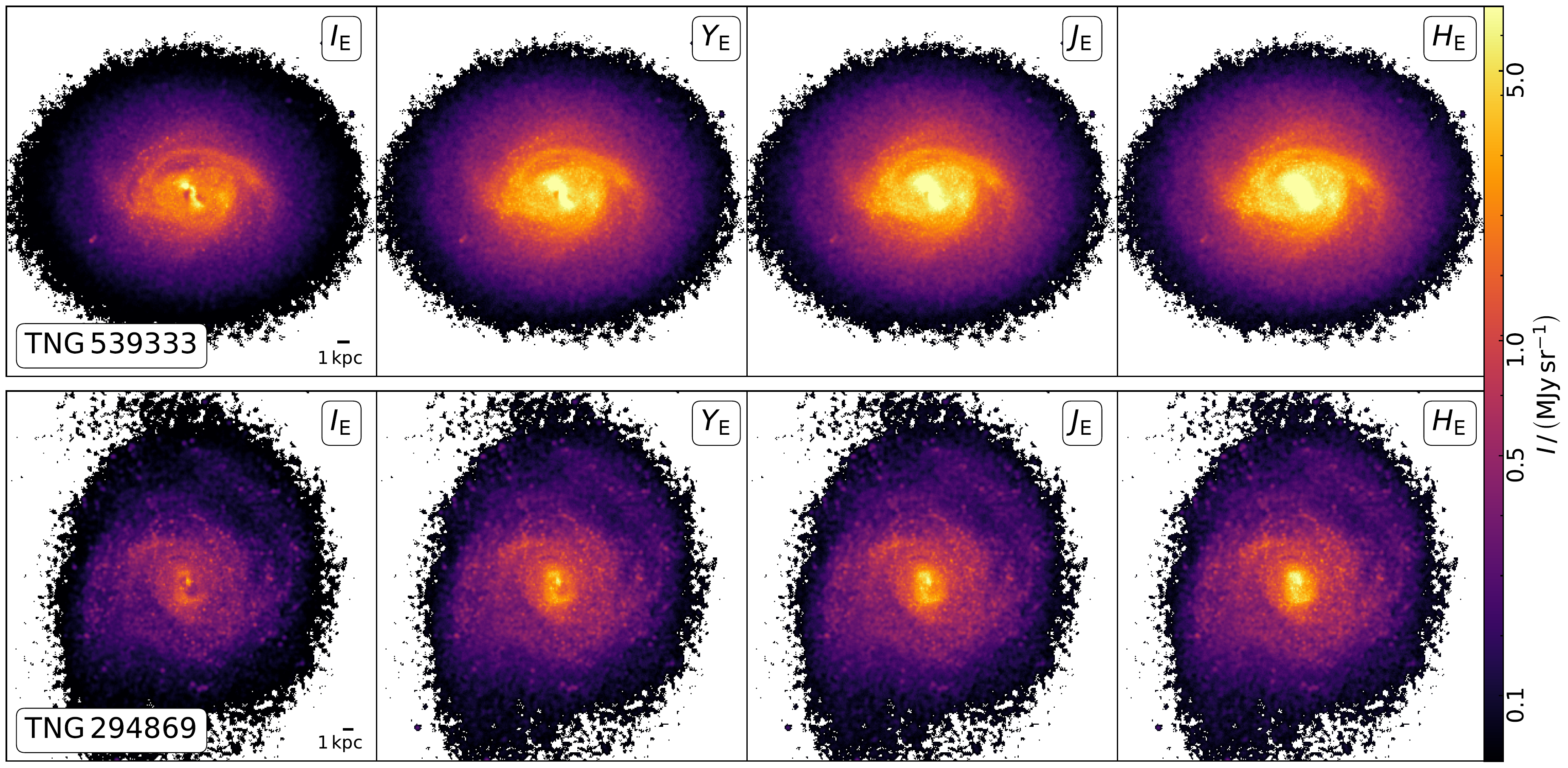}

    \caption{Mock observations for our example galaxies TNG\,539333 (top) and TNG\,294869 (bottom) in four \Euclid\/\ bands. From left to right: \IE, \YE, \JE, and \HE. The bottom-right corner of the leftmost image for each galaxy shows the 1\,kpc scale, which is equal to ten pixels with a 100\,pc side.}
    \label{fig:galaxy_obs}
\end{figure*}

\begin{table*}[]
    \caption{Accuracy scores (as defined in Sect.\,\ref{sec:ML_metrics}) for the three galaxy properties ($\mass$, $Z_{\star}$, and $t_{\star}$) using four \Euclid\/\ bands and a DNN for the two example galaxies TNG\,539333 and TNG\,294869.}
    
    \smallskip
\label{tab:table_single_galaxies}
\smallskip
\centering
\begin{tabular}{l l l c c c c c c}
\hline \hline
  & & &  & & \\[-0.3cm]

Galaxy & Pixel count & Target  &  $R^2$  & percentile & RMSE  & NMAD & $\left< \Delta Y \right>$ & $\rho$\\
 & & & & && & \\[-0.3cm]
 \hline
 & & & & & \\[-0.3cm]
 && $\mass /{\rm  \left(M_{\odot}\, pc^{-2}\right)}$  & $0.973$& 87& $0.078$  & $0.0518$ & $-0.0016$ &0.995\\
  TNG\,539333 & 54\,030& $Z_{\star}$  & $0.682$& 90 &$0.053$ & $0.0502$ & $-0.0259$&0.878 \\
 && $t_{\star} /{\rm Gyr}$  & $-2.653$& 26 &$0.059$ & $0.0628$ & $0.0349$ & 0.013\\

  & & & && &  & \\[-0.3cm]

        \hline
 & & &&& & & \\[-0.3cm]
 &  &$\mass /{\rm  \left(M_{\odot}\, pc^{-2}\right)}$  & $0.890$& 12&$0.115$ & $0.1140$ & $0.0702$ & 0.968\\
  TNG\,294869 & 76\,151& $Z_{\star}$  & $0.192$& 41 &$0.062$  & $0.0629$ & $-0.0337$ &0.693 \\
 && $t_{\star} /{\rm Gyr}$  & $-2.520$& 28 &$0.117$ & $0.1185$ & $0.0717$ & 0.082\\

  & & & & && & \\[-0.3cm]

        \hline \hline
    \end{tabular}

    \tablefoot{
   Left to right: Galaxy; number of pixels; target; coefficient of determination ($R^2$); the percentile of the coefficient of determination for that galaxy; root-mean-square error (${\rm RMSE}$); normalised median absolute deviation (${\rm NMAD}$); and bias ($\left< \Delta Y \right>$, where $Y$ represents the target galaxy property); Pearson $\rho$ coefficient. }
\end{table*}

\begin{figure*}
    \centering
    \includegraphics[width=\textwidth]{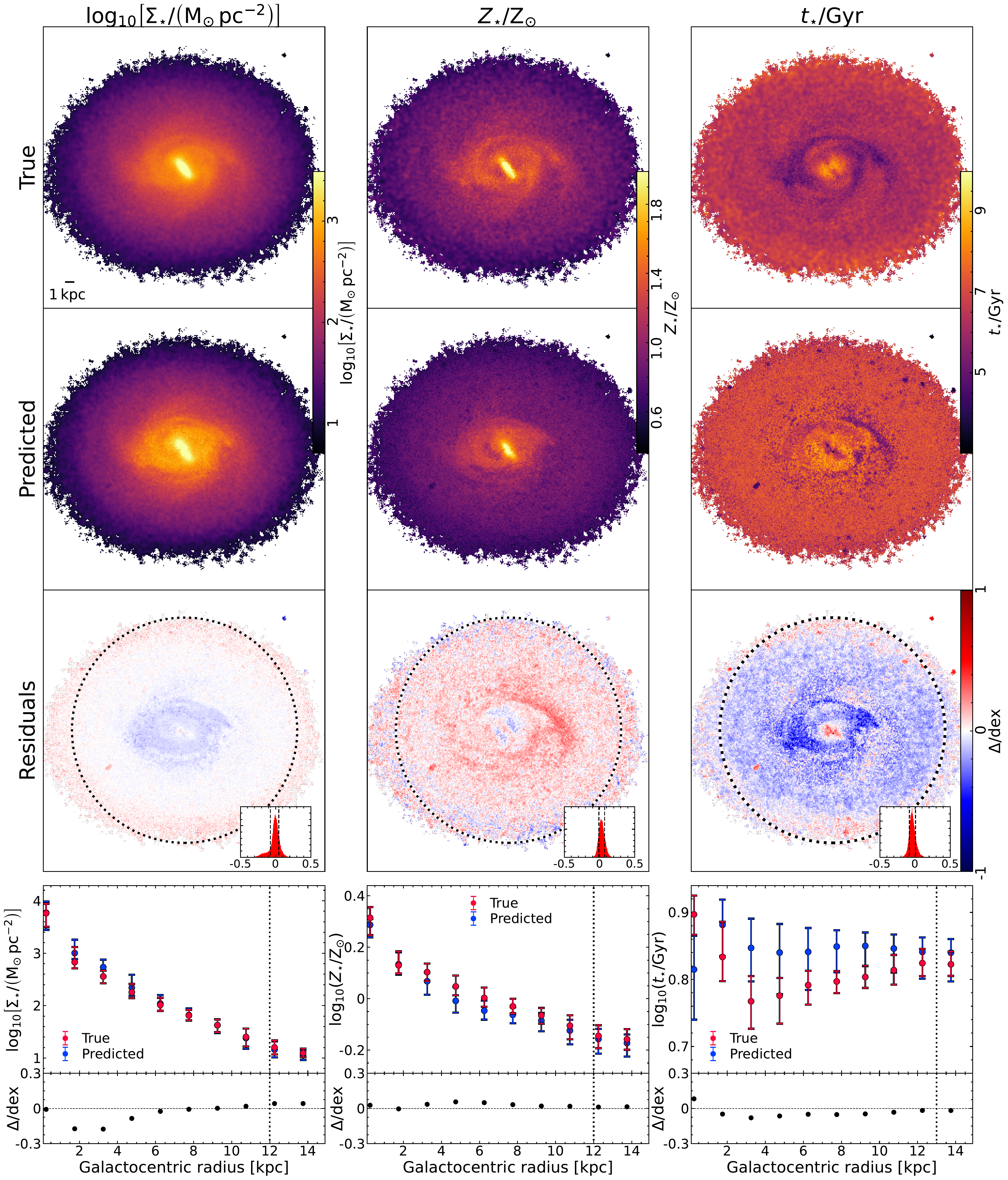}

    \caption{Full disc map of galaxy TNG\,539333 (O5 projection) for $\logten \left[\mass /{\rm \left(M_{\odot}\,pc^{-2}\right)}\right]$\ (left), $\logten \left(Z_{\star} / {\rm Z_{\odot}}\right) $ (middle), and $\logten \left(t_{\star} / {\rm Gyr}\right)$ (right) as an example of a more accurate model. From top to bottom: The top row shows the actual values for each property followed by the model created by a DNN algorithm on four \Euclid\/\ bands, and the third row shows the residuals, where points in red (blue) represent pixels where the actual values are higher (lower) than those in the model. Histograms contained in the lower right corner of the residual maps show the distribution of error prediction ($\logten$ $y_i$ - $\logten$ $\hat{y}_i$, where $y_i$ is the actual $i$-th value, and $\hat{y}_i$ is the predicted $i$-th value of the pixel), dashed lines mark the 16$^\text{th}$ and 84$^\text{th}$ percentile of the distribution for the pixels contained within the dashed circle of a 12\,kpc radius. Bottom row shows radial profiles for the true (red) and predicted (blue) values where large points mark the median, and error bars mark the 16$^\text{th}$ and 84$^\text{th}$ percentile of the distribution contained within the radial bin, as well as their residuals plotted underneath.}
    \label{fig:galaxy_good}
\end{figure*}

\begin{figure*}
    \centering
    \includegraphics[width=\textwidth]{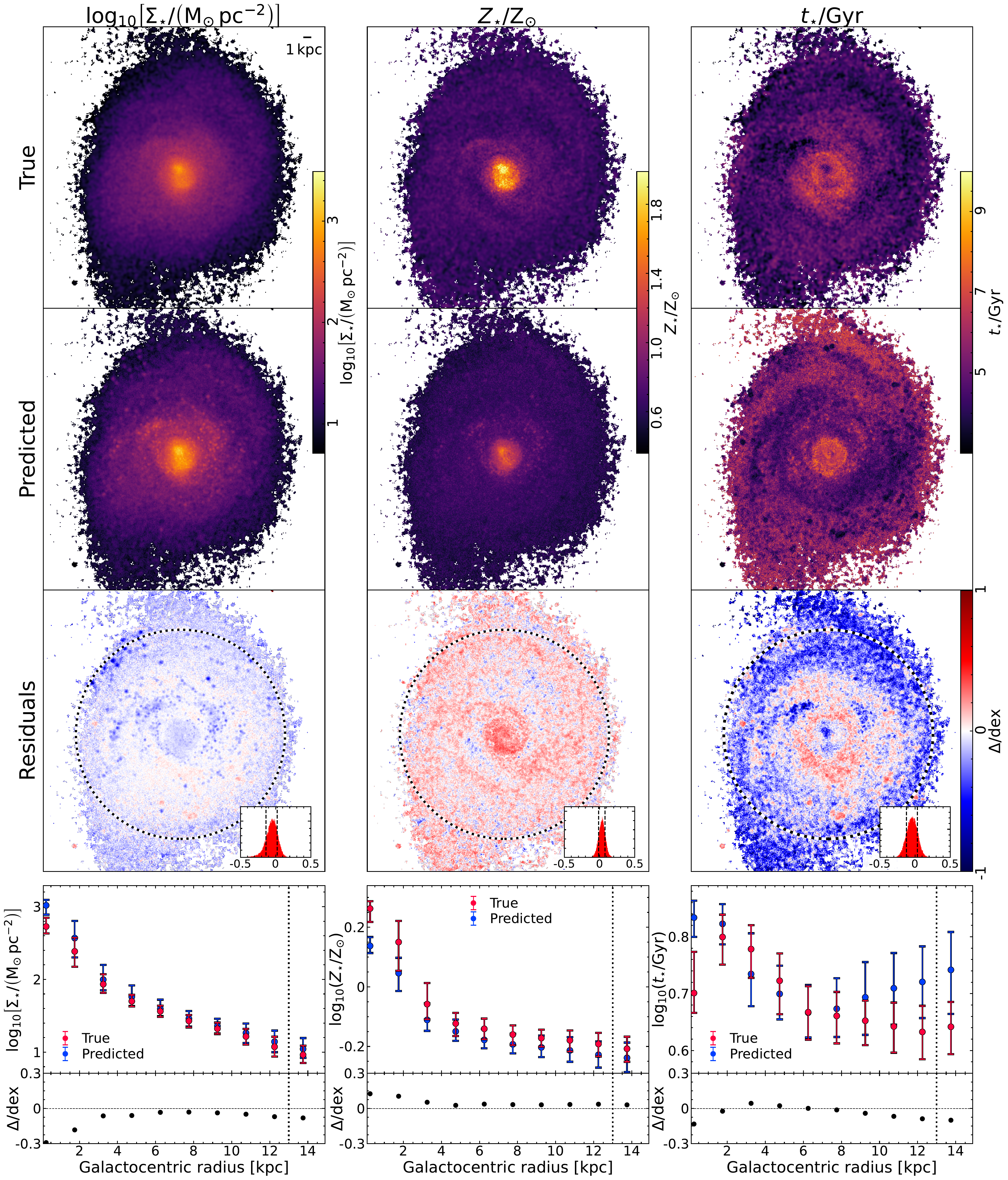}
    \caption{Same as Fig.\,\ref{fig:galaxy_good} but for the galaxy TNG\,294869. The dashed circle marks the 13\,kpc radius. }
    \label{fig:galaxy_bad}
\end{figure*}

In this section we present two galaxies taken from the test set whose properties were inferred with a DNN using the four \Euclid\/\ bands. As a reminder, when splitting our data set into training and test sets, we ensured to select the whole galaxies, meaning the test set consists of all pixels belonging to 115 whole galaxies, which were not involved in training of any of the algorithms.

We present these two galaxies, TNG\,539333 as an example of a more accurately predicted galaxy, and TNG\,294869 as a less accurately predicted galaxy. To select these two galaxies, we used the following procedure: first we chose the top and the bottom quantile of $R^2$ values for $\mass$, and visually inspected them, opting for galaxies that are somewhat face-on, which also have complex features visible; after we located at least 10 of such galaxies for each of the two categories, we randomly selected one. The full disc mock observations in the four \Euclid\/\ bands for these two galaxies can be seen in Fig.\,\ref{fig:galaxy_obs}. We show a full disc map for both the true and predicted values of $\mass$, $Z_{\star}$ and $t_{\star}$, as well as their residuals, and radial profiles which were extracted with circular annuli as both galaxies are relatively face on. We focus our analysis in the inner region marked by the dashed circle and line in the residuals map and radial profile respectively. This area spans the radius equalling 12\,kpc for TNG\,539333 and 13\,kpc for TNG\,294869. Each point in the radial profile represents a median with error bars that mark the $16^\text{th}$ and $84^\text{th}$ percentile of values within a 1.5\,kpc bin of the galactocentric radius. The metrics of accuracy (as described in Sect.\,\ref{sec:methods}) for each galaxy are presented in Table\,\ref{tab:table_single_galaxies}.

\subsubsection{TNG\,539333}

In Fig.\,\ref{fig:galaxy_good}, we plot the full disc of TNG\,539333 as an example of a good prediction by the algorithm. This galaxy is projected face-on, with a total stellar mass of $5.6 \times 10^{10} {\rm \,M_{\odot}}$ and stellar half-mass radius of 4.4\,kpc. It is a star-forming (SFR $\approx 5.54 {\rm \,M_{\odot}\, yr^{-1}}$) galaxy. It has a bright central bar where $\mass$\ is predicted quite accurately (median with 16$^\text{th}$ and 84$^\text{th}$ percentile scatter being $-0.007^{+0.052}_{-0.067}{\rm \,dex}$ for the inner 12\,kpc). The $\mass$\ is mildly overestimated throughout the bright inner region of the disc, overestimating the complex spiral structure seen in the inner part of the disc. This then reverses into a very a mild underestimation for the outer disc, which increases with radius, towards the sparser regions more affected by the Monte Carlo noise. With an $R^2 \approx 0.973$, it is in fact predicted slightly better than the general sample (where $R^2 \approx 0.927$). The radial profile is accurately recovered, with a bright central region, followed by a stronger drop towards the inner $\sim$\,3\,kpc, but it has a generally linear decrease throughout the remainder of the disc. The central region of the modelled profile is very well recovered, followed by an over-prediction that shows the strongest difference with a maximum of 0.18\,dex at around $3\,$kpc. After $\sim 6\,$kpc, the radial profiles mostly begin to overlap, with the difference between them increasing mildly towards the noisy edges of the full disc map.

$Z_{\star}$ is under-predicted through most of the galactic disc ($-0.028^{+0.043}_{-0.039}{\rm \,dex}$), with the inner more metal rich spiral arms not being as prominent in the model as they are on the actual map. The actual map of TNG\,539333 shows the entire disc to be of a higher metallicity than the model, except for the area directly around the galactic bar, which is predicted to be more metal rich. At the very outer, noisy regions, as with $\mass$, the relationship starts to mildly reverse. In essence, while the $Z_{\star}$ map manages to recover a number of galactic features seen in the actual map, they are not as prominent, nor do they extend as far into the disc, and in fact assume a shape more reminiscent of the predicted $\mass$\ map than the features seen in the actual $Z_{\star}$ map. With $R^2 \approx 0.682$, the $Z_{\star}$ is predicted with a higher accuracy than that shown in the general sample ($R^2 \approx 0.595$, as seen in Table\,\ref{tab:table_all_results}). The radial profile is somewhat similar to $\mass$, with a slightly stronger dip separating the central bin from the remainder of the disc where the profile shows a steady decline. The $Z_{\star}$ is systematically under-predicted through the entire profile, with the highest difference being $0.056{\rm \,dex}$ at $\sim 6 {\rm \, kpc}$, prominent as a spiral feature that is not well recovered in the full disc maps.

Of the three properties, $t_{\star}$ is predicted with the lowest accuracy ($-0.041^{+0.048}_{-0.039} {\rm \,dex}$), in fact, with $R^2 \approx -2.653$ (Table\,\ref{tab:table_single_galaxies}), the prediction is worse than that expected of a random chance. Most features in the actual map are not being mimicked by the model. For example, the central region shows several large spiral arms expanding beyond the galactic bar filled with young stars, but the model only manages to mimic the shape of the galactic bar somewhat, and fails at the remainder of the structure, with the only older feature visible being the most prominent, extended spiral arm. This makes the model predict $t_{\star}$ rather accurately around the galactic bar, but it over-predicts the ages strongly outside of this small ring, and under-predicts the ages inside the bar. The algorithm manages to mildly mimic the spiral structure outside of the central region populated with old stars, since the structure is visible both for $\mass$ and $Z_{\star}$, but $t_{\star}$ is over-predicted and very flat throughout the entire disc. The shape of the radial profile is not that well recovered either. We expect an older central bin, followed by a steady, mild decrease to $\sim 4{\rm \,kpc}$, then a mild reversal. The modelled radial profile shows a younger centre which jumps to higher values at around $2\,$kpc, and remains fairly flat throughout the disc. The highest difference ($\sim 0.082{\rm \, dex}$) is seen at the very centre.

We stress once more that $\mass$ spans a wider range compared to $Z_{\star}$ and $t_{\star}$, as described in Sect.\,\ref{sec:preprocessing}. These differences should be considered in regard to the results, meaning that $\mass$ is recovered more accurately than $Z_{\star}$ and $t_{\star}$ even if it presents with a higher scatter. In fact, the Pearson $\rho$ coefficient (Table\,\ref{tab:table_single_galaxies}) shows a strong correlation with $\mass$ ($\rho \approx 0.995$), followed by $Z_{\star}$ ($\rho \approx 0.878$) which is still recovered better than the general sample. However, $\rho \approx 0.013$ for $t_{\star}$ points to stellar ages not being well recovered for that galaxy at all.

\subsubsection{TNG\,294869}

TNG\,294869 (Fig.\,\ref{fig:galaxy_bad}) is an example of a galaxy in which some of the parameters are recovered less accurately. It is a moderately star-forming (SFR $\approx 3.04{\rm \, M_{\odot}\, yr^{-1}}$) galaxy with total stellar mass $2.37 \times 10^{10} {\rm \,M_{\odot}}$ and stellar half-mass radius 8.6\,kpc. This galaxy has a bright centre with visible spiral arms, a feature where the algorithm generally struggles more in other galaxies as well. Both arms extend out of the disc; however, the bottom arm extends further. The $\mass$\ is predicted rather accurately ($-0.052^{+0.074}_{-0.086}{\rm \,dex}$ scatter for the inner 13\,kpc), but on a somewhat lower scale in comparison to TNG\,539333, and with $R^2 \approx 0.890$ (Table\,\ref{tab:table_single_galaxies}), it shows lower accuracy in comparison to the entire test sample. In most regions, except for a few inner spiral features, the algorithm over-predicts $\mass$ and this inaccuracy grows further out, a feature visible in the radial profiles as well. The $\mass$ takes the maximum in the central bin both for the true values, as well as the model, followed by a stronger decrease until $\sim 4 {\rm\,kpc}$ where the decline somewhat evens out. The highest difference ($\sim 0.29{\rm \,dex}$) is seen in the central bin, as the algorithm seems to struggle the most with the central region of the galaxy. With $\rho \approx 0.968$, this galaxy shows a very similar correlation between predicted and true values for $\mass$\ as compared to the full sample.

$Z_{\star}$ model shows a somewhat larger scatter to TNG\,539333 ($0.033^{+0.045}_{-0.047}{\rm \,dex}$). However, as with TNG\,539333, $Z_{\star}$ shows somewhat of a reversal in the predicted values when compared to $\mass$. The innermost area of the galaxy shows a spherical feature populated with high metallicity stars. The shape of this feature is predicted by the algorithm; however, the values of $Z_{\star}$ belong to the most under-predicted values in the entire galaxy. The stark contrast between the central bright feature, followed by a much darker, metal poor disc is not recovered, as it seems that the $Z_{\star}$ map is again a scaled version of the $\mass$\ map, where this contrast between the centre and the remainder of the disc is not as prominent. In fact, a number of bright, overestimated pixels around the bright centre seen on the predicted $\mass$ map are also visible on the $Z_{\star}$ map, albeit scaled down. In all, $Z_{\star}$ is not well predicted, with $R^2 \approx 0.192$ (Table\,\ref{tab:table_single_galaxies}), which is significantly lower than the entire test sample. The radial profile for $Z_{\star}$ assumes a shape similar to $\mass$ but with a steeper drop from the metal-rich central region; however, the area with $r \geq 4{\rm \,kpc}$ is somewhat constant for the rest of the radial profile. The $Z_{\star}$ radial profile is rather well recovered in this area as well, and the highest difference (0.13\,dex) belongs to the central bin. Similar to $\mass$, the Pearson correlation coefficient (Table\,\ref{tab:table_single_galaxies}) $\rho \approx 0.693$ is fairly close to the full test set ($\rho \approx 0.773$).

Surprisingly, the actual spiral structure outside of the very centre seen in the $t_{\star}$ map is copied quite well, albeit mostly in reverse. The algorithm over-predicts the age in the outer parts of the galaxy, while under-predicting it in the spiral arms, with a $R^2 \approx -2.520$ (Table\,\ref{tab:table_single_galaxies}), pointing to the algorithm not being able to predict $t_{\star}$ with any accuracy. Though interestingly, this value is somewhat better than for TNG\,539333. Stellar age seems rather flat in the central region, while the actual map has visible structure, with younger stars concentrated in the very centre, surrounded by a thick disc of older stars. The algorithm shows a younger, albeit much smaller area at the very centre in a few pixels; however, it struggles with the shape of the entire structure. The distant spiral arms in the outermost area of the galaxy show some of the youngest ages in the entire galaxy that the algorithm seems to struggle the most with, showing some of the largest under-predictions. However, it is worth noting, as seen in Fig.\,\ref{fig:models_all}, lower values of $t_{\star}$, especially $t_{\star} \leq$\,2.5\,Gyr, are very badly recovered regardless of the ML algorithm employed. In the case of this galaxy, the outermost regions fall within that range. $t_{\star}$ for TNG\,294869 shows the highest scatter ($-0.044^{+0.076}_{-0.078}{\rm \,dex}$) out of all six cases discussed here, pointing to age being recovered the least accurately when taking into account that it also spans the smallest range. The radial profile is not that well recovered. The central bin is strongly over-predicted, with the highest difference (0.13\,dex) in the entire disc. The model shows a young centre, with age that decreases consistently until $\sim 6{\rm\,kpc}$, where it then reverses and continues increasing, but with a very high scatter in the outer bins. The true values, on the other hand, show a young central bin, followed by a jump in age at the second bin ($\sim 2{\rm\,kpc}$), and a decrease similar to the model. However, this decrease continues throughout the disc with a somewhat less steep fall at $r \geq 4 {\rm\, kpc}$, but there is no reverse as there is in the model at $r \approx 7 {\rm\, kpc}$. With $\rho \approx 0.082$, this galaxy, as well as TNG\,539333 show a much lower correlation for $t_{\star}$ when compared to the full test set.

It is worth noting that $Z_{\star}$ is fairly under-predicted, while $\mass$\ and $t_{\star}$ are both under-predicted for both of our example galaxies. In general, the trend for the entire test set, in case of RF and DNN for when at least all four \Euclid\/\ bands are used, shows the reversal in the bias for $Z_{\star}$ in comparison to $\mass$\ and $t_{\star} $ (Table\,\ref{tab:table_all_results}). Not shown here, but when visually inspecting full disc maps of other test set galaxies, this trend was noticeable, though sometimes it is reversed, that is $Z_{\star}$ may be generally over-predicted, which is followed with $\mass$\ and $t_{\star}$ being under-predicted for the same galaxy.

\subsection{Effects of the projection angle: Face-on versus edge-on}

\begin{table*}[]
    \caption{Accuracy scores (as defined in Sect.\,\ref{sec:ML_metrics}) for the three galaxy properties ($\mass$, $Z_{\star}$, and $t_{\star}$) using four \Euclid\/\ bands and a DNN for the galaxy TNG\,539333 O2 projection and TNG\,294869 O1 projection as an example of edge-on cases.}
    
    \smallskip
\label{tab:edge_on}
\smallskip
\centering
\begin{tabular}{l l l c c c c c}
\hline \hline
  & & &  & & \\[-0.3cm]

Galaxy & Pixel count & Target  &  $R^2$  & RMSE  & NMAD & $\left< \Delta Y \right>$ & $\rho$\\
 & & & & && & \\[-0.3cm]
 \hline
 & & & & & \\[-0.3cm]
 && $\mass /{\rm  \left(M_{\odot}\, pc^{-2}\right)}$  & $0.971$&  $0.098$  & $0.0852$ & $-0.0489$ &0.994\\
  TNG\,539333 & 30\,171& $Z_{\star}$  & $0.665$& $0.081$ & $0.0673$ & $0.0394$&0.893 \\
 && $t_{\star} /{\rm Gyr}$ & $-1.522$&  $0.097$  & $0.0878$ & $-0.0155$ &-0.345\\

  & & & && &  & \\[-0.3cm]

        \hline
 & & &&& & & \\[-0.3cm]
 &  &$\mass /{\rm  \left(M_{\odot}\, pc^{-2}\right)}$  & $0.963$& $0.093$ & $0.0991$ & $0.0298$ & 0.983\\
  TNG\,294869 & 38\,116& $Z_{\star}$  & $0.175$& $0.089$  & $0.1041$ & $0.0664$ &0.769 \\
 && $t_{\star} /{\rm Gyr}$  & $-0.918$&$0.130$ & $0.1479$ & $0.0778$ & 0.015\\

  & & & & && & \\[-0.3cm]

        \hline \hline
    \end{tabular}
    \tablefoot{
   Left to right: Galaxy; number of pixels; target; coefficient of determination ($R^2$); root-mean-square error (${\rm RMSE}$); normalised median absolute deviation (${\rm NMAD}$); and bias ($\left< \Delta Y \right>$, where $Y$ represents the target galaxy property); Pearson $\rho$ coefficient.}
\end{table*}

\begin{figure*}
    \centering
    \includegraphics[width=\textwidth]{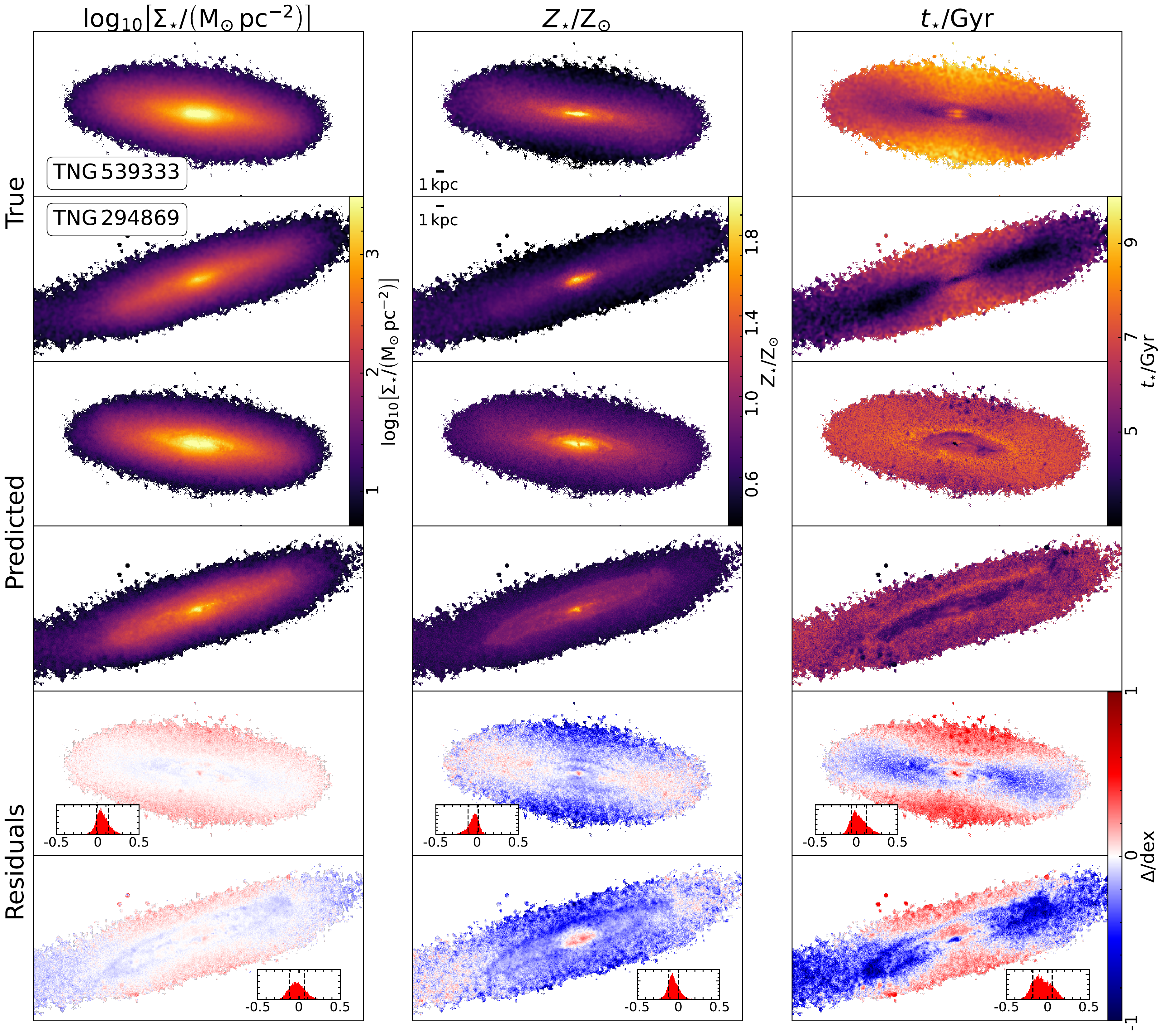}

    \caption{Full disc map of galaxy TNG\,539333 O2 projection (upper) and TNG\,294869 O1 projection (lower) for $\logten \left[\mass /{\rm \left(M_{\odot}\,pc^{-2}\right)}\right]$\ (left), $Z_{\star} / {\rm Z_{\odot}} $ (middle), and $t_{\star} / {\rm Gyr}$ (right) as an example of edge-on projection. From top to bottom: The top row shows the actual values for each property followed by the model created by a DNN algorithm on four \Euclid\/\ bands, and the bottom row shows the residuals, where points in red (blue) represent pixels where the actual values are higher (lower) than those in the model. Histograms contained in the lower left or right corner of the residual maps show the distribution of error prediction ($\logten$ $y_i$ - $\logten$ $\hat{y}_i$, where $y_i$ is the actual $i$-th value, and $\hat{y}_i$ is the predicted $i$-th value of the pixel), dashed lines mark the 16$^\text{th}$ and 84$^\text{th}$ percentile of the distribution.}
    \label{fig:edge_on}
\end{figure*}

When inferring galaxy properties, the orientation of the observed galaxy should be taken into account. Each of the galaxies in the TNG50-SKIRT atlas is projected at five random angles fixed with respect to the simulation volume. When training our machine learning algorithm, we only use one image per galaxy; however, in case of our two example galaxies, we can directly compare inferred galaxy properties both in face-on and edge-on cases, as shown in Fig.\,\ref{fig:edge_on}.

The total accuracy scores for the edge-on case (Table\,\ref{tab:edge_on}) do not deviate much from the face-on case for either galaxy; however, the distribution of the pixels throughout the galactic disc does. For example, in TNG\,539333, $Z_{\star}$ is under-predicted throughout most of the disc in the face-on projection, but in the edge-on case, it shows a strong over-prediction above and below the galactic disc. Similarly, $t_{\star}$ is over-predicted through most of the disc, which slowly turns into an under-prediction at the very edges in the face-on case, but in the edge-on case we can also observe a strong under-prediction above and below the disc.

TNG\,294869 shows a similar behaviour as well, where the edge-on case prediction behaves quite similarly to its face-on counterpart, while showing a strong deviation from the fit in the area above and below the galactic disc. In these two example galaxies we can conclude that, even if the DNN can predict galaxy properties quite well, as is the case for $\mass$, for example, it struggles in the outer parts of the edge-on cases. In fact, it is the edge-on cases where it becomes very prominent that the $Z_{\star}$ and $t_{\star}$ maps behave in a manner very similar to a scaled version of the $\mass$\ map. This shows quite strongly in the residuals maps for $Z_{\star}$ and $t_{\star}$, specifically as the true values for these two properties have a much stronger gradient throughout the galactic disc when compared to $\mass$, causing a very strong under- or over-prediction in the edge-on cases of both galaxies.

\section{Discussion}\label{sec:discussion}

\subsection{Non-linearity and the importance of colour}

We use three different methods to predict $\mass$, $Z_{\star}$, and $t_{\star}$, and to test the linearity of their relationship in regard to \Euclid\/\ observations. We compare our results in three different approaches: using a single (best) \Euclid\/\ band which in all cases proves to be $\HE$ (though in the case of stellar age, a single band cannot be used to model it, with $R^2 \approx 0$ in all cases), using all four bands, and using ten features, consisting of four bands and six colours modelled from them. The choice for this approach was partly inspired by \citet{Acquaviva_2015}, where the author shows that it is possible to recover gas-phase metallicity from SDSS five-band photometry with a ${\rm RMSE} < 0.1\,{\rm \,dex}$, regardless of the ML algorithm chosen. They attempt to include both colours and squared colours into their input features, and show that, after $M_{\star}$, $(g-i)$ colour is the most important feature for low-$z$, and $(g-z)^2$ for high-$z$ test sets. Our results behave differently, as the inclusion of colour does not offer any additional benefit for any of the galaxy parameters (at least in the case of RF and DNN, which out-compete LR in every case). We believe that the difference lies in the fact that \citet{Acquaviva_2015} was using optical colours, which contain information on metallicity \citep{Sanders_2013}, while in our case, we are using IR colours which do not seem to contain any additional information that the algorithms cannot already glean from the information contained in the four \Euclid\/\ bands. While in this work we focus on the four \Euclid\/\ bands, as well as colours modelled from them, we did also test whether additional information can be inferred from squared colours. The results are shown in Appendix\,\ref{app:c2}. The improvements to the accuracy of our method are negligible, and for that reason we do not include them in the main analysis.

Unfortunately, it can be difficult to disentangle the underlying processes that drive a ML algorithm to make the predictions that it makes. However, in this case, we assume that the reason why the six additional colours do not offer any additional benefit comes from the fact that they are easily derived from the four \Euclid\/\ bands (essentially by dividing two input features), which should make them easily inferred by the algorithm, and thus they do not offer any new information that cannot already be extracted from the four bands during training. We test this by building a model that predicts \Euclid\/\ colours from the four bands, and in all cases the algorithm converges very quickly with the worst accuracy ($R^2 = 0.9981$, $ {\rm RMSE} = 0.0113 {\rm \, dex}$) belonging to $\IE-\HE$. The results can be seen in Appendix\,\ref{app:colors}. Either way, the addition of six colours offers negligible benefits for RF at best, and while in the case of DNN it helps the network to converge faster during training, their inclusion becomes negligible during deployment, and then it even comes with an additional computational cost, however minor, that would have to be expended on the creation of the colour maps for the new data. That is why, in the following discussion, we focus solely on the case of the four \Euclid\/\ bands.

Regardless of the choice of ML algorithm, $\mass$ of the simulated galaxy is predicted with a very high level of accuracy ($R^2 = 0.927$, $ {\rm RMSE} = 0.130 {\rm\,dex}$, $\rho = 0.963$), followed by $Z_{\star}$ ($R^2 = 0.595$, ${\rm RMSE} = 0.076 {\rm \,dex}$, $\rho = 0.773$), then $t_{\star}$ ($R^2 = 0.382$, ${\rm RMSE} = 0.128 {\rm \,dex}$, $\rho = 0.623$) which neither of the approaches is able to predict accurately. Of course, we would like to stress once more that $\mass$ range spans several orders of magnitude more than the other two properties (Sect.\,\ref{sec:preprocessing}), making the $Z_{\star}$ and $t_{\star}$ models seem better in comparison in regard to the values of RMSE if that is not taken into account. Additionally, the Pearson correlation coefficient, $\rho$ shows a clearer picture, pointing to a strong correlation between inferred and real $\mass$, followed by $Z_{\star}$, and then $t_{\star}$ which is very hard to infer with ML.

$\mass$\ shows a very strong linear relationship with the $\HE$-band surface brightness, and this is a preferred band regardless of the choice of the algorithm. The accuracy increases with the addition of the other three \Euclid\/\ bands, but the increase is higher in the case of RF and DNN, pointing to the fact that their relationship to $\mass$ comes with more non-linear features. The case is similar for $Z_{\star}$; however, the accuracy using only the $\HE$-band surface brightness is low enough that the addition of the other three bands creates a much better prediction, pointing to a stronger non-linear relationship between the features. ML performs poorly inferring $t_{\star}$, regardless of the choice of algorithm.

\subsection{Mass-metallicity relation}

\begin{figure*}
    \centering
    \includegraphics[width=0.95\textwidth]{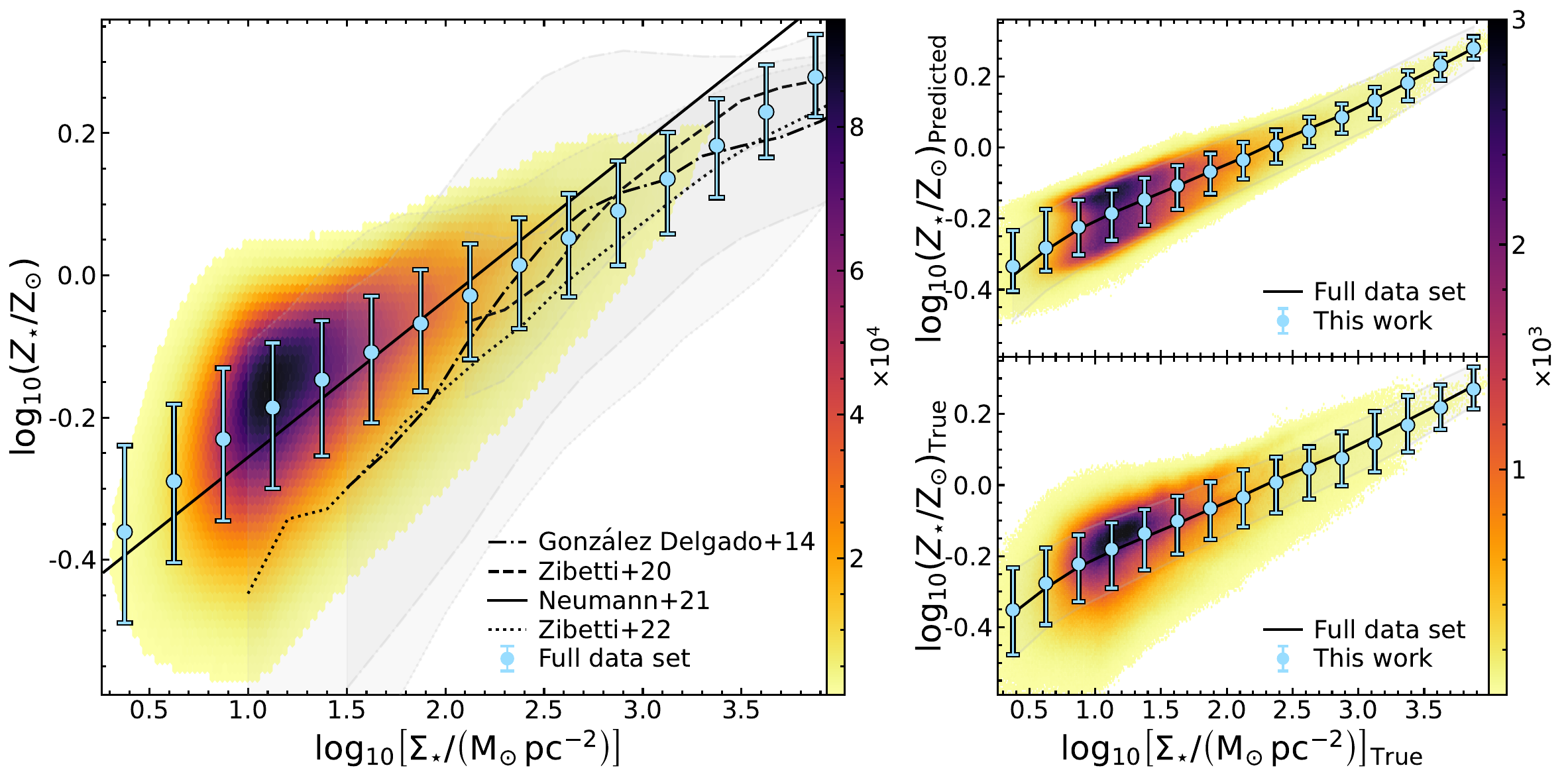}
    \caption{Density plots of the distribution of $\logten \left(Z_{\star} /{\rm Z_{\odot}}\right)$ as a function of $\logten \left[\Sigma_{\star} / {\rm \left(M_{\odot}~pc^{-2}\right)}\right]
    $. Large light-blue points with error bars represent the median, 16$^\text{th}$ and 84$^\text{th}$ percentile of the values of $\logten Z_{\star}$ per each $\logten \Sigma_{\star}$ bin. \emph{Left}: The entire data set, consisting of 71\,120\,654 individual pixels. Only areas with $\geq$\,1000 pixels are included. Full, dashed, dash-dotted, and dotted lines represent fits from \citet{Delgado_2014}, \citet{Zibetti_2019}, \citet{Neumann_2021}, and \citet{Zibetti_2022}, respectively. \emph{Right}: The test data set (consisting of 9\,035\,270 pixels from 150 galaxies), predicted (top) and true (bottom) $Z_{\star}$ as a function of true $\mass$. Only areas with $\geq$\,10 pixels are included. Black line and the grey area marks the median, 16$^\text{th}$ and 84$^\text{th}$ percentile of the whole TNG50 data set (the light-blue points with error bars from the plot on the left), while the light-blue points with error bars represent the median, 16$^\text{th}$ and 84$^\text{th}$ percentile of $\logten Z_{\star}$ for each $\logten \mass$\ bin in the test set.}
    \label{fig:mass_metallicity}
\end{figure*}

Generally, stellar metallicity is a difficult quantity to model, as it requires high-resolution observations in the optical range. The fact that any of the ML algorithms are capable of modelling it with such accuracy from \Euclid\/ observations requires a more focused discussion.

Stellar metallicity scales primarily with stellar mass (and quantities strongly related to it). This so-called stellar mass-metallicity relation \citep[sMZR; for an overview, see Sect.~5.1.1 in][]{Maiolino_2019} is originally based on global values due to the historical lack of spatially resolved data \citep[e.g.][]{Delgado_2014}, which makes it difficult to determine the influence of local processes within galaxies. This relationship also appears when modelling stellar metallicity using ML. For example, \citet{Simet_2021} use a neural network to predict the distribution of metallicity on semi-analytic galaxies tuned to match CANDELS observations \citep{Lu_2014}, and they find a strong relationship between stellar mass and stellar metallicity in the simulations, which influences the accuracy of their stellar metallicity predictions, even though the algorithm itself is not aware of the complexity of underlying processes which guide this relationship. \citet{Acquaviva_2015}, building on the work by \citet{Sanders_2013} on the SDSS-DR7 \citep{Abazajian_2009} data, test the non-linearity of a relationship between a number of input features and gas-phase metallicity on the same data set. In the performance diagnostics of their non-linear extremely randomised trees (ERT) estimator, they show that mass is the most important feature for predicting gas-phase metallicity; however, information about luminosity, colours, and squared colours is required to better constrain the relationship.

Our data is spatially resolved at $100 {\rm \, pc} \times100 {\rm \, pc}$ pixel scale, and yet we notice a similarly strong relationship between $\mass$ and $Z_{\star}$ that positively affects the accuracy of predicting stellar metallicity. While the relationship between stellar mass and gas-phase metallicity in the TNG has been discussed before \citep{Torrey_2019}, in this section we focus on the relationship between $\mass$ and $Z_{\star}$ found in the simulation. We plot $Z_{\star}$ against $\mass$\ in Fig.\,\ref{fig:mass_metallicity} for our entire data set (left), as well as for the test data set, plotting the predicted $Z_{\star}$ values (top right), as well as the true $Z_{\star}$ values (bottom right), against true $\mass$. We calculate a running median, 16$^\text{th}$ and 84$^\text{th}$ percentile (light-blue points with error bars) for each of the $\mass$\ bins, showing a strong relationship within the TNG50 simulation itself. We notice a tighter relationship between the predicted $Z_{\star}$, when compared to the true $Z_{\star}$, and $\mass$, giving credence to our assumption that the surprisingly high accuracy of modelling $Z_{\star}$ is explained with this underlying $\mass$-$Z_{\star}$ relation.

We also plot a number of examples from the literature over the entire data set. For example, \citet{Delgado_2014} investigate the relationship between stellar metallicity and local mass surface density in 300 CALIFA integral survey \citep{S_nchez_2012} spheroidal and disc dominated galaxies. They find that metallicity in galaxies is driven by both global ($M_{\star}$) and local ($\Sigma_{\star}$) processes, and conclude that the local $\Sigma_{\star}$ is an important driver of $Z_{\star}$ in galaxies everywhere but their innermost, densest regions. \citet{Zibetti_2022} use more than 600\,000 individual regions of $\sim$\,1\,kpc linear size from a sample of 362 galaxies from the CALIFA integral survey as well, and investigate how local properties are affected by both the local stellar mass surface density, as well as the global stellar mass. The authors derive a local stellar mass surface density-\textit{r}-band light weighted metallicity relation with a $< 0.1{\rm \,dex}$ scatter, independent of stellar age, beyond a slightly higher ($\leq 0.1 {\rm \,dex}$) metallicity at younger regions. In an earlier work, \citet{Zibetti_2019} already analysed a sample of 69 early-type galaxies from the CALIFA survey, noting that the stellar metallicity and local mass surface density relation is tighter ($0.07 {\rm\,dex}$) than the relation between galactocentric radius and $Z_{\star}$ ($0.10 {\rm \,dex}$), which points to stellar mass surface density being a better predictor of stellar metallicity than radial distance. \citet{Neumann_2021} analyse more than 2.6 million spatial bins from 7439 nearby galaxies in the SDSS-IV MaNGA survey \citep{Bundy_2015} and derive a local stellar mass surface density-metallicity relation (rMZR), while also noting a positive trend of stellar age with both stellar mass surface density and stellar metallicity. \citet{Baker_2022} discuss local mass-gas metallicity relation as well, and mention gas-stellar metallicity correlation derived by \citet{Kauffmann_2003}. However, \citet{S_nchez_2020} warn that the local stellar metallicity-$\mass$\ depends strongly on galaxy mass and morphology, properties which we do not consider in this procedure.

MZR has been taken as prior in some SED fitting codes on global scales (e.g. \citealp{Webb_2020, Bellstedt_2021, Thorne_2022}; Euclid Collaboration: Abdurro'uf et al., in prep), with some promising results, especially in cases where photometry is too poor to measure metallicity. However, while the derived relations may show marginally improved results in extracting stellar metallicity, a caution should be taken since the relation itself can be affected by sample selection and methodology used.

While taking into account a number of issues with photometric based SED fitting \citep[see, for example][and references within]{Nersesian_2023}, the strong relationship we find between $\mass$\ and $Z_{\star}$ on local scales in TNG50, as well as in several surveys outlined earlier makes for an interesting case for the addition of a rMZR prior to the SED fitting procedure to better constrain stellar metallicities on resolved scales.

\subsection{Linear regression as a baseline}\label{sec:ml_alg_dependence}

\begin{table}[]
    \caption{Coefficients for the ordinary least squares LR, as used in Eq.\,(\ref{eq:lr}), with \Euclid\/\ bands in units of ${\rm MJy \,sr^{-1}}$.}
    \centering
    \begin{tabular}{l r r r r r}
    \hline \hline
  & & & & & \\[-0.3cm]

         $\hat{Y}$ &  $I_\text{\IE} $ & $I_\text{\YE}$ &$I_\text{\JE}$ &$I_\text{\HE}$&    $b$ \\
  & & & & & \\[-0.3cm]
         \hline 
  & & & & & \\[-0.3cm]
        $\mass /{\rm \left(M_{\odot}\, pc^{-2}\right)}$ &$-11.113$ & $4.941$& $2.156$&$7.924$& $1.446$ \\
          $Z_{\star}$ & $-4.823$& $0.549$&$-0.318$ &$3.178$ & $0.045$\\
          $t_{\star} /{\rm Gyr}$ & $-8.035 $&$1.058$ &  $0.031$ &$3.870$ & $3.650$\\
  & & & & & \\[-0.3cm]
          \hline \hline
    \end{tabular}
    \label{tab:coeff}
   
\end{table}

Using ordinary least squares LR, which assumes a simple linear relationship between the input features and the output target, we can extract the slopes for each of our input parameters, as well as the intercept, used to model our target feature
\begin{equation}
    \logten \hat{Y} = \sum_{i=1}^{4} {\rm \,} a_i \logten X_i + b \quad ,
    \label{eq:lr}
\end{equation}
where $\hat{Y}$ is the modelled galaxy property (target: $\mass /{\rm \, \left(M_{\odot} \, kpc^{-2}\right)}$, $Z_{\star}$, or $t_{\star}/{\rm  Gyr }$), and $X$ are the observational surface brightnesses of the four \Euclid\/\ bands (input features: $\IE$, $\YE$, $\JE$, and $\HE$) in units of ${\rm MJy \, sr^{-1}}$. We show these values in Table\,\ref{tab:coeff}. $\HE$ most strongly correlated with all three galaxy properties, while $\IE$ shows the strongest anti-correlation. However, as discussed in Sect.\,\ref{sec:results_general}, $\HE$ shows a strong linear relationship with $\mass$, while other relationships fall into a more complex, non-linear regime.

\subsection{Interpretability of machine learning predictions}\label{sect:interpr}

Machine learning could offer a fast alternative to traditional methods when predicting galaxy parameters from mock observations, and with modern libraries does not come with a steep learning curve. Prior to our choice of RF, and DNN as the algorithms used to constrain galactic parameters from \Euclid\/\ observations, we tested a number of other ML algorithms (AdaBoost, Extremely randomised trees, Ridge regression, Support vector machine, and MLPRegressor from the {\tt{scikit-learn}} Python library) on a subset of our data set, and they all showed a similar level of accuracy with the default values of their hyper-parameters. It is highly unlikely that the choice of algorithm would have a drastic effect on the final result, with the exception of the linear methods, as a number of relationships within our data set are non-linear, an effect that most likely stems from the fact that galaxies are complex systems with a complex formation and evolution history.

To test the non-linear component we choose RF, as it is a fairly beginner-friendly algorithm well guarded from user error, and is inherently good at avoiding over-fitting. In our case it also shows really high accuracy that can compete with a far more complex and computationally demanding DNN. RF takes up to several hours to train on an average desktop machine on a sample of over 1000 resolved galaxies when used out-of-the-box, and in this work we dedicated an effort to refine its hyper-parameters until the learning process took less than an hour without any losses to the accuracy.

In fact, due to its speed and interpretability, it should be a preferred algorithm for this work, since a DNN introduces a lot of complexity that is much harder to disentangle with the current tools. One of the main issues with using ML in general arises due to the opaque way in which ML algorithms operate. The goal of this work also extends to understanding the relationship between the observations and the galaxy properties that were mapped to them. 

\subsubsection{Random forest and feature importance}\label{sec:feature_imp}

\begin{figure}
    \centering
    \includegraphics[width=\columnwidth]{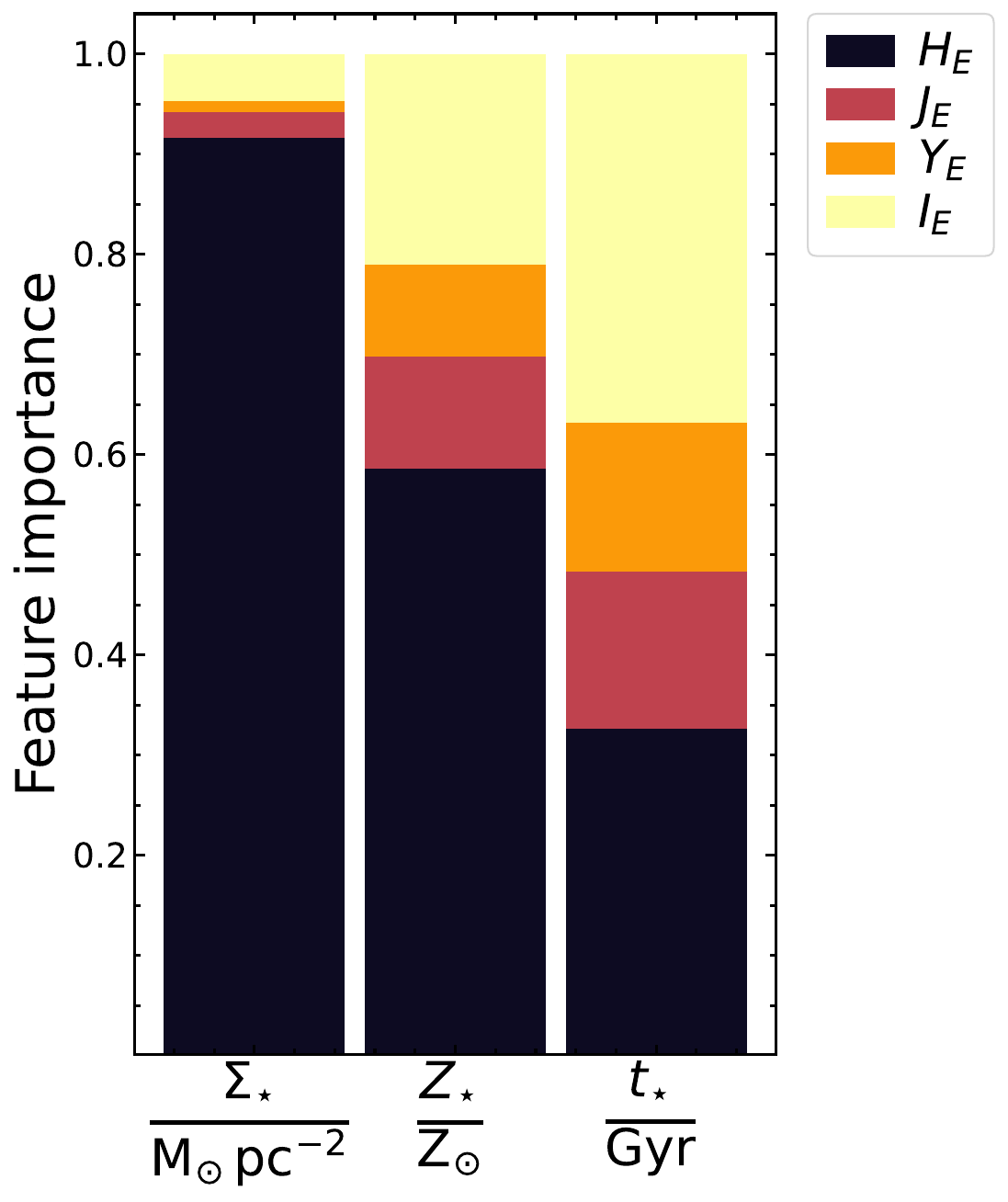}
    \caption{Feature ranking for RF predictions of $\mass$, $Z_{\star}$, and $t_{\star}$. $\HE$ is ranked as the most important feature for $\mass$ and $Z_{\star}$ (91.5\% and 58.6\% respectively), while $\IE$ (36.8\%) and $\HE$ (32.7\%) are the two most important features for $t_{\star}$.}
    \label{fig:features}
\end{figure}

As an ensemble method, RF offers feature importance \citep[Gini index;][]{Jost_2006}, which ranks the features according to how much they contribute to the final prediction of the algorithm. The algorithm also introduces a certain level of randomness into feature selection, meaning that at each start of training, the algorithm chooses a new, random subset of input features, thus avoiding the possibility of a strong feature taking over during training, but also removes the requirement of having to shuffle features when loading them into the algorithm to avoid the risk of the first loaded feature being assigned higher importance than the ones loaded later. The values for feature importance are shown in Fig.\,\ref{fig:features}. $\HE$ ranks as by far the most important feature for $\mass$, and as we have seen earlier, mostly relates to $\mass$\ in a linear regime; however, the rest of the bands do offer an additional non-linear information that increases the predictive power for RF. The situation is somewhat similar, albeit weaker, for $Z_{\star}$, which is an expected result due to the resolved mass-metallicity relation inherent in the cosmological simulation. None of the \Euclid\/\ bands show a strong importance in modelling $t_{\star}$, and $t_{\star}$ cannot be recovered accurately with ML from \Euclid\/\ observations.

\subsubsection{Choice of algorithm and missing data}

In addition to feature importance offered by the RF, we also used the leave-one-feature-out (LOFO) method to assess the amount of information contributed to our result from each of the four \Euclid\/\ surface brightnesses. In this method, we repeat the ML algorithm training process for each of the three galaxy parameters, but each time we leave out one of the four input features and compare the changes in the accuracy on a metric of choice. In our case, we decided on RMSE:

\begin{equation}\label{eq:lofo}
    {\rm LOFO_{X}} = \frac{{\rm RMSE_{X}} - {\rm RMSE}}{{\rm RMSE}}\quad ,
\end{equation}
where X is one of the input features left out during training (\IE, \YE, \JE, or \HE). We show the results in Table\,\ref{tab:lofo}. We repeat the same procedure for both our ML algorithms, RF and DNN. For RF, the accuracy of the prediction is most strongly affected with the removal of \IE, with ${\rm LOFO}_{\IE}$ being at least several times higher than the next highest value), and then \HE, but in the case of the latter, the decrease is very minor. This is followed by \JE, and then \YE, whose removal has the smallest effect on RMSE for each of the three galaxy parameters. This may seem to contradict the feature importance values shown in Sect.\,\ref{sec:feature_imp}, pointing to the necessity of using several methods when trying to interpret a ML model.

\begin{figure*}
        \centering
    \includegraphics[width=\textwidth]{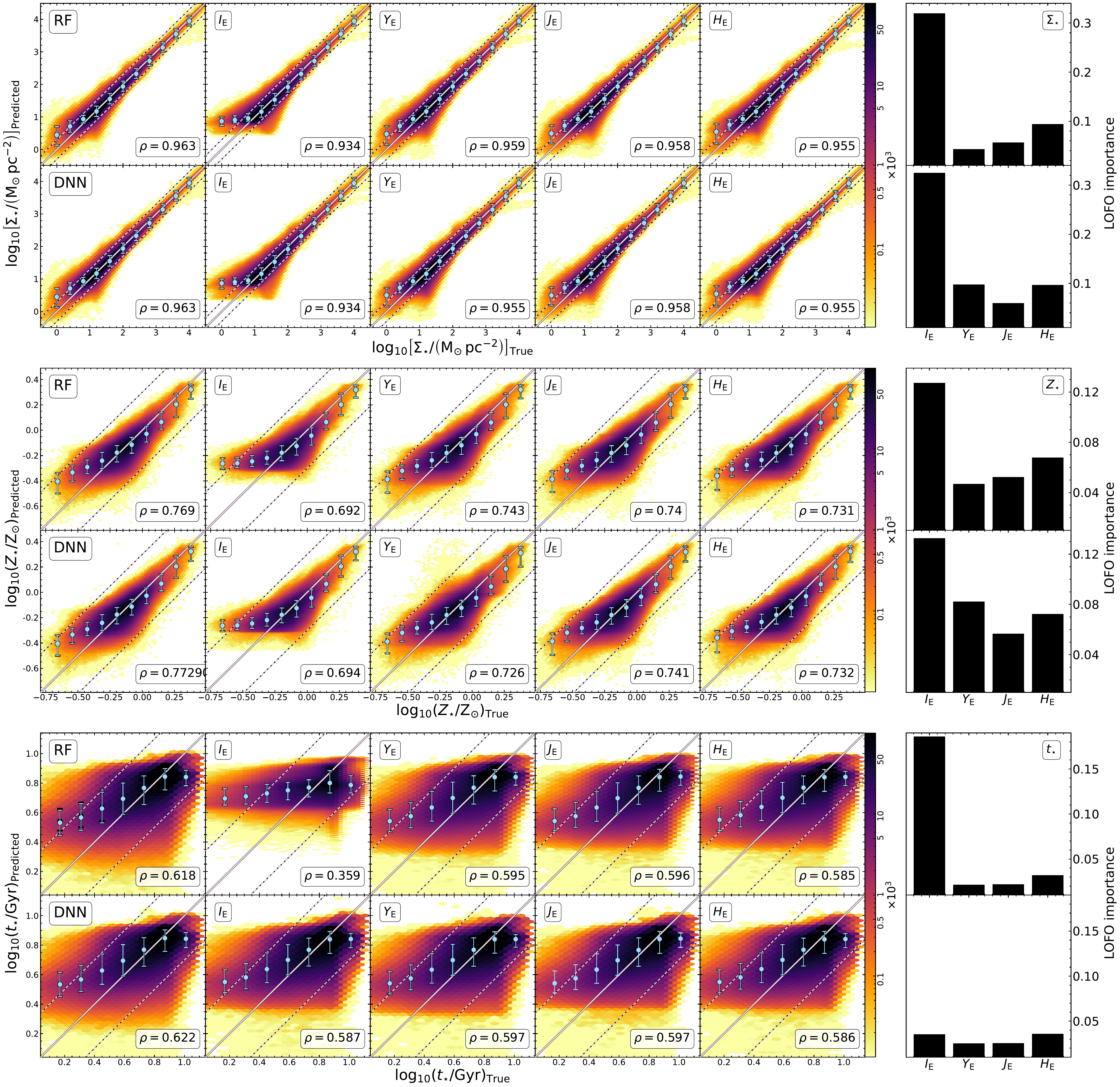}
    \caption{Same as Fig.\,\ref{fig:models_all} but for LOFO analysis. Left to right: Predicted value when all four \Euclid\/\ bands are used followed by four cases where the model was trained without one of the \Euclid\/\ bands (marked in the upper-left corner). The LOFO importance is shown at the far right, calculated as Eq.\,(\ref{eq:lofo}). For each of the three target values, the top row shows the case when using RF and the bottom row when using DNN.}
    \label{fig:lofo}
\end{figure*}

We explore this further in Fig.\,\ref{fig:lofo}, where we plot the distribution of both the true and the inferred values for our entire data set. There it becomes obvious that the removal of \IE mostly affects the lower part of the distribution in case of all three galaxy parameters, while the effect of other three \Euclid\/\ bands seems very minor and spread through the distribution. Only in case of $Z_{\star}$ does \HE have a prominent effect mostly on the lowest end.

DNN behaves quite similar to RF for most of our target values, except in the case where we try to predict $t_{\star}$ without including the \IE. As mentioned earlier, $t_{\star}$ cannot be inferred accurately even when using the entire data set, and the removal of \YE, \JE, or \HE\ has only a minor effect on the already quite low accuracy. However, as most information on the lower end of the distribution seems to be contained within \IE, this effect is very strong in the case of RF, where the Pearson $\rho$ coefficient drops from 0.6 to 0.36, while in the case of DNN $\rho$ remains quite high at 0.59. We believe this makes a strong case for considering different algorithms when trying to infer data with ML. While both RF and DNN show similar results when using the entire data set and, as we mentioned earlier (Sect.\,\ref{sect:interpr}), due to its speed and better interpretability, RF should be the better choice of an algorithm for our work, LOFO analysis shows the superior power of neural networks when managing missing data.

\begin{table*}[]
    \caption{Accuracy scores (as defined in Sect.\,\ref{sec:ML_metrics} for the three galaxy properties ($\mass$, $Z_{\star}$, and $t_{\star}$) for the LOFO analysis for the two ML algorithms (RF DNN).}
    
    \smallskip
\label{tab:lofo}
\smallskip
\centering
\resizebox{\textwidth}{!}{
\begin{tabular}{l l c c c c c c c c c c c c c c c }
\hline \hline
  & & & & & \\[-0.3cm]
  &  & \multicolumn{7}{c}{RF} &  \multicolumn{7}{c}{DNN} \\ 
    & & & & & \\[-0.3cm]
 \cline{3-9} \cline{11-17} 
   & & & & & \\[-0.3cm]
Target & Feature &  $R^2$  & RMSE  & LOFO &  $f_{\text{out}}$ & NMAD & $\left< \Delta Y \right>$ & $\rho$ &  & $R^2$  & RMSE & LOFO & $f_{\text{out}}$ & NMAD & $\left< \Delta Y \right>$ & $\rho$ \\
 & & & &  & \\[-0.3cm]
 \hline
 & & & &  & \\[-0.3cm]
 $\mass /{\rm \left(M_{\odot}\, pc^{-2}\right)}$ & $I_\text{E}$ &  $0.872$ &  $0.1727 $ & 0.3199& $0.0712$ & $0.1502$ & $-0.0066$ & 0.934&&0.872&0.1724& 0.3250&0.0706&0.1498&$-0.0063$&0.934 \\
  & $Y_\text{E}$ &  $0.920$ &  $0.1365 $ & 0.0431& $0.0379$ & $0.1143$ & $-0.0010$ & 0.959 &&0.912&0.1428&0.0975&0.0450&0.1173&$-0.0034$&0.955\\
 & $J_\text{E}$ &  $0.918$ &  $0.1383 $&0.0570 & $0.0400$ & $0.1143$ & $-0.0010$ & 0.958&&0.918&0.1379&0.0597&0.0397&0.1140&$-0.0009$&0.958 \\
& $H_\text{E}$ &  $0.912$ &  $0.1432 $& 0.0944& $0.0450$ & $0.1171$ & $-0.0013$ & 0.955 &&0.913&0.1427&0.0966&0.0445&0.1169&$-0.0020$&0.955\\

      & & & && & \\[-0.3cm]
 \hline
  & & & &  & & \\[-0.3cm]

 $Z_{\star}$ & $I_\text{E}$ &  $0.477$ &  $0.0866 $&0.1274 & $0.0029$ & $0.0806$ & $0.0007$ & 0.692&&0.479&0.0864&0.1327&0.0029&0.0802&0.0011&0.694 \\
  & $Y_\text{E}$ &  $0.549$ &  $0.0804 $ &0.0469& $0.0009$ & $0.0758$ & $0.0000$ & 0.743 &&0.524&0.0826&0.0823&0.0012&0.0784&$-0.0023$&0.726 \\
 & $J_\text{E}$ &  $0.544$ &  $0.0808 $ &0.0524& $0.0010$ & $0.0763$ & $0.0000$ & 0.740 &&0.546&0.0806&0.0567&0.0010&0.0761&0.0003&0.741\\
& $H_\text{E}$ &  $0.531$ &  $0.0820 $ &0.0678& $0.0012$ & $0.0769$ & $0.0000$ & 0.731 &&0.533&0.0818&0.0723&0.0012&0.0768&0.0011&0.732\\
 
& & & &  & & \\[-0.3cm]   
\hline
& & & &  & & \\[-0.3cm]   

 $t_{\star} {\rm \,/\, Gyr}$ & $I_\text{E}$ &  $0.123$ &  $0.1531 $ &0.1861& $0.0462$ & $0.1252$ & $-0.0123$ & 0.359&&0.338&0.1330& 0.0355&0.0317&0.1046&$-0.0072$&0.587 \\
  & $Y_\text{E}$ &  $0.349$ &  $0.1319 $ & 0.0215& $0.0305$ & $0.1045$ & $-0.0081$ & 0.595 &&0.351&0.1317&0.0255& 0.0303&0.1043&$-0.0079$&0.597\\
 & $J_\text{E}$ &  $0.349$ &  $0.1319 $ & 0.0219& $0.0307$ & $0.1042$ & $-0.0073$ & 0.596 &&0.351&0.1317&0.0256 &0.0306&0.1039&$-0.0070$&0.597\\
& $H_\text{E}$ &  $0.336$ &  $0.1332 $&0.0321 & $0.0319$ & $0.1049$ & $-0.0073$ & 0.585 &&0.337&0.1331& 0.0360&0.0318&0.1047&$-0.0070$&0.586\\

      & & & &  & & \\[-0.3cm]
        \hline \hline
    \end{tabular}
    }
    \tablefoot{Left to right: Target; \Euclid\/\ surface brightness removed for LOFO analysis; coefficient of determination ($R^2$); root-mean-square error (${\rm RMSE}$); LOFO importance; fraction of catastrophic outliers ($f_\text{out}$); normalised median absolute deviation (${\rm NMAD}$); and bias ($\left< \Delta Y \right>$, where $Y$ represents the target galaxy property); Pearson $\rho$ coefficient.}
\end{table*}

Unlike RF, DNN requires a dedicated computer architecture. Still, the algorithm was trained in under three days on the VSC Tier dedicated GPU machine. Once trained, the model can easily be saved and deployed on new galaxies, as well as new computer infrastructure. In fact, it only takes several minutes to finish modelling a map of a single galaxy on a sub-kpc ($100{\rm \, pc} \times100 {\rm\,pc}$) resolution on an 11$^\text{th}$ generation Intel i7 2.30\,GHz 16-core laptop, something that is unfortunately still not possible with traditional methods, such as SED fitting. As the most complex method of the three used in this work, DNN is also hardest to interpret, but with the recent advances in ML, a number of methods have been developed to offer better interpretability of various algorithms, which should make using neural networks to solve astrophysical problems more feasible in the future \citep[for an overview on interpreting neural networks, see e.g.][]{make3030032,LONGO2024102301}.

Regardless, as our results show, the choice of the algorithm only has a very minor effect on the accuracy of inferring galaxy parameters from \Euclid\/\ observations on the full data set, with the differences only becoming prominent in the case of missing data and hard to infer values. All of these properties need to be taken into account when choosing the best machine learning algorithm for the task.

\section{Summary}\label{sec:summary}

In this work, we have tested the viability of using supervised ML to map mock \Euclid\/\ observations of nearby galaxies with galaxy parameters extracted from the Illustris TNG50 simulation and processed by the Monte Carlo radiative transfer code SKIRT on sub-kiloparsec scales. We tested three distinct methods, an ordinary least squares LR and two ML algorithms (RF and DNN) with three different choices of input parameters, single \Euclid\/\ band, all four \Euclid\/\ bands, and ten input parameters consisting of both the four \Euclid\/\ bands and six \Euclid\/\ colours modelled from them, for three galaxy parameters, stellar mass surface density, mass-averaged stellar metallicity, and stellar age.

We showed that when using a single band, the $\HE$ band performs the best when predicting $\mass$\ and $Z_{\star}$ in the case of all three algorithms. It is the most important input feature according to RF and has the second-strongest effect on the accuracy during the LOFO analysis. On the other hand, \IE\ is shown to be the second most important feature when building a tree in RF, but it has the strongest effect in the LOFO analysis, showcasing the importance of using several methods when trying to analyse ML results. Specifically, \IE\ seems to carry the most information from the lower end of the distribution for all three galaxy parameters, while \HE\ affects the entire range. Any single band cannot be used to model stellar age (with $R^2\approx 0$ in all four cases).

We showed that the accuracy of modelling $\mass$\ with only the $\HE$ band is very high, with similar values of $R^2$ for all three algorithms, pointing to a strong linear relationship between $\mass$\ and the $\HE$-band surface brightness. However, the accuracy increases slightly with the introduction of the three other bands, with this increase being higher in the case of RF and DNN, indicating that the information gained from the $\IE$, $\YE$, and $\JE$ bands belongs to a more non-linear regime. The introduction of colour does not offer any significant increase to accuracy, pointing to the fact that ML can extract all the necessary information from the four \Euclid\/\ bands.

We also presented full disc maps of our model for two example galaxies: TNG\,539333, which represents a better outcome, and TNG\,294869, as an example of a worse outcome. We note that the model performance struggles around more complex features (for example, spiral arms or dusty features) and on the outskirts of the galactic disc with low surface brightness, where the effects of Monte Carlo noise become more prominent. Additionally, we compared both face-on and edge-on projections for these galaxies and showed that even in cases of good prediction, the algorithm struggles in the outer parts of the edge-on cases, that is, the most distant regions above and below the galactic disc.

As expected, we cannot model stellar age from the \Euclid\/\ observations; however, we do get an unexpectedly high accuracy when modelling $Z_{\star}$, and the behaviour of the predictions is similar to $\mass$. We would not expect to be able to extract the information on $Z_{\star}$ from the \Euclid\/\ observations due to the fact that $Z_{\star}$ is extracted from spectral absorption features in the optical regime, which only the $\IE$-band covers, but the $\IE$-band itself is too broad. However, we noticed a strong relationship between $\mass$\ and $Z_{\star}$ within the TNG50 simulation, which explains the unexpected quality of the $Z_{\star}$ prediction. We compared this finding to earlier studies done on simulated (EAGLE) and observational (CALIFA, MaNGA) data. While the MZR is a well-known galactic relation on global data, in this work we hypothesise that this relation could be used on resolved (sub-kiloparsec) data as well.

\section{Implications for \Euclid\/}

With the observations of thousands of nearby galaxies that we expect to get access to with \Euclid\/, traditional methods used to extract galaxy parameters, such as SED fitting, become unfeasible due to their large computational time requirements. In this work, we offer an alternative method that has become standard in analysing large amounts of data. We trained two distinct ML algorithms on mock images of 1154 nearby galaxies extracted from the TNG50 simulation, compared them to an ordinary least squares LR, and showed that it is possible to map \Euclid\/\ observations to stellar mass surface density on the sub-kiloparsec level with a high accuracy in a short amount of time. We also showed that though mapping \Euclid\/\ observations to stellar metallicity and age is less robust, the information still holds value as a first-order approximation of stellar metallicities across large samples.

It is nevertheless important to stress that this work was performed on idealised data. The ML procedure offers a very limited ability to analyse the underlying relationship between the input parameters and the output targets. The goal of this work was to focus on the pure physical properties, as this is the necessary first step in trying to apply machine learning algorithms to real life data. The inclusion of noise could potentially introduce additional complex underlying relationships that could be difficult to properly account for without the baseline set here. For this reason, we could not determine the aleatoric uncertainty of our algorithms. As a reminder, there are two types of uncertainties possible in an ML procedure: epistemic uncertainty, which is caused by the stochasticity present in the algorithm itself and informs one of how uncertain the algorithm is in its prediction, and aleatoric uncertainty, which is caused by the presence of observational noise in the data set. For a discussion on epistemic uncertainty, which has been assessed in this work, we refer to Sect.\,\ref{sec:ML_errors}. We plan to assess the aleatoric uncertainty in the future, for which we will have to perform the entire ML procedure first on idealised data and then on data with noise added to it. This will require adding a field of stars to mimic the presence of contaminants, Poissonian noise to mimic effects of photon shot noise, Gaussian noise to simulate observational background, and RMS noise to assign uncertainties to the images themselves. For details on this procedure, we refer to \citet{Merlin_2023}. The resolution of the images will differ from the current $100{\rm \, pc} \times 100 {\rm \,pc}$ once the \Euclid\/\ PSF is taken into account, potentially affecting the accuracy of our method. However, we have performed some initial tests on different resolutions, and they have shown that the effects of resolution should be very minor at best.

Additionally, we have treated each pixel as an individual data point, ignoring the spatial information that could be garnered from the neighbouring pixels. This information could be accessed, for instance, via a CNN setup, which we plan to deploy in a future work. That way, having both the information contained in individual pixels as well as when including the spatial information from the neighbouring pixels, we should be able to gain a better understanding of how much spatial information contributes to the accuracy of inferring galaxy parameters with ML.

Another potential way to improve these results would be to take advantage of the relationships between our three targets and use chained regression (e.g. Euclid Collaboration: Humphrey et al., in prep). In our initial tests, the inclusion of $\mass$ in the input features increased the accuracy in modelling $Z_{\star}$, and the inclusion of $Z_{\star}$ increased the accuracy in $t_{\star}$. We also plan to combine the \Euclid\/\ data with other (specifically optical) data to test how much we can increase the accuracy of ML at predicting galaxy properties from observations. We are also aware of the fact that algorithms on simulated data might not transfer well into actual observations \citep[e.g.][where the authors notice a discrepancy between simulated galaxies from TNG50/TNG100 simulations, and SDSS]{Zanisi_2021}.
And finally, we will test our model on actual observations and compare it with traditional methods currently being developed (Euclid Collaboration: Abdurro'uf et al., in prep; Euclid Collaboration: Nersesian et al., in prep) in order to see how feasible it is to use supervised ML to accurately infer galaxy parameters, such as $\mass$, $Z_{\star}$, and $t_{\star}$, from actual \Euclid\/\ observations.


\begin{acknowledgements}

We thank the referee for useful comments that helped to improve the quality of the manuscript. Co-funded by the European Union. Views and opinions expressed are however those of the author(s) only and do not necessarily reflect those of the European Union. Neither the European Union nor the granting authority can be held responsible for them. 
IK, MB and AN acknowledge support from the Belgian Science Policy Office (BELSPO) through the PRODEX project “Belgian Euclid Science Exploitation (BESE)” (No. 4000143202).
JHK acknowledges grant PID2022-136505NB-I00 funded by MCIN/AEI/10.13039/501100011033 and EU, ERDF. We wish to thank the “Summer School for Astrostatistics in Crete” for providing training on the statistical methods adopted in this work.
\AckEC

\end{acknowledgements}

\bibliography{mybib}

\begin{thebibliography}{155}
\expandafter\ifx\csname natexlab\endcsname\relax\def\natexlab#1{#1}\fi

\bibitem[{Abadi {et~al.}(2015)Abadi, Agarwal, Barham, Brevdo, Chen, Citro,
  Corrado, Davis, Dean, Devin, Ghemawat, Goodfellow, Harp, Irving, Isard, Jia,
  Jozefowicz, Kaiser, Kudlur, Levenberg, Man\'{e}, Monga, Moore, Murray, Olah,
  Schuster, Shlens, Steiner, Sutskever, Talwar, Tucker, Vanhoucke, Vasudevan,
  Vi\'{e}gas, Vinyals, Warden, Wattenberg, Wicke, Yu, \&
  Zheng}]{abadi2016tensorflow}
Abadi, M., Agarwal, A., Barham, P., {et~al.} 2015, {TensorFlow}: Large-Scale
  Machine Learning on Heterogeneous Systems, software available from
  tensorflow.org

\bibitem[{{Abazajian} {et~al.}(2009){Abazajian}, {Adelman-McCarthy},
  {Ag{\"u}eros}, {Allam}, {Allende Prieto}, {An}, {Anderson}, {Anderson},
  {Annis}, {Bahcall}, {Bailer-Jones}, {Barentine}, {Bassett}, {Becker},
  {Beers}, {Bell}, {Belokurov}, {Berlind}, {Berman}, {Bernardi}, {Bickerton},
  {Bizyaev}, {Blakeslee}, {Blanton}, {Bochanski}, {Boroski}, {Brewington},
  {Brinchmann}, {Brinkmann}, {Brunner}, {Budav{\'a}ri}, {Carey}, {Carliles},
  {Carr}, {Castander}, {Cinabro}, {Connolly}, {Csabai}, {Cunha}, {Czarapata},
  {Davenport}, {de Haas}, {Dilday}, {Doi}, {Eisenstein}, {Evans}, {Evans},
  {Fan}, {Friedman}, {Frieman}, {Fukugita}, {G{\"a}nsicke}, {Gates},
  {Gillespie}, {Gilmore}, {Gonzalez}, {Gonzalez}, {Grebel}, {Gunn},
  {Gy{\"o}ry}, {Hall}, {Harding}, {Harris}, {Harvanek}, {Hawley}, {Hayes},
  {Heckman}, {Hendry}, {Hennessy}, {Hindsley}, {Hoblitt}, {Hogan}, {Hogg},
  {Holtzman}, {Hyde}, {Ichikawa}, {Ichikawa}, {Im}, {Ivezi{\'c}}, {Jester},
  {Jiang}, {Johnson}, {Jorgensen}, {Juri{\'c}}, {Kent}, {Kessler}, {Kleinman},
  {Knapp}, {Konishi}, {Kron}, {Krzesinski}, {Kuropatkin}, {Lampeitl},
  {Lebedeva}, {Lee}, {Lee}, {French Leger}, {L{\'e}pine}, {Li}, {Lima}, {Lin},
  {Long}, {Loomis}, {Loveday}, {Lupton}, {Magnier}, {Malanushenko},
  {Malanushenko}, {Mandelbaum}, {Margon}, {Marriner}, {Mart{\'\i}nez-Delgado},
  {Matsubara}, {McGehee}, {McKay}, {Meiksin}, {Morrison}, {Mullally}, {Munn},
  {Murphy}, {Nash}, {Nebot}, {Neilsen}, {Newberg}, {Newman}, {Nichol},
  {Nicinski}, {Nieto-Santisteban}, {Nitta}, {Okamura}, {Oravetz}, {Ostriker},
  {Owen}, {Padmanabhan}, {Pan}, {Park}, {Pauls}, {Peoples}, {Percival}, {Pier},
  {Pope}, {Pourbaix}, {Price}, {Purger}, {Quinn}, {Raddick}, {Re Fiorentin},
  {Richards}, {Richmond}, {Riess}, {Rix}, {Rockosi}, {Sako}, {Schlegel},
  {Schneider}, {Scholz}, {Schreiber}, {Schwope}, {Seljak}, {Sesar}, {Sheldon},
  {Shimasaku}, {Sibley}, {Simmons}, {Sivarani}, {Allyn Smith}, {Smith},
  {Smol{\v{c}}i{\'c}}, {Snedden}, {Stebbins}, {Steinmetz}, {Stoughton},
  {Strauss}, {SubbaRao}, {Suto}, {Szalay}, {Szapudi}, {Szkody}, {Tanaka},
  {Tegmark}, {Teodoro}, {Thakar}, {Tremonti}, {Tucker}, {Uomoto}, {Vanden
  Berk}, {Vandenberg}, {Vidrih}, {Vogeley}, {Voges}, {Vogt}, {Wadadekar},
  {Watters}, {Weinberg}, {West}, {White}, {Wilhite}, {Wonders}, {Yanny},
  {Yocum}, {York}, {Zehavi}, {Zibetti}, \& {Zucker}}]{Abazajian_2009}
{Abazajian}, K.~N., {Adelman-McCarthy}, J.~K., {Ag{\"u}eros}, M.~A., {et~al.}
  2009, \apjs, 182, 543

\bibitem[{{Abdalla} {et~al.}(2011){Abdalla}, {Banerji}, {Lahav}, \&
  {Rashkov}}]{Abdalla_2011}
{Abdalla}, F.~B., {Banerji}, M., {Lahav}, O., \& {Rashkov}, V. 2011, \mnras,
  417, 1891

\bibitem[{{Abdurro'uf} {et~al.}(2022{\natexlab{a}}){Abdurro'uf}, {Accetta},
  {Aerts}, {Silva Aguirre}, {Ahumada}, {Ajgaonkar}, {Filiz Ak}, {Alam},
  {Allende Prieto}, {Almeida}, {Anders}, {Anderson}, {Andrews}, {Anguiano},
  {Aquino-Ort{\'\i}z}, {Arag{\'o}n-Salamanca}, {Argudo-Fern{\'a}ndez}, {Ata},
  {Aubert}, {Avila-Reese}, {Badenes}, {Barb{\'a}}, {Barger},
  {Barrera-Ballesteros}, {Beaton}, {Beers}, {Belfiore}, {Bender}, {Bernardi},
  {Bershady}, {Beutler}, {Bidin}, {Bird}, {Bizyaev}, {Blanc}, {Blanton},
  {Boardman}, {Bolton}, {Boquien}, {Borissova}, {Bovy}, {Brandt}, {Brown},
  {Brownstein}, {Brusa}, {Buchner}, {Bundy}, {Burchett}, {Bureau}, {Burgasser},
  {Cabang}, {Campbell}, {Cappellari}, {Carlberg}, {Wanderley}, {Carrera},
  {Cash}, {Chen}, {Chen}, {Cherinka}, {Chiappini}, {Choi}, {Chojnowski},
  {Chung}, {Clerc}, {Cohen}, {Comerford}, {Comparat}, {da Costa}, {Covey},
  {Crane}, {Cruz-Gonzalez}, {Culhane}, {Cunha}, {Dai}, {Damke}, {Darling},
  {Davidson}, {Davies}, {Dawson}, {De Lee}, {Diamond-Stanic}, {Cano-D{\'\i}az},
  {S{\'a}nchez}, {Donor}, {Duckworth}, {Dwelly}, {Eisenstein}, {Elsworth},
  {Emsellem}, {Eracleous}, {Escoffier}, {Fan}, {Farr}, {Feng},
  {Fern{\'a}ndez-Trincado}, {Feuillet}, {Filipp}, {Fillingham}, {Frinchaboy},
  {Fromenteau}, {Galbany}, {Garc{\'\i}a}, {Garc{\'\i}a-Hern{\'a}ndez}, {Ge},
  {Geisler}, {Gelfand}, {G{\'e}ron}, {Gibson}, {Goddy}, {Godoy-Rivera},
  {Grabowski}, {Green}, {Greener}, {Grier}, {Griffith}, {Guo}, {Guy},
  {Hadjara}, {Harding}, {Hasselquist}, {Hayes}, {Hearty}, {Hern{\'a}ndez},
  {Hill}, {Hogg}, {Holtzman}, {Horta}, {Hsieh}, {Hsu}, {Hsu}, {Huber},
  {Huertas-Company}, {Hutchinson}, {Hwang}, {Ibarra-Medel}, {Chitham}, {Ilha},
  {Imig}, {Jaekle}, {Jayasinghe}, {Ji}, {Johnson}, {Jones}, {J{\"o}nsson},
  {Katkov}, {Khalatyan}, {Kinemuchi}, {Kisku}, {Knapen}, {Kneib}, {Kollmeier},
  {Kong}, {Kounkel}, {Kreckel}, {Krishnarao}, {Lacerna}, {Lane}, {Langgin},
  {Lavender}, {Law}, {Lazarz}, {Leung}, {Leung}, {Lewis}, {Li}, {Li}, {Lian},
  {Liang}, {Lin}, {Lin}, {Lin}, {Lintott}, {Long}, {Longa-Pe{\~n}a},
  {L{\'o}pez-Cob{\'a}}, {Lu}, {Lundgren}, {Luo}, {Mackereth}, {de la Macorra},
  {Mahadevan}, {Majewski}, {Manchado}, {Mandeville}, {Maraston},
  {Margalef-Bentabol}, {Masseron}, {Masters}, {Mathur}, {McDermid}, {Mckay},
  {Merloni}, {Merrifield}, {Meszaros}, {Miglio}, {Di Mille}, {Minniti},
  {Minsley}, {Monachesi}, {Moon}, {Mosser}, {Mulchaey}, {Muna}, {Mu{\~n}oz},
  {Myers}, {Myers}, {Nadathur}, {Nair}, {Nandra}, {Neumann}, {Newman},
  {Nidever}, {Nikakhtar}, {Nitschelm}, {O'Connell}, {Garma-Oehmichen}, {Luan
  Souza de Oliveira}, {Olney}, {Oravetz}, {Ortigoza-Urdaneta}, {Osorio},
  {Otter}, {Pace}, {Padilla}, {Pan}, {Pan}, {Parikh}, {Parker}, {Peirani},
  {Pe{\~n}a Ram{\'\i}rez}, {Penny}, {Percival}, {Perez-Fournon},
  {Pinsonneault}, {Poidevin}, {Poovelil}, {Price-Whelan}, {B{\'a}rbara de
  Andrade Queiroz}, {Raddick}, {Ray}, {Rembold}, {Riddle}, {Riffel}, {Riffel},
  {Rix}, {Robin}, {Rodr{\'\i}guez-Puebla}, {Roman-Lopes},
  {Rom{\'a}n-Z{\'u}{\~n}iga}, {Rose}, {Ross}, {Rossi}, {Rubin}, {Salvato},
  {S{\'a}nchez}, {S{\'a}nchez-Gallego}, {Sanderson}, {Santana Rojas},
  {Sarceno}, {Sarmiento}, {Sayres}, {Sazonova}, {Schaefer}, {Schiavon},
  {Schlegel}, {Schneider}, {Schultheis}, {Schwope}, {Serenelli}, {Serna},
  {Shao}, {Shapiro}, {Sharma}, {Shen}, {Shetrone}, {Shu}, {Simon}, {Skrutskie},
  {Smethurst}, {Smith}, {Sobeck}, {Spoo}, {Sprague}, {Stark}, {Stassun},
  {Steinmetz}, {Stello}, {Stone-Martinez}, {Storchi-Bergmann}, {Stringfellow},
  {Stutz}, {Su}, {Taghizadeh-Popp}, {Talbot}, {Tayar}, {Telles}, {Teske},
  {Thakar}, {Theissen}, {Tkachenko}, {Thomas}, {Tojeiro}, {Hernandez Toledo},
  {Troup}, {Trump}, {Trussler}, {Turner}, {Tuttle}, {Unda-Sanzana},
  {V{\'a}zquez-Mata}, {Valentini}, {Valenzuela}, {Vargas-Gonz{\'a}lez},
  {Vargas-Maga{\~n}a}, {Alfaro}, {Villanova}, {Vincenzo}, {Wake}, {Warfield},
  {Washington}, {Weaver}, {Weijmans}, {Weinberg}, {Weiss}, {Westfall}, {Wild},
  {Wilde}, {Wilson}, {Wilson}, {Wilson}, {Wolf}, {Wood-Vasey}, {Yan}, {Zamora},
  {Zasowski}, {Zhang}, {Zhao}, {Zheng}, {Zheng}, \& {Zhu}}]{Abdurrouf_2022}
{Abdurro'uf}, {Accetta}, K., {Aerts}, C., {et~al.} 2022{\natexlab{a}}, \apjs,
  259, 35

\bibitem[{{Abdurro'uf} {et~al.}(2023){Abdurro'uf}, {Coe}, {Jung}, {Ferguson},
  {Brammer}, {Iyer}, {Bradley}, {Dayal}, {Windhorst}, {Zitrin}, {Meena},
  {Oguri}, {Diego}, {Kokorev}, {Dimauro}, {Adamo}, {Conselice}, {Welch},
  {Vanzella}, {Hsiao}, {Xu}, {Roy}, \& {Mulcahey}}]{Abdurro'uf_2023}
{Abdurro'uf}, {Coe}, D., {Jung}, I., {et~al.} 2023, \apj, 945, 117

\bibitem[{{Abdurro'uf} {et~al.}(2022{\natexlab{b}}){Abdurro'uf}, {Lin},
  {Hirashita}, {Morishita}, {Tacchella}, {Wu}, {Akiyama}, \&
  {Takeuchi}}]{Abdurro'uf_2022}
{Abdurro'uf}, {Lin}, Y.-T., {Hirashita}, H., {et~al.} 2022{\natexlab{b}}, \apj,
  935, 98

\bibitem[{{Abdurro'uf} {et~al.}(2021){Abdurro'uf}, {Lin}, {Wu}, \&
  {Akiyama}}]{Abdurro'uf_2021}
{Abdurro'uf}, {Lin}, Y.-T., {Wu}, P.-F., \& {Akiyama}, M. 2021, \apjs, 254, 15

\bibitem[{Acquaviva(2015)}]{Acquaviva_2015}
Acquaviva, V. 2015, \mnras, 456, 1618

\bibitem[{{Alsing} {et~al.}(2023){Alsing}, {Peiris}, {Mortlock}, {Leja}, \&
  {Leistedt}}]{Alsing_2023}
{Alsing}, J., {Peiris}, H., {Mortlock}, D., {Leja}, J., \& {Leistedt}, B. 2023,
  \apjs, 264, 29

\bibitem[{{Alsing} {et~al.}(2024){Alsing}, {Thorp}, {Deger}, {Peiris},
  {Leistedt}, {Mortlock}, \& {Leja}}]{Alsing_2024}
{Alsing}, J., {Thorp}, S., {Deger}, S., {et~al.} 2024, \apjs, 274, 12

\bibitem[{{Baes} \& {Dejonghe}(2001)}]{Baes_2001a}
{Baes}, M. \& {Dejonghe}, H. 2001, \mnras, 326, 722

\bibitem[{{Baes} {et~al.}(2024){Baes}, {Gebek}, {Tr{\v{c}}ka}, {Camps}, {van
  der Wel}, {Abdurro'uf}, {Andreadis}, {Tulu}, {Emana}, {Fritz}, {Kelly},
  {Kova{\v{c}}i{\'c}}, {La Marca}, {Martorano}, {Mosenkov}, {Nersesian},
  {Rodriguez-Gomez}, {Tortora}, {Vander Meulen}, \&
  {Wang}}]{baes2024tng50skirt}
{Baes}, M., {Gebek}, A., {Tr{\v{c}}ka}, A., {et~al.} 2024, \aap, 683, A181

\bibitem[{{Baes} {et~al.}(2011){Baes}, {Verstappen}, {De Looze}, {Fritz},
  {Saftly}, {Vidal P{\'e}rez}, {Stalevski}, \& {Valcke}}]{Baes_2011}
{Baes}, M., {Verstappen}, J., {De Looze}, I., {et~al.} 2011, \apjs, 196, 22

\bibitem[{Baker {et~al.}(2022)Baker, Maiolino, Belfiore, Curti, Bluck, Lin,
  Ellison, Thorp, \& Pan}]{Baker_2022}
Baker, W.~M., Maiolino, R., Belfiore, F., {et~al.} 2022, \mnras, 519, 1149

\bibitem[{{Baron}(2019)}]{Baron_2019}
{Baron}, D. 2019, arXiv:1904.07248

\bibitem[{{Bell} \& {de Jong}(2001)}]{Bell_2001}
{Bell}, E.~F. \& {de Jong}, R.~S. 2001, \apj, 550, 212

\bibitem[{{Bellstedt} {et~al.}(2021){Bellstedt}, {Robotham}, {Driver},
  {Thorne}, {Davies}, {Holwerda}, {Hopkins}, {Lara-Lopez},
  {L{\'o}pez-S{\'a}nchez}, \& {Phillipps}}]{Bellstedt_2021}
{Bellstedt}, S., {Robotham}, A. S.~G., {Driver}, S.~P., {et~al.} 2021, \mnras,
  503, 3309

\bibitem[{{Breiman}(2001)}]{Breiman_2001}
{Breiman}, L. 2001, Machine Learning, 45, 5

\bibitem[{{Brescia} {et~al.}(2013){Brescia}, {Cavuoti}, {D'Abrusco}, {Longo},
  \& {Mercurio}}]{Brescia_2013}
{Brescia}, M., {Cavuoti}, S., {D'Abrusco}, R., {Longo}, G., \& {Mercurio}, A.
  2013, \apj, 772, 140

\bibitem[{{Brinchmann} {et~al.}(2004){Brinchmann}, {Charlot}, {White},
  {Tremonti}, {Kauffmann}, {Heckman}, \& {Brinkmann}}]{Brinchmann_2004}
{Brinchmann}, J., {Charlot}, S., {White}, S.~D.~M., {et~al.} 2004, \mnras, 351,
  1151

\bibitem[{{Bruzual} \& {Charlot}(1993)}]{Bruzual_1993}
{Bruzual}, G. \& {Charlot}, S. 1993, \apj, 405, 538

\bibitem[{{Bruzual} \& {Charlot}(2003)}]{Charlot_2003}
{Bruzual}, G. \& {Charlot}, S. 2003, \mnras, 344, 1000

\bibitem[{{Bundy} {et~al.}(2015){Bundy}, {Bershady}, {Law}, {Yan}, {Drory},
  {MacDonald}, {Wake}, {Cherinka}, {S{\'a}nchez-Gallego}, {Weijmans}, {Thomas},
  {Tremonti}, {Masters}, {Coccato}, {Diamond-Stanic}, {Arag{\'o}n-Salamanca},
  {Avila-Reese}, {Badenes}, {Falc{\'o}n-Barroso}, {Belfiore}, {Bizyaev},
  {Blanc}, {Bland-Hawthorn}, {Blanton}, {Brownstein}, {Byler}, {Cappellari},
  {Conroy}, {Dutton}, {Emsellem}, {Etherington}, {Frinchaboy}, {Fu}, {Gunn},
  {Harding}, {Johnston}, {Kauffmann}, {Kinemuchi}, {Klaene}, {Knapen},
  {Leauthaud}, {Li}, {Lin}, {Maiolino}, {Malanushenko}, {Malanushenko}, {Mao},
  {Maraston}, {McDermid}, {Merrifield}, {Nichol}, {Oravetz}, {Pan}, {Parejko},
  {Sanchez}, {Schlegel}, {Simmons}, {Steele}, {Steinmetz}, {Thanjavur},
  {Thompson}, {Tinker}, {van den Bosch}, {Westfall}, {Wilkinson}, {Wright},
  {Xiao}, \& {Zhang}}]{Bundy_2015}
{Bundy}, K., {Bershady}, M.~A., {Law}, D.~R., {et~al.} 2015, \apj, 798, 7

\bibitem[{{Calzetti} {et~al.}(2000){Calzetti}, {Armus}, {Bohlin}, {Kinney},
  {Koornneef}, \& {Storchi-Bergmann}}]{Calzetti_2000}
{Calzetti}, D., {Armus}, L., {Bohlin}, R.~C., {et~al.} 2000, \apj, 533, 682

\bibitem[{{Camps} \& {Baes}(2018)}]{Camps_2018}
{Camps}, P. \& {Baes}, M. 2018, \apj, 861, 80

\bibitem[{{Camps} \& {Baes}(2020)}]{Camps_Baes_2020}
{Camps}, P. \& {Baes}, M. 2020, Astronomy and Computing, 31, 100381

\bibitem[{{Camps} {et~al.}(2015){Camps}, {Misselt}, {Bianchi}, {Lunttila},
  {Pinte}, {Natale}, {Juvela}, {Fischera}, {Fitzgerald}, {Gordon}, {Baes}, \&
  {Steinacker}}]{Camps_2015}
{Camps}, P., {Misselt}, K., {Bianchi}, S., {et~al.} 2015, \aap, 580, A87

\bibitem[{{Carliles} {et~al.}(2010){Carliles}, {Budav{\'a}ri}, {Heinis},
  {Priebe}, \& {Szalay}}]{Carliles_2010}
{Carliles}, S., {Budav{\'a}ri}, T., {Heinis}, S., {Priebe}, C., \& {Szalay},
  A.~S. 2010, \apj, 712, 511

\bibitem[{{Cavuoti} {et~al.}(2012){Cavuoti}, {Brescia}, {Longo}, \&
  {Mercurio}}]{Cavuoti_2012}
{Cavuoti}, S., {Brescia}, M., {Longo}, G., \& {Mercurio}, A. 2012, \aap, 546,
  A13

\bibitem[{{Chabrier}(2003)}]{Chabrier_2003}
{Chabrier}, G. 2003, \pasp, 115, 763

\bibitem[{{Chamba} {et~al.}(2022){Chamba}, {Trujillo}, \&
  {Knapen}}]{Chamba_2022}
{Chamba}, N., {Trujillo}, I., \& {Knapen}, J.~H. 2022, \aap, 667, A87

\bibitem[{{Charlot} \& {Bruzual}(1991)}]{Charlot_1991}
{Charlot}, S. \& {Bruzual}, G. 1991, \apj, 367, 126

\bibitem[{{Charlot} \& {Fall}(2000)}]{Charlot_2000}
{Charlot}, S. \& {Fall}, S.~M. 2000, \apj, 539, 718

\bibitem[{Chicco {et~al.}(2021)Chicco, Warrens, \& Jurman}]{Chicco_2021}
Chicco, D., Warrens, M., \& Jurman, G. 2021, PeerJ Computer Science, 486, 5104

\bibitem[{{Collister} \& {Lahav}(2004)}]{Collister_2004}
{Collister}, A.~A. \& {Lahav}, O. 2004, \pasp, 116, 345

\bibitem[{{Conroy}(2013)}]{Conroy_2013}
{Conroy}, C. 2013, \araa, 51, 393

\bibitem[{{Cropper} {et~al.}(2018){Cropper}, {Pottinger}, {Azzollini},
  {Szafraniec}, {Awan}, {Mellier}, {Berth{\'e}}, {Martignac}, {Cara}, {Di
  Giorgio}, {Sciortino}, {Bozzo}, {Genolet}, {Philippon}, {Hailey}, {Hunt},
  {Swindells}, {Holland}, {Gow}, {Murray}, {Hall}, {Skottfelt}, {Amiaux},
  {Laureijs}, {Racca}, {Salvignol}, {Short}, {Lorenzo Alvarez}, {Kitching},
  {Hoekstra}, {Galli}, {Willis}, {Hu}, {Candini}, {Boucher}, {Al Bahlawan},
  {Chaudery}, {de Lacy}, {Pendem}, {Smit}, {Dubois}, {Horeau}, {Carty},
  {Fontignie}, {Doumayrou}, {Larcheveque}, {Castelli}, {Cole}, {Niemi},
  {Denniston}, {Massey}, {Kohley}, {Ferrando}, \& {Conversi}}]{Cropper_2018}
{Cropper}, M., {Pottinger}, S., {Azzollini}, R., {et~al.} 2018, in Society of
  Photo-Optical Instrumentation Engineers (SPIE) Conference Series, Vol. 10698,
  Space Telescopes and Instrumentation 2018: Optical, Infrared, and Millimeter
  Wave, ed. M.~{Lystrup}, H.~A. {MacEwen}, G.~G. {Fazio}, N.~{Batalha},
  N.~{Siegler}, \& E.~C. {Tong}, 1069828

\bibitem[{{D'Abrusco} {et~al.}(2007){D'Abrusco}, {Staiano}, {Longo}, {Brescia},
  {Paolillo}, {De Filippis}, \& {Tagliaferri}}]{Dabrusco_2007}
{D'Abrusco}, R., {Staiano}, A., {Longo}, G., {et~al.} 2007, \apj, 663, 752

\bibitem[{Dahl {et~al.}(2013)Dahl, Sainath, \& Hinton}]{6639346}
Dahl, G.~E., Sainath, T.~N., \& Hinton, G.~E. 2013, in 2013 IEEE International
  Conference on Acoustics, Speech and Signal Processing, 8609--8613

\bibitem[{{Dale} \& {Helou}(2002)}]{Dale_2002}
{Dale}, D.~A. \& {Helou}, G. 2002, \apj, 576, 159

\bibitem[{{Decleir} {et~al.}(2019){Decleir}, {De Looze}, {Boquien}, {Baes},
  {Verstocken}, {Calzetti}, {Ciesla}, {Fritz}, {Kennicutt}, {Nersesian}, \&
  {Page}}]{Decleir_2019}
{Decleir}, M., {De Looze}, I., {Boquien}, M., {et~al.} 2019, \mnras, 486, 743

\bibitem[{{Delli Veneri} {et~al.}(2019){Delli Veneri}, {Cavuoti}, {Brescia},
  {Longo}, \& {Riccio}}]{Delli_2019}
{Delli Veneri}, M., {Cavuoti}, S., {Brescia}, M., {Longo}, G., \& {Riccio}, G.
  2019, \mnras, 486, 1377

\bibitem[{{D'Isanto} \& {Polsterer}(2018)}]{Disanto_2018}
{D'Isanto}, A. \& {Polsterer}, K.~L. 2018, \aap, 609, A111

\bibitem[{{Ditrani} {et~al.}(2023){Ditrani}, {Longhetti}, {La Barbera},
  {Iovino}, {Costantin}, {Zibetti}, {Gallazzi}, {Fossati}, {Angthopo},
  {Ascasibar}, {Poggianti}, {S{\'a}nchez-Bl{\'a}zquez}, {Balcells}, {Bianconi},
  {Bolzonella}, {Cassar{\`a}}, {Cucciati}, {Dalton}, {Ferr{\'e}-Mateu},
  {Garc{\'\i}a-Benito}, {Granett}, {Gullieuszik}, {Ikhsanova}, {Jin}, {Knapen},
  {McGee}, {Mercurio}, {Morelli}, {Moretti}, {Murphy}, {Pizzella}, {Pozzetti},
  {Spiniello}, {Tortora}, {Trager}, {Vazdekis}, {Vergani}, \&
  {Vulcani}}]{ditrani2023stellar}
{Ditrani}, F.~R., {Longhetti}, M., {La Barbera}, F., {et~al.} 2023, \aap, 677,
  A93

\bibitem[{{Dobbels} \& {Baes}(2021)}]{Dobbels_2021}
{Dobbels}, W. \& {Baes}, M. 2021, \aap, 655, A34

\bibitem[{{Dobbels} {et~al.}(2020){Dobbels}, {Baes}, {Viaene}, {Bianchi},
  {Davies}, {Casasola}, {Clark}, {Fritz}, {Galametz}, {Galliano}, {Mosenkov},
  {Nersesian}, \& {Tr{\v{c}}ka}}]{Dobbels_2020}
{Dobbels}, W., {Baes}, M., {Viaene}, S., {et~al.} 2020, \aap, 634, A57

\bibitem[{{Draine}(2003)}]{Draine_2003}
{Draine}, B.~T. 2003, \araa, 41, 241

\bibitem[{Duchi {et~al.}(2011)Duchi, Hazan, \& Singer}]{optimization_2011}
Duchi, J., Hazan, E., \& Singer, Y. 2011, Journal of Machine Learning Research,
  12, 2121

\bibitem[{{Euclid Collaboration: Aussel} {et~al.}(2024){Euclid Collaboration:
  Aussel}, {Kruk}, {Walmsley}, {et~al.}}]{Aussel_2024}
{Euclid Collaboration: Aussel}, B., {Kruk}, S., {Walmsley}, M., {et~al.} 2024,
  \aap, 689, A274

\bibitem[{{Euclid Collaboration: Bisigello} {et~al.}(2023){Euclid
  Collaboration: Bisigello}, {Conselice}, {Baes}, {et~al.}}]{Bisigello_2023}
{Euclid Collaboration: Bisigello}, L., {Conselice}, C.~J., {Baes}, M., {et~al.}
  2023, \mnras, 520, 3529

\bibitem[{{Euclid Collaboration: Bretonni{\`e}re} {et~al.}(2022){Euclid
  Collaboration: Bretonni{\`e}re}, {Huertas-Company}, {Boucaud},
  {et~al.}}]{Bretonniere_2022}
{Euclid Collaboration: Bretonni{\`e}re}, H., {Huertas-Company}, M., {Boucaud},
  A., {et~al.} 2022, \aap, 657, A90

\bibitem[{{Euclid Collaboration: Bretonni{\`e}re} {et~al.}(2023){Euclid
  Collaboration: Bretonni{\`e}re}, {Kuchner}, {Huertas-Company},
  {et~al.}}]{Bretonniere_2023}
{Euclid Collaboration: Bretonni{\`e}re}, H., {Kuchner}, U., {Huertas-Company},
  M., {et~al.} 2023, \aap, 671, A102

\bibitem[{{Euclid Collaboration: Desprez} {et~al.}(2020){Euclid Collaboration:
  Desprez}, {Paltani}, {Coupon}, {et~al.}}]{Desprez_2020}
{Euclid Collaboration: Desprez}, G., {Paltani}, S., {Coupon}, J., {et~al.}
  2020, \aap, 644, A31

\bibitem[{{Euclid Collaboration: Enia} {et~al.}(2024){Euclid Collaboration:
  Enia}, {Bolzonella}, {Pozzetti}, {et~al.}}]{EP-Enia}
{Euclid Collaboration: Enia}, A., {Bolzonella}, M., {Pozzetti}, L., {et~al.}
  2024, \aap, 691, A175

\bibitem[{{Euclid Collaboration: Humphrey} {et~al.}(2023){Euclid Collaboration:
  Humphrey}, {Bisigello}, {Cunha}, {et~al.}}]{Humphrey_2023}
{Euclid Collaboration: Humphrey}, A., {Bisigello}, L., {Cunha}, P.~A.~C.,
  {et~al.} 2023, \aap, 671, A99

\bibitem[{{Euclid Collaboration: Jahnke} {et~al.}(2024){Euclid Collaboration:
  Jahnke}, {Gillard}, {Schirmer}, {et~al.}}]{2024arXiv240513493E}
{Euclid Collaboration: Jahnke}, K., {Gillard}, W., {Schirmer}, M., {et~al.}
  2024, \aap, accepted, arXiv:2405.13493

\bibitem[{{Euclid Collaboration: Leuzzi} {et~al.}(2024){Euclid Collaboration:
  Leuzzi}, {Meneghetti}, {Angora}, {et~al.}}]{Leuzzi_2023}
{Euclid Collaboration: Leuzzi}, L., {Meneghetti}, M., {Angora}, G., {et~al.}
  2024, \aap, 681, A68

\bibitem[{{Euclid Collaboration: Mellier} {et~al.}(2024){Euclid Collaboration:
  Mellier}, {Abdurro'uf}, {Acevedo~Barroso}, {et~al.}}]{2024arXiv240513491E}
{Euclid Collaboration: Mellier}, Y., {Abdurro'uf}, {Acevedo~Barroso}, J.,
  {et~al.} 2024, \aap, accepted, arXiv:2405.13491

\bibitem[{{Euclid Collaboration: Merlin} {et~al.}(2023){Euclid Collaboration:
  Merlin}, {Castellano}, {Bretonni{\`e}re}, {et~al.}}]{Merlin_2023}
{Euclid Collaboration: Merlin}, E., {Castellano}, M., {Bretonni{\`e}re}, H.,
  {et~al.} 2023, \aap, 671, A101

\bibitem[{{Euclid Collaboration: Scaramella} {et~al.}(2022){Euclid
  Collaboration: Scaramella}, {Amiaux}, {Mellier}, {et~al.}}]{Scaramella_2022}
{Euclid Collaboration: Scaramella}, R., {Amiaux}, J., {Mellier}, Y., {et~al.}
  2022, \aap, 662, A112

\bibitem[{{Euclid Collaboration: Schirmer} {et~al.}(2022){Euclid Collaboration:
  Schirmer}, {Jahnke}, {Seidel}, {et~al.}}]{Schirmer_2022}
{Euclid Collaboration: Schirmer}, M., {Jahnke}, K., {Seidel}, G., {et~al.}
  2022, \aap, 662, A92

\bibitem[{{Faber}(1973)}]{Faber_1973}
{Faber}, S.~M. 1973, \apj, 179, 731

\bibitem[{{Gallazzi} {et~al.}(2005){Gallazzi}, {Charlot}, {Brinchmann},
  {White}, \& {Tremonti}}]{Gallazzi_2005}
{Gallazzi}, A., {Charlot}, S., {Brinchmann}, J., {White}, S. D.~M., \&
  {Tremonti}, C.~A. 2005, \mnras, 362, 41

\bibitem[{{Galliano} {et~al.}(2021){Galliano}, {Nersesian}, {Bianchi}, {De
  Looze}, {Roychowdhury}, {Baes}, {Casasola}, {Cassar{\'a}}, {Dobbels},
  {Fritz}, {Galametz}, {Jones}, {Madden}, {Mosenkov}, {Xilouris}, \&
  {Ysard}}]{Galliano_2021}
{Galliano}, F., {Nersesian}, A., {Bianchi}, S., {et~al.} 2021, \aap, 649, A18

\bibitem[{{Gao} {et~al.}(2018){Gao}, {Bao}, {Yuan}, {Kong}, {Zou}, {Zhou},
  {Gu}, {Lin}, {Liang}, \& {Huang}}]{Gao_2018}
{Gao}, Y., {Bao}, M., {Yuan}, Q., {et~al.} 2018, \apj, 869, 15

\bibitem[{Genel {et~al.}(2014)Genel, Vogelsberger, Springel, Sijacki, Nelson,
  Snyder, Rodriguez-Gomez, Torrey, \& Hernquist}]{Genel_2014}
Genel, S., Vogelsberger, M., Springel, V., {et~al.} 2014, \mnras, 445, 175

\bibitem[{{Gilda} {et~al.}(2021){Gilda}, {Lower}, \& {Narayanan}}]{Gilda_2021}
{Gilda}, S., {Lower}, S., \& {Narayanan}, D. 2021, \apj, 916, 43

\bibitem[{Glorot {et~al.}(2011)Glorot, Bordes, \& Bengio}]{pmlr-v15-glorot11a}
Glorot, X., Bordes, A., \& Bengio, Y. 2011, in Proceedings of Machine Learning
  Research, Vol.~15, Proceedings of the Fourteenth International Conference on
  Artificial Intelligence and Statistics, ed. G.~Gordon, D.~Dunson, \&
  M.~Dudík (Fort Lauderdale, FL, USA: PMLR), 315--323

\bibitem[{{Gonz{\'a}lez Delgado} {et~al.}(2014){Gonz{\'a}lez Delgado}, {Cid
  Fernandes}, {Garc{\'\i}a-Benito}, {P{\'e}rez}, {de Amorim},
  {Cortijo-Ferrero}, {Lacerda}, {L{\'o}pez Fern{\'a}ndez}, {S{\'a}nchez}, {Vale
  Asari}, {Alves}, {Bland-Hawthorn}, {Galbany}, {Gallazzi}, {Husemann},
  {Bekeraite}, {Jungwiert}, {L{\'o}pez-S{\'a}nchez}, {de Lorenzo-C{\'a}ceres},
  {Marino}, {Mast}, {Moll{\'a}}, {del Olmo}, {S{\'a}nchez-Bl{\'a}zquez}, {van
  de Ven}, {V{\'\i}lchez}, {Walcher}, {Wisotzki}, {Ziegler}, \& {CALIFA
  Collaboration}}]{Delgado_2014}
{Gonz{\'a}lez Delgado}, R.~M., {Cid Fernandes}, R., {Garc{\'\i}a-Benito}, R.,
  {et~al.} 2014, \apjl, 791, L16

\bibitem[{Goodfellow {et~al.}(2016)Goodfellow, Bengio, \&
  Courville}]{Goodfellow2016}
Goodfellow, I., Bengio, Y., \& Courville, A. 2016, Deep Learning (MIT Press),
  \url{http://www.deeplearningbook.org}

\bibitem[{Hastie {et~al.}(2009)Hastie, Tibshirani, \&
  Friedman}]{Hastie2009elements}
Hastie, T., Tibshirani, R., \& Friedman, J. 2009, The elements of statistical
  learning: Data mining, inference, and prediction, second edition (Springer
  Science \& Business Media)

\bibitem[{{Hildebrandt} {et~al.}(2010){Hildebrandt}, {Arnouts}, {Capak},
  {Moustakas}, {Wolf}, {Abdalla}, {Assef}, {Banerji}, {Ben{\'\i}tez},
  {Brammer}, {Budav{\'a}ri}, {Carliles}, {Coe}, {Dahlen}, {Feldmann}, {Gerdes},
  {Gillis}, {Ilbert}, {Kotulla}, {Lahav}, {Li}, {Miralles}, {Purger},
  {Schmidt}, \& {Singal}}]{Hildebrandt_2010}
{Hildebrandt}, H., {Arnouts}, S., {Capak}, P., {et~al.} 2010, \aap, 523, A31

\bibitem[{Höhle \& Höhle(2009)}]{Hohle_2009}
Höhle, J. \& Höhle, M. 2009, ISPRS Journal of Photogrammetry and Remote
  Sensing, 64, 398

\bibitem[{Hüllermeier \& Waegeman(2021)}]{H_llermeier_2021}
Hüllermeier, E. \& Waegeman, W. 2021, Machine Learning, 110, 457

\bibitem[{{Iglesias-Navarro} {et~al.}(2024){Iglesias-Navarro},
  {Huertas-Company}, {Mart{\'\i}n-Navarro}, {Knapen}, \&
  {Pernet}}]{Iglesias_Navarro_2024}
{Iglesias-Navarro}, P., {Huertas-Company}, M., {Mart{\'\i}n-Navarro}, I.,
  {Knapen}, J.~H., \& {Pernet}, E. 2024, \aap, 689, A58

\bibitem[{{Ivezi{\'c}} {et~al.}(2019){Ivezi{\'c}}, {Kahn}, {Tyson}, {Abel},
  {Acosta}, {Allsman}, {Alonso}, {AlSayyad}, {Anderson}, {Andrew}, {Angel},
  {Angeli}, {Ansari}, {Antilogus}, {Araujo}, {Armstrong}, {Arndt}, {Astier},
  {Aubourg}, {Auza}, {Axelrod}, {Bard}, {Barr}, {Barrau}, {Bartlett}, {Bauer},
  {Bauman}, {Baumont}, {Bechtol}, {Bechtol}, {Becker}, {Becla}, {Beldica},
  {Bellavia}, {Bianco}, {Biswas}, {Blanc}, {Blazek}, {Blandford}, {Bloom},
  {Bogart}, {Bond}, {Booth}, {Borgland}, {Borne}, {Bosch}, {Boutigny},
  {Brackett}, {Bradshaw}, {Brandt}, {Brown}, {Bullock}, {Burchat}, {Burke},
  {Cagnoli}, {Calabrese}, {Callahan}, {Callen}, {Carlin}, {Carlson},
  {Chandrasekharan}, {Charles-Emerson}, {Chesley}, {Cheu}, {Chiang}, {Chiang},
  {Chirino}, {Chow}, {Ciardi}, {Claver}, {Cohen-Tanugi}, {Cockrum}, {Coles},
  {Connolly}, {Cook}, {Cooray}, {Covey}, {Cribbs}, {Cui}, {Cutri}, {Daly},
  {Daniel}, {Daruich}, {Daubard}, {Daues}, {Dawson}, {Delgado}, {Dellapenna},
  {de Peyster}, {de Val-Borro}, {Digel}, {Doherty}, {Dubois},
  {Dubois-Felsmann}, {Durech}, {Economou}, {Eifler}, {Eracleous}, {Emmons},
  {Fausti Neto}, {Ferguson}, {Figueroa}, {Fisher-Levine}, {Focke}, {Foss},
  {Frank}, {Freemon}, {Gangler}, {Gawiser}, {Geary}, {Gee}, {Geha}, {Gessner},
  {Gibson}, {Gilmore}, {Glanzman}, {Glick}, {Goldina}, {Goldstein}, {Goodenow},
  {Graham}, {Gressler}, {Gris}, {Guy}, {Guyonnet}, {Haller}, {Harris},
  {Hascall}, {Haupt}, {Hernandez}, {Herrmann}, {Hileman}, {Hoblitt}, {Hodgson},
  {Hogan}, {Howard}, {Huang}, {Huffer}, {Ingraham}, {Innes}, {Jacoby}, {Jain},
  {Jammes}, {Jee}, {Jenness}, {Jernigan}, {Jevremovi{\'c}}, {Johns}, {Johnson},
  {Johnson}, {Jones}, {Juramy-Gilles}, {Juri{\'c}}, {Kalirai}, {Kallivayalil},
  {Kalmbach}, {Kantor}, {Karst}, {Kasliwal}, {Kelly}, {Kessler}, {Kinnison},
  {Kirkby}, {Knox}, {Kotov}, {Krabbendam}, {Krughoff}, {Kub{\'a}nek},
  {Kuczewski}, {Kulkarni}, {Ku}, {Kurita}, {Lage}, {Lambert}, {Lange},
  {Langton}, {Le Guillou}, {Levine}, {Liang}, {Lim}, {Lintott}, {Long},
  {Lopez}, {Lotz}, {Lupton}, {Lust}, {MacArthur}, {Mahabal}, {Mandelbaum},
  {Markiewicz}, {Marsh}, {Marshall}, {Marshall}, {May}, {McKercher}, {McQueen},
  {Meyers}, {Migliore}, {Miller}, {Mills}, {Miraval}, {Moeyens}, {Moolekamp},
  {Monet}, {Moniez}, {Monkewitz}, {Montgomery}, {Morrison}, {Mueller},
  {Muller}, {Mu{\~n}oz Arancibia}, {Neill}, {Newbry}, {Nief}, {Nomerotski},
  {Nordby}, {O'Connor}, {Oliver}, {Olivier}, {Olsen}, {O'Mullane}, {Ortiz},
  {Osier}, {Owen}, {Pain}, {Palecek}, {Parejko}, {Parsons}, {Pease},
  {Peterson}, {Peterson}, {Petravick}, {Libby Petrick}, {Petry},
  {Pierfederici}, {Pietrowicz}, {Pike}, {Pinto}, {Plante}, {Plate}, {Plutchak},
  {Price}, {Prouza}, {Radeka}, {Rajagopal}, {Rasmussen}, {Regnault}, {Reil},
  {Reiss}, {Reuter}, {Ridgway}, {Riot}, {Ritz}, {Robinson}, {Roby}, {Roodman},
  {Rosing}, {Roucelle}, {Rumore}, {Russo}, {Saha}, {Sassolas}, {Schalk},
  {Schellart}, {Schindler}, {Schmidt}, {Schneider}, {Schneider}, {Schoening},
  {Schumacher}, {Schwamb}, {Sebag}, {Selvy}, {Sembroski}, {Seppala}, {Serio},
  {Serrano}, {Shaw}, {Shipsey}, {Sick}, {Silvestri}, {Slater}, {Smith},
  {Smith}, {Sobhani}, {Soldahl}, {Storrie-Lombardi}, {Stover}, {Strauss},
  {Street}, {Stubbs}, {Sullivan}, {Sweeney}, {Swinbank}, {Szalay}, {Takacs},
  {Tether}, {Thaler}, {Thayer}, {Thomas}, {Thornton}, {Thukral}, {Tice},
  {Trilling}, {Turri}, {Van Berg}, {Vanden Berk}, {Vetter}, {Virieux},
  {Vucina}, {Wahl}, {Walkowicz}, {Walsh}, {Walter}, {Wang}, {Wang}, {Warner},
  {Wiecha}, {Willman}, {Winters}, {Wittman}, {Wolff}, {Wood-Vasey}, {Wu},
  {Xin}, {Yoachim}, \& {Zhan}}]{Ivezic_2019}
{Ivezi{\'c}}, {\v{Z}}., {Kahn}, S.~M., {Tyson}, J.~A., {et~al.} 2019, \apj,
  873, 111

\bibitem[{James {et~al.}(2023)James, Witten, Hasatie, Tibshirani, \&
  Taylor}]{ISLP}
James, G., Witten, D., Hasatie, T., Tibshirani, R., \& Taylor, J. 2023, An
  Introduction to Statistical Learning: with Applications in Python (Springer)

\bibitem[{Jost(2006)}]{Jost_2006}
Jost, L. 2006, Oikos, 113, 363

\bibitem[{Kauffmann {et~al.}(2003)Kauffmann, Heckman, White, Charlot, Tremonti,
  Brinchmann, Bruzual, Peng, Seibert, Bernardi, Blanton, Brinkmann, Castander,
  Cs{\'{a}}bai, Fukugita, Ivezic, Munn, Nichol, Padmanabhan, Thakar, Weinberg,
  \& York}]{Kauffmann_2003}
Kauffmann, G., Heckman, T.~M., White, D. M.~S., {et~al.} 2003, \mnras, 341, 33

\bibitem[{Kendall \& Gal(2017)}]{kendall2017uncertainties}
Kendall, A. \& Gal, Y. 2017, in Proceedings of the 31st International
  Conference on Neural Information Processing Systems, NIPS'17 (Red Hook, NY,
  USA: Curran Associates Inc.), 5580–5590

\bibitem[{Kingma \& Ba(2015)}]{kingma2017adam}
Kingma, D.~P. \& Ba, J. 2015, in 3rd International Conference on Learning
  Representations, {ICLR} 2015, San Diego, CA, USA, May 7-9, 2015, Conference
  Track Proceedings, ed. Y.~Bengio \& Y.~LeCun

\bibitem[{Krizhevsky {et~al.}(2017)Krizhevsky, Sutskever, \&
  Hinton}]{Krizhevsky_relu}
Krizhevsky, A., Sutskever, I., \& Hinton, G.~E. 2017, Commun. ACM, 60, 84–90

\bibitem[{{Lagache} {et~al.}(2004){Lagache}, {Dole}, {Puget},
  {P{\'e}rez-Gonz{\'a}lez}, {Le Floc'h}, {Rieke}, {Papovich}, {Egami},
  {Alonso-Herrero}, {Engelbracht}, {Gordon}, {Misselt}, \&
  {Morrison}}]{Lagache_2004}
{Lagache}, G., {Dole}, H., {Puget}, J.~L., {et~al.} 2004, \apjs, 154, 112

\bibitem[{{Laureijs} {et~al.}(2011){Laureijs}, {Amiaux}, {Arduini},
  {Augu{\`e}res}, {Brinchmann}, {Cole}, {Cropper}, {Dabin}, {Duvet}, {Ealet},
  {Garilli}, {Gondoin}, {Guzzo}, {Hoar}, {Hoekstra}, {Holmes}, {Kitching},
  {Maciaszek}, {Mellier}, {Pasian}, {Percival}, {Rhodes}, {Saavedra Criado},
  {Sauvage}, {Scaramella}, {Valenziano}, {Warren}, {Bender}, {Castander},
  {Cimatti}, {Le F{\`e}vre}, {Kurki-Suonio}, {Levi}, {Lilje}, {Meylan},
  {Nichol}, {Pedersen}, {Popa}, {Rebolo Lopez}, {Rix}, {Rottgering},
  {Zeilinger}, {Grupp}, {Hudelot}, {Massey}, {Meneghetti}, {Miller}, {Paltani},
  {Paulin-Henriksson}, {Pires}, {Saxton}, {Schrabback}, {Seidel}, {Walsh},
  {Aghanim}, {Amendola}, {Bartlett}, {Baccigalupi}, {Beaulieu}, {Benabed},
  {Cuby}, {Elbaz}, {Fosalba}, {Gavazzi}, {Helmi}, {Hook}, {Irwin}, {Kneib},
  {Kunz}, {Mannucci}, {Moscardini}, {Tao}, {Teyssier}, {Weller}, {Zamorani},
  {Zapatero Osorio}, {Boulade}, {Foumond}, {Di Giorgio}, {Guttridge}, {James},
  {Kemp}, {Martignac}, {Spencer}, {Walton}, {Bl{\"u}mchen}, {Bonoli},
  {Bortoletto}, {Cerna}, {Corcione}, {Fabron}, {Jahnke}, {Ligori}, {Madrid},
  {Martin}, {Morgante}, {Pamplona}, {Prieto}, {Riva}, {Toledo}, {Trifoglio},
  {Zerbi}, {Abdalla}, {Douspis}, {Grenet}, {Borgani}, {Bouwens}, {Courbin},
  {Delouis}, {Dubath}, {Fontana}, {Frailis}, {Grazian}, {Koppenh{\"o}fer},
  {Mansutti}, {Melchior}, {Mignoli}, {Mohr}, {Neissner}, {Noddle}, {Poncet},
  {Scodeggio}, {Serrano}, {Shane}, {Starck}, {Surace}, {Taylor},
  {Verdoes-Kleijn}, {Vuerli}, {Williams}, {Zacchei}, {Altieri}, {Escudero
  Sanz}, {Kohley}, {Oosterbroek}, {Astier}, {Bacon}, {Bardelli}, {Baugh},
  {Bellagamba}, {Benoist}, {Bianchi}, {Biviano}, {Branchini}, {Carbone},
  {Cardone}, {Clements}, {Colombi}, {Conselice}, {Cresci}, {Deacon}, {Dunlop},
  {Fedeli}, {Fontanot}, {Franzetti}, {Giocoli}, {Garcia-Bellido}, {Gow},
  {Heavens}, {Hewett}, {Heymans}, {Holland}, {Huang}, {Ilbert}, {Joachimi},
  {Jennins}, {Kerins}, {Kiessling}, {Kirk}, {Kotak}, {Krause}, {Lahav}, {van
  Leeuwen}, {Lesgourgues}, {Lombardi}, {Magliocchetti}, {Maguire}, {Majerotto},
  {Maoli}, {Marulli}, {Maurogordato}, {McCracken}, {McLure}, {Melchiorri},
  {Merson}, {Moresco}, {Nonino}, {Norberg}, {Peacock}, {Pello}, {Penny},
  {Pettorino}, {Di Porto}, {Pozzetti}, {Quercellini}, {Radovich}, {Rassat},
  {Roche}, {Ronayette}, {Rossetti}, {Sartoris}, {Schneider}, {Semboloni},
  {Serjeant}, {Simpson}, {Skordis}, {Smadja}, {Smartt}, {Spano}, {Spiro},
  {Sullivan}, {Tilquin}, {Trotta}, {Verde}, {Wang}, {Williger}, {Zhao},
  {Zoubian}, \& {Zucca}}]{laureijs2011euclid}
{Laureijs}, R., {Amiaux}, J., {Arduini}, S., {et~al.} 2011, arXiv:1110.3193

\bibitem[{{Law} {et~al.}(2018){Law}, {Gordon}, \& {Misselt}}]{Law_2018}
{Law}, K.-H., {Gordon}, K.~D., \& {Misselt}, K.~A. 2018, \apjs, 236, 32

\bibitem[{{Leistedt} {et~al.}(2023){Leistedt}, {Alsing}, {Peiris}, {Mortlock},
  \& {Leja}}]{Leistedt_2023}
{Leistedt}, B., {Alsing}, J., {Peiris}, H., {Mortlock}, D., \& {Leja}, J. 2023,
  \apjs, 264, 23

\bibitem[{{Leja} {et~al.}(2019){Leja}, {Carnall}, {Johnson}, {Conroy}, \&
  {Speagle}}]{Leja_2019}
{Leja}, J., {Carnall}, A.~C., {Johnson}, B.~D., {Conroy}, C., \& {Speagle},
  J.~S. 2019, \apj, 876, 3

\bibitem[{{Leja} {et~al.}(2017){Leja}, {Johnson}, {Conroy}, {van Dokkum}, \&
  {Byler}}]{Leja_2017}
{Leja}, J., {Johnson}, B.~D., {Conroy}, C., {van Dokkum}, P.~G., \& {Byler}, N.
  2017, \apj, 837, 170

\bibitem[{{Lequeux} {et~al.}(1979){Lequeux}, {Peimbert}, {Rayo}, {Serrano}, \&
  {Torres-Peimbert}}]{Lequeux_1979}
{Lequeux}, J., {Peimbert}, M., {Rayo}, J.~F., {Serrano}, A., \&
  {Torres-Peimbert}, S. 1979, \aap, 80, 155

\bibitem[{{Lo Faro} {et~al.}(2017){Lo Faro}, {Buat}, {Roehlly},
  {Alvarez-Marquez}, {Burgarella}, {Silva}, \& {Efstathiou}}]{LoFaro_2017}
{Lo Faro}, B., {Buat}, V., {Roehlly}, Y., {et~al.} 2017, \mnras, 472, 1372

\bibitem[{Longo {et~al.}(2024)Longo, Brcic, Cabitza, Choi, Confalonieri, Ser,
  Guidotti, Hayashi, Herrera, Holzinger, Jiang, Khosravi, Lecue, Malgieri,
  Páez, Samek, Schneider, Speith, \& Stumpf}]{LONGO2024102301}
Longo, L., Brcic, M., Cabitza, F., {et~al.} 2024, Information Fusion, 106,
  102301

\bibitem[{{Lovell} {et~al.}(2019){Lovell}, {Acquaviva}, {Thomas}, {Iyer},
  {Gawiser}, \& {Wilkins}}]{Lovell_2019}
{Lovell}, C.~C., {Acquaviva}, V., {Thomas}, P.~A., {et~al.} 2019, \mnras, 490,
  5503

\bibitem[{Lu {et~al.}(2014)Lu, Wechsler, Somerville, Croton, Porter, Primack,
  Behroozi, Ferguson, Koo, Guo, Safarzadeh, Finlator, Castellano, White,
  Sommariva, \& Moody}]{Lu_2014}
Lu, Y., Wechsler, R.~H., Somerville, R.~S., {et~al.} 2014, \apj, 795, 123

\bibitem[{{Luhman} {et~al.}(2003){Luhman}, {Satyapal}, {Fischer}, {Wolfire},
  {Sturm}, {Dudley}, {Lutz}, \& {Genzel}}]{Luhman_FIR}
{Luhman}, M.~L., {Satyapal}, S., {Fischer}, J., {et~al.} 2003, \apj, 594, 758

\bibitem[{{Ly} {et~al.}(2016){Ly}, {Malkan}, {Rigby}, \& {Nagao}}]{Ly_2016}
{Ly}, C., {Malkan}, M.~A., {Rigby}, J.~R., \& {Nagao}, T. 2016, \apj, 828, 67

\bibitem[{{Maiolino} \& {Mannucci}(2019)}]{Maiolino_2019}
{Maiolino}, R. \& {Mannucci}, F. 2019, \aapr, 27, 3

\bibitem[{Marinacci {et~al.}(2018)Marinacci, Vogelsberger, Pakmor, Torrey,
  Springel, Hernquist, Nelson, Weinberger, Pillepich, Naiman, \&
  Genel}]{Marinacci_2018}
Marinacci, F., Vogelsberger, M., Pakmor, R., {et~al.} 2018, \mnras, 480, 5113

\bibitem[{{Matteucci}(2008)}]{Matteucci_2008}
{Matteucci}, F. 2008, in Massive Stars as Cosmic Engines, ed. F.~{Bresolin},
  P.~A. {Crowther}, \& J.~{Puls}, Vol. 250, 391--400

\bibitem[{Murphy(2012)}]{murphy2012machine}
Murphy, K. 2012, Machine Learning: A Probabilistic Perspective, Adaptive
  Computation and Machine Learning series (MIT Press)

\bibitem[{Naiman {et~al.}(2018)Naiman, Pillepich, Springel, Ramirez-Ruiz,
  Torrey, Vogelsberger, Pakmor, Nelson, Marinacci, Hernquist, Weinberger, \&
  Genel}]{Naiman_2018}
Naiman, J.~P., Pillepich, A., Springel, V., {et~al.} 2018, \mnras, 477, 1206

\bibitem[{Nair \& Hinton(2010)}]{Nair2010RectifiedLU}
Nair, V. \& Hinton, G.~E. 2010, in Proceedings of the 27th International
  Conference on International Conference on Machine Learning, ICML'10 (Madison,
  WI, USA: Omnipress), 807–814

\bibitem[{{Nelson} {et~al.}(2019{\natexlab{a}}){Nelson}, {Pillepich},
  {Springel}, {Pakmor}, {Weinberger}, {Genel}, {Torrey}, {Vogelsberger},
  {Marinacci}, \& {Hernquist}}]{Nelson_2019}
{Nelson}, D., {Pillepich}, A., {Springel}, V., {et~al.} 2019{\natexlab{a}},
  \mnras, 490, 3234

\bibitem[{{Nelson} {et~al.}(2019{\natexlab{b}}){Nelson}, {Springel},
  {Pillepich}, {Rodriguez-Gomez}, {Torrey}, {Genel}, {Vogelsberger}, {Pakmor},
  {Marinacci}, {Weinberger}, {Kelley}, {Lovell}, {Diemer}, \&
  {Hernquist}}]{Nelson_2018}
{Nelson}, D., {Springel}, V., {Pillepich}, A., {et~al.} 2019{\natexlab{b}},
  Computational Astrophysics and Cosmology, 6, 2

\bibitem[{{Nersesian} {et~al.}(2024){Nersesian}, {van der Wel}, {Gallazzi},
  {Leja}, {Bezanson}, {Bell}, {D'Eugenio}, {de Graaff}, {Kaushal}, {Martorano},
  {Maseda}, \& {Zibetti}}]{Nersesian_2023}
{Nersesian}, A., {van der Wel}, A., {Gallazzi}, A., {et~al.} 2024, \aap, 681,
  A94

\bibitem[{{Nersesian} {et~al.}(2019){Nersesian}, {Xilouris}, {Bianchi},
  {Galliano}, {Jones}, {Baes}, {Casasola}, {Cassar{\`a}}, {Clark}, {Davies},
  {Decleir}, {Dobbels}, {De Looze}, {De Vis}, {Fritz}, {Galametz}, {Madden},
  {Mosenkov}, {Tr{\v{c}}ka}, {Verstocken}, {Viaene}, \&
  {Lianou}}]{Nersesian_2019}
{Nersesian}, A., {Xilouris}, E.~M., {Bianchi}, S., {et~al.} 2019, \aap, 624,
  A80

\bibitem[{Neumann {et~al.}(2021)Neumann, Thomas, Maraston, Goddard, Lian, Hill,
  S{\'{a}}nchez, Bernardi, Margalef-Bentabol, Barrera-Ballesteros, Bizyaev,
  Boardman, Drory, Fern{\'{a}}ndez-Trincado, \& Lane}]{Neumann_2021}
Neumann, J., Thomas, D., Maraston, C., {et~al.} 2021, \mnras, 508, 4844

\bibitem[{{Pacifici} {et~al.}(2023){Pacifici}, {Iyer}, {Mobasher}, {da Cunha},
  {Acquaviva}, {Burgarella}, {Calistro Rivera}, {Carnall}, {Chang}, {Chartab},
  {Cooke}, {Fairhurst}, {Kartaltepe}, {Leja}, {Ma{\l}ek}, {Salmon}, {Torelli},
  {Vidal-Garc{\'\i}a}, {Boquien}, {Brammer}, {Brown}, {Capak}, {Chevallard},
  {Circosta}, {Croton}, {Davidzon}, {Dickinson}, {Duncan}, {Faber}, {Ferguson},
  {Fontana}, {Guo}, {Haeussler}, {Hemmati}, {Jafariyazani}, {Kassin}, {Larson},
  {Lee}, {Mantha}, {Marchi}, {Nayyeri}, {Newman}, {Pandya}, {Pforr}, {Reddy},
  {Sanders}, {Shah}, {Shahidi}, {Stevans}, {Triani}, {Tyler}, {Vanderhoof}, {de
  la Vega}, {Wang}, \& {Weston}}]{Pacifici_2023}
{Pacifici}, C., {Iyer}, K.~G., {Mobasher}, B., {et~al.} 2023, \apj, 944, 141

\bibitem[{{Pearson}(1895)}]{pearson}
{Pearson}, K. 1895, Proceedings of the Royal Society of London Series I, 58,
  240

\bibitem[{Pedregosa {et~al.}(2011)Pedregosa, Varoquaux, Gramfort, Michel,
  Thirion, Grisel, Blondel, Prettenhofer, Weiss, Dubourg, Vanderplas, Passos,
  Cournapeau, Brucher, Perrot, \& Duchesnay}]{scikit-learn}
Pedregosa, F., Varoquaux, G., Gramfort, A., {et~al.} 2011, Journal of Machine
  Learning Research, 12, 2825

\bibitem[{{Pellerin} \& {Finkelstein}(2010)}]{Pellerin_2009}
{Pellerin}, A. \& {Finkelstein}, S.~L. 2010, Proceedings of the International
  Astronomical Union, 262, 283

\bibitem[{Pillepich {et~al.}(2017)Pillepich, Nelson, Hernquist, Springel,
  Pakmor, Torrey, Weinberger, Genel, Naiman, Marinacci, \&
  Vogelsberger}]{Pillepich_2017}
Pillepich, A., Nelson, D., Hernquist, L., {et~al.} 2017, \mnras, 475, 648

\bibitem[{{Pillepich} {et~al.}(2019){Pillepich}, {Nelson}, {Springel},
  {Pakmor}, {Torrey}, {Weinberger}, {Vogelsberger}, {Marinacci}, {Genel}, {van
  der Wel}, \& {Hernquist}}]{Pillepich_2019}
{Pillepich}, A., {Nelson}, D., {Springel}, V., {et~al.} 2019, \mnras, 490, 3196

\bibitem[{{Pillepich} {et~al.}(2018){Pillepich}, {Springel}, {Nelson}, {Genel},
  {Naiman}, {Pakmor}, {Hernquist}, {Torrey}, {Vogelsberger}, {Weinberger}, \&
  {Marinacci}}]{Pillepich_2018}
{Pillepich}, A., {Springel}, V., {Nelson}, D., {et~al.} 2018, \mnras, 473, 4077

\bibitem[{{Planck Collaboration: Ade} {et~al.}(2016){Planck Collaboration:
  Ade}, {Aghanim}, {Arnaud}, {Ashdown}, {Aumont}, {Baccigalupi}, {Banday},
  {Barreiro}, {Bartlett}, {Bartolo}, {Battaner}, {Battye}, {Benabed},
  {Beno{\^\i}t}, {Benoit-L{\'e}vy}, {Bernard}, {Bersanelli}, {Bielewicz},
  {Bock}, {Bonaldi}, {Bonavera}, {Bond}, {Borrill}, {Bouchet}, {Boulanger},
  {Bucher}, {Burigana}, {Butler}, {Calabrese}, {Cardoso}, {Catalano},
  {Challinor}, {Chamballu}, {Chary}, {Chiang}, {Chluba}, {Christensen},
  {Church}, {Clements}, {Colombi}, {Colombo}, {Combet}, {Coulais}, {Crill},
  {Curto}, {Cuttaia}, {Danese}, {Davies}, {Davis}, {de Bernardis}, {de Rosa},
  {de Zotti}, {Delabrouille}, {D{\'e}sert}, {Di Valentino}, {Dickinson},
  {Diego}, {Dolag}, {Dole}, {Donzelli}, {Dor{\'e}}, {Douspis}, {Ducout},
  {Dunkley}, {Dupac}, {Efstathiou}, {Elsner}, {En{\ss}lin}, {Eriksen},
  {Farhang}, {Fergusson}, {Finelli}, {Forni}, {Frailis}, {Fraisse},
  {Franceschi}, {Frejsel}, {Galeotta}, {Galli}, {Ganga}, {Gauthier}, {Gerbino},
  {Ghosh}, {Giard}, {Giraud-H{\'e}raud}, {Giusarma}, {Gjerl{\o}w},
  {Gonz{\'a}lez-Nuevo}, {G{\'o}rski}, {Gratton}, {Gregorio}, {Gruppuso},
  {Gudmundsson}, {Hamann}, {Hansen}, {Hanson}, {Harrison}, {Helou},
  {Henrot-Versill{\'e}}, {Hern{\'a}ndez-Monteagudo}, {Herranz}, {Hildebrandt},
  {Hivon}, {Hobson}, {Holmes}, {Hornstrup}, {Hovest}, {Huang}, {Huffenberger},
  {Hurier}, {Jaffe}, {Jaffe}, {Jones}, {Juvela}, {Keih{\"a}nen}, {Keskitalo},
  {Kisner}, {Kneissl}, {Knoche}, {Knox}, {Kunz}, {Kurki-Suonio}, {Lagache},
  {L{\"a}hteenm{\"a}ki}, {Lamarre}, {Lasenby}, {Lattanzi}, {Lawrence}, {Leahy},
  {Leonardi}, {Lesgourgues}, {Levrier}, {Lewis}, {Liguori}, {Lilje},
  {Linden-V{\o}rnle}, {L{\'o}pez-Caniego}, {Lubin}, {Mac{\'\i}as-P{\'e}rez},
  {Maggio}, {Maino}, {Mandolesi}, {Mangilli}, {Marchini}, {Maris}, {Martin},
  {Martinelli}, {Mart{\'\i}nez-Gonz{\'a}lez}, {Masi}, {Matarrese}, {McGehee},
  {Meinhold}, {Melchiorri}, {Melin}, {Mendes}, {Mennella}, {Migliaccio},
  {Millea}, {Mitra}, {Miville-Desch{\^e}nes}, {Moneti}, {Montier}, {Morgante},
  {Mortlock}, {Moss}, {Munshi}, {Murphy}, {Naselsky}, {Nati}, {Natoli},
  {Netterfield}, {N{\o}rgaard-Nielsen}, {Noviello}, {Novikov}, {Novikov},
  {Oxborrow}, {Paci}, {Pagano}, {Pajot}, {Paladini}, {Paoletti}, {Partridge},
  {Pasian}, {Patanchon}, {Pearson}, {Perdereau}, {Perotto}, {Perrotta},
  {Pettorino}, {Piacentini}, {Piat}, {Pierpaoli}, {Pietrobon}, {Plaszczynski},
  {Pointecouteau}, {Polenta}, {Popa}, {Pratt}, {Pr{\'e}zeau}, {Prunet},
  {Puget}, {Rachen}, {Reach}, {Rebolo}, {Reinecke}, {Remazeilles}, {Renault},
  {Renzi}, {Ristorcelli}, {Rocha}, {Rosset}, {Rossetti}, {Roudier},
  {Rouill{\'e} d'Orfeuil}, {Rowan-Robinson}, {Rubi{\~n}o-Mart{\'\i}n},
  {Rusholme}, {Said}, {Salvatelli}, {Salvati}, {Sandri}, {Santos},
  {Savelainen}, {Savini}, {Scott}, {Seiffert}, {Serra}, {Shellard}, {Spencer},
  {Spinelli}, {Stolyarov}, {Stompor}, {Sudiwala}, {Sunyaev}, {Sutton},
  {Suur-Uski}, {Sygnet}, {Tauber}, {Terenzi}, {Toffolatti}, {Tomasi},
  {Tristram}, {Trombetti}, {Tucci}, {Tuovinen}, {T{\"u}rler}, {Umana},
  {Valenziano}, {Valiviita}, {Van Tent}, {Vielva}, {Villa}, {Wade}, {Wandelt},
  {Wehus}, {White}, {White}, {Wilkinson}, {Yvon}, {Zacchei}, \&
  {Zonca}}]{Planck_2016}
{Planck Collaboration: Ade}, P.~A.~R., {Aghanim}, N., {Arnaud}, M., {et~al.}
  2016, \aap, 594, A13

\bibitem[{{Poggianti} \& {Barbaro}(1997)}]{Poggianti_1997}
{Poggianti}, B.~M. \& {Barbaro}, G. 1997, \aap, 325, 1025

\bibitem[{{Popesso} {et~al.}(2023){Popesso}, {Concas}, {Cresci}, {Belli},
  {Rodighiero}, {Inami}, {Dickinson}, {Ilbert}, {Pannella}, \&
  {Elbaz}}]{Popesso_2023}
{Popesso}, P., {Concas}, A., {Cresci}, G., {et~al.} 2023, \mnras, 519, 1526

\bibitem[{{Rosales-Ortega} {et~al.}(2012){Rosales-Ortega}, {S{\'a}nchez},
  {Iglesias-P{\'a}ramo}, {D{\'\i}az}, {V{\'\i}lchez}, {Bland-Hawthorn},
  {Husemann}, \& {Mast}}]{Rosales-Ortega_2012}
{Rosales-Ortega}, F.~F., {S{\'a}nchez}, S.~F., {Iglesias-P{\'a}ramo}, J.,
  {et~al.} 2012, \apjl, 756, L31

\bibitem[{Sadeh {et~al.}(2016)Sadeh, Abdalla, \& Lahav}]{Sadeh_2016}
Sadeh, I., Abdalla, F.~B., \& Lahav, O. 2016, PASP, 128, 104502

\bibitem[{{Salim} {et~al.}(2018){Salim}, {Boquien}, \& {Lee}}]{Salim_2018}
{Salim}, S., {Boquien}, M., \& {Lee}, J.~C. 2018, \apj, 859, 11

\bibitem[{{Salim} \& {Narayanan}(2020)}]{Salim_2020}
{Salim}, S. \& {Narayanan}, D. 2020, \araa, 58, 529

\bibitem[{{Salucci}(2019)}]{Salucci_2019}
{Salucci}, P. 2019, \aapr, 27, 2

\bibitem[{{S{\'a}nchez}(2020)}]{S_nchez_2020}
{S{\'a}nchez}, S.~F. 2020, \araa, 58, 99

\bibitem[{{S{\'a}nchez} {et~al.}(2012){S{\'a}nchez}, {Kennicutt}, {Gil de Paz},
  {van de Ven}, {V{\'\i}lchez}, {Wisotzki}, {Walcher}, {Mast}, {Aguerri},
  {Albiol-P{\'e}rez}, {Alonso-Herrero}, {Alves}, {Bakos}, {Bart{\'a}kov{\'a}},
  {Bland-Hawthorn}, {Boselli}, {Bomans}, {Castillo-Morales}, {Cortijo-Ferrero},
  {de Lorenzo-C{\'a}ceres}, {Del Olmo}, {Dettmar}, {D{\'\i}az}, {Ellis},
  {Falc{\'o}n-Barroso}, {Flores}, {Gallazzi}, {Garc{\'\i}a-Lorenzo},
  {Gonz{\'a}lez Delgado}, {Gruel}, {Haines}, {Hao}, {Husemann},
  {Igl{\'e}sias-P{\'a}ramo}, {Jahnke}, {Johnson}, {Jungwiert}, {Kalinova},
  {Kehrig}, {Kupko}, {L{\'o}pez-S{\'a}nchez}, {Lyubenova}, {Marino},
  {M{\'a}rmol-Queralt{\'o}}, {M{\'a}rquez}, {Masegosa}, {Meidt},
  {Mendez-Abreu}, {Monreal-Ibero}, {Montijo}, {Mour{\~a}o}, {Palacios-Navarro},
  {Papaderos}, {Pasquali}, {Peletier}, {P{\'e}rez}, {P{\'e}rez}, {Quirrenbach},
  {Rela{\~n}o}, {Rosales-Ortega}, {Roth}, {Ruiz-Lara},
  {S{\'a}nchez-Bl{\'a}zquez}, {Sengupta}, {Singh}, {Stanishev}, {Trager},
  {Vazdekis}, {Viironen}, {Wild}, {Zibetti}, \& {Ziegler}}]{S_nchez_2012}
{S{\'a}nchez}, S.~F., {Kennicutt}, R.~C., {Gil de Paz}, A., {et~al.} 2012,
  \aap, 538, A8

\bibitem[{{S{\'a}nchez} {et~al.}(2013){S{\'a}nchez}, {Rosales-Ortega},
  {Jungwiert}, {Iglesias-P{\'a}ramo}, {V{\'\i}lchez}, {Marino}, {Walcher},
  {Husemann}, {Mast}, {Monreal-Ibero}, {Cid Fernandes}, {P{\'e}rez},
  {Gonz{\'a}lez Delgado}, {Garc{\'\i}a-Benito}, {Galbany}, {van de Ven},
  {Jahnke}, {Flores}, {Bland-Hawthorn}, {L{\'o}pez-S{\'a}nchez}, {Stanishev},
  {Miralles-Caballero}, {D{\'\i}az}, {S{\'a}nchez-Blazquez}, {Moll{\'a}},
  {Gallazzi}, {Papaderos}, {Gomes}, {Gruel}, {P{\'e}rez}, {Ruiz-Lara},
  {Florido}, {de Lorenzo-C{\'a}ceres}, {Mendez-Abreu}, {Kehrig}, {Roth},
  {Ziegler}, {Alves}, {Wisotzki}, {Kupko}, {Quirrenbach}, {Bomans}, \& {CALIFA
  Collaboration}}]{Sanchez_2013}
{S{\'a}nchez}, S.~F., {Rosales-Ortega}, F.~F., {Jungwiert}, B., {et~al.} 2013,
  \aap, 554, A58

\bibitem[{{Sanders} {et~al.}(2013){Sanders}, {Levesque}, \&
  {Soderberg}}]{Sanders_2013}
{Sanders}, N.~E., {Levesque}, E.~M., \& {Soderberg}, A.~M. 2013, \apj, 775, 125

\bibitem[{{Schmidt}(1959)}]{Schmidt_1959}
{Schmidt}, M. 1959, \apj, 129, 243

\bibitem[{{Siebenmorgen} \& {Kr{\"u}gel}(2007)}]{Siebenmorgen_2007}
{Siebenmorgen}, R. \& {Kr{\"u}gel}, E. 2007, \aap, 461, 445

\bibitem[{{Simet} {et~al.}(2021){Simet}, {Chartab}, {Lu}, \&
  {Mobasher}}]{Simet_2021}
{Simet}, M., {Chartab}, N., {Lu}, Y., \& {Mobasher}, B. 2021, \apj, 908, 47

\bibitem[{{Sorba} \& {Sawicki}(2015)}]{Sorba_2015}
{Sorba}, R. \& {Sawicki}, M. 2015, \mnras, 452, 235

\bibitem[{{Speagle} {et~al.}(2014){Speagle}, {Steinhardt}, {Capak}, \&
  {Silverman}}]{Speagle_2014}
{Speagle}, J.~S., {Steinhardt}, C.~L., {Capak}, P.~L., \& {Silverman}, J.~D.
  2014, \apjs, 214, 15

\bibitem[{{Springel}(2010)}]{Springel_2010}
{Springel}, V. 2010, \mnras, 401, 791

\bibitem[{Springel {et~al.}(2017)Springel, Pakmor, Pillepich, Weinberger,
  Nelson, Hernquist, Vogelsberger, Genel, Torrey, Marinacci, \&
  Naiman}]{Springel_2017}
Springel, V., Pakmor, R., Pillepich, A., {et~al.} 2017, \mnras, 475, 676

\bibitem[{{Steinacker} {et~al.}(2013){Steinacker}, {Baes}, \&
  {Gordon}}]{Steinacker_2013}
{Steinacker}, J., {Baes}, M., \& {Gordon}, K.~D. 2013, \araa, 51, 63

\bibitem[{{Stensbo-Smidt} {et~al.}(2017){Stensbo-Smidt}, {Gieseke}, {Igel},
  {Zirm}, \& {Steenstrup Pedersen}}]{Stensbo_2017}
{Stensbo-Smidt}, K., {Gieseke}, F., {Igel}, C., {Zirm}, A., \& {Steenstrup
  Pedersen}, K. 2017, \mnras, 464, 2577

\bibitem[{{Surana} {et~al.}(2020){Surana}, {Wadadekar}, {Bait}, \&
  {Bhosale}}]{Surana_2020}
{Surana}, S., {Wadadekar}, Y., {Bait}, O., \& {Bhosale}, H. 2020, \mnras, 493,
  4808

\bibitem[{{Tagliaferri} {et~al.}(2003){Tagliaferri}, {Longo}, {Andreon},
  {Capozziello}, {Donalek}, \& {Giordano}}]{Tagliaferri_2003}
{Tagliaferri}, R., {Longo}, G., {Andreon}, S., {et~al.} 2003, in Lecture Notes
  in Computer Science, Vol. 2859, 226--234

\bibitem[{Tammes(1930)}]{tammes1930origin}
Tammes, P. M.~L. 1930, Recueil des travaux botaniques n{\'e}erlandais, 27, 1

\bibitem[{{Thorne} {et~al.}(2022){Thorne}, {Robotham}, {Bellstedt}, {Davies},
  {Cook}, {Cortese}, {Holwerda}, {Phillipps}, \& {Siudek}}]{Thorne_2022}
{Thorne}, J.~E., {Robotham}, A. S.~G., {Bellstedt}, S., {et~al.} 2022, \mnras,
  517, 6035

\bibitem[{{Torrey} {et~al.}(2019){Torrey}, {Vogelsberger}, {Marinacci},
  {Pakmor}, {Springel}, {Nelson}, {Naiman}, {Pillepich}, {Genel}, {Weinberger},
  \& {Hernquist}}]{Torrey_2019}
{Torrey}, P., {Vogelsberger}, M., {Marinacci}, F., {et~al.} 2019, \mnras, 484,
  5587

\bibitem[{{Tremonti} {et~al.}(2004){Tremonti}, {Heckman}, {Kauffmann},
  {Brinchmann}, {Charlot}, {White}, {Seibert}, {Peng}, {Schlegel}, {Uomoto},
  {Fukugita}, \& {Brinkmann}}]{Tremonti_2004}
{Tremonti}, C.~A., {Heckman}, T.~M., {Kauffmann}, G., {et~al.} 2004, \apj, 613,
  898

\bibitem[{{Viaene} {et~al.}(2014){Viaene}, {Fritz}, {Baes}, {Bendo},
  {Blommaert}, {Boquien}, {Boselli}, {Ciesla}, {Cortese}, {De Looze}, {Gear},
  {Gentile}, {Hughes}, {Jarrett}, {Karczewski}, {Smith}, {Spinoglio}, {Tamm},
  {Tempel}, {Thilker}, \& {Verstappen}}]{Viaene_2014}
{Viaene}, S., {Fritz}, J., {Baes}, M., {et~al.} 2014, \aap, 567, A71

\bibitem[{Vilone \& Longo(2021)}]{make3030032}
Vilone, G. \& Longo, L. 2021, Machine Learning and Knowledge Extraction, 3, 615

\bibitem[{{Vogelsberger} {et~al.}(2014){Vogelsberger}, {Genel}, {Springel},
  {Torrey}, {Sijacki}, {Xu}, {Snyder}, {Nelson}, \&
  {Hernquist}}]{Vogelsberger_2014}
{Vogelsberger}, M., {Genel}, S., {Springel}, V., {et~al.} 2014, \mnras, 444,
  1518

\bibitem[{{Walcher} {et~al.}(2011){Walcher}, {Groves}, {Budav{\'a}ri}, \&
  {Dale}}]{Walcher_2010}
{Walcher}, J., {Groves}, B., {Budav{\'a}ri}, T., \& {Dale}, D. 2011, \apss,
  331, 1

\bibitem[{{Watkins} {et~al.}(2022){Watkins}, {Salo}, {Laurikainen},
  {D{\'\i}az-Garc{\'\i}a}, {Comer{\'o}n}, {Janz}, {Su}, {Buta}, {Athanassoula},
  {Bosma}, {Ho}, {Holwerda}, {Kim}, {Knapen}, {Laine},
  {Men{\'e}ndez-Delmestre}, {Peletier}, {Sheth}, \& {Zaritsky}}]{Watkins_2022}
{Watkins}, A.~E., {Salo}, H., {Laurikainen}, E., {et~al.} 2022, \aap, 660, A69

\bibitem[{{Webb} {et~al.}(2020){Webb}, {Balogh}, {Leja}, {van der Burg},
  {Rudnick}, {Muzzin}, {Boak}, {Cerulo}, {Gilbank}, {Lidman}, {Old},
  {Pintos-Castro}, {McGee}, {Shipley}, {Biviano}, {Chan}, {Cooper}, {De Lucia},
  {Demarco}, {Forrest}, {Jablonka}, {Kukstas}, {McCarthy}, {McNab}, {Nantais},
  {Noble}, {Poggianti}, {Reeves}, {Vulcani}, {Wilson}, {Yee}, \&
  {Zaritsky}}]{Webb_2020}
{Webb}, K., {Balogh}, M.~L., {Leja}, J., {et~al.} 2020, \mnras, 498, 5317

\bibitem[{Weinberger {et~al.}(2016)Weinberger, Springel, Hernquist, Pillepich,
  Marinacci, Pakmor, Nelson, Genel, Vogelsberger, Naiman, \&
  Torrey}]{Weinberger_2016}
Weinberger, R., Springel, V., Hernquist, L., {et~al.} 2016, \mnras, 465, 3291

\bibitem[{{Whitaker} {et~al.}(2012){Whitaker}, {van Dokkum}, {Brammer}, \&
  {Franx}}]{Whitaker_2012}
{Whitaker}, K.~E., {van Dokkum}, P.~G., {Brammer}, G., \& {Franx}, M. 2012,
  \apjl, 754, L29

\bibitem[{Wright(1921)}]{Wright_1921}
Wright, S. 1921, Journal of Agricultural Research, 20, 557

\bibitem[{{Zahid} {et~al.}(2017){Zahid}, {Kudritzki}, {Conroy}, {Andrews}, \&
  {Ho}}]{Zahid_2017}
{Zahid}, H.~J., {Kudritzki}, R.-P., {Conroy}, C., {Andrews}, B., \& {Ho}, I.~T.
  2017, \apj, 847, 18

\bibitem[{{Zanisi} {et~al.}(2021){Zanisi}, {Huertas-Company}, {Lanusse},
  {Bottrell}, {Pillepich}, {Nelson}, {Rodriguez-Gomez}, {Shankar}, {Hernquist},
  {Dekel}, {Margalef-Bentabol}, {Vogelsberger}, \& {Primack}}]{Zanisi_2021}
{Zanisi}, L., {Huertas-Company}, M., {Lanusse}, F., {et~al.} 2021, \mnras, 501,
  4359

\bibitem[{Zeiler {et~al.}(2013)Zeiler, Ranzato, Monga, Mao, Yang, Le, Nguyen,
  Senior, Vanhoucke, Dean, \& Hinton}]{6638312}
Zeiler, M., Ranzato, M., Monga, R., {et~al.} 2013, in 2013 IEEE International
  Conference on Acoustics, Speech and Signal Processing, 3517--3521

\bibitem[{Zibetti {et~al.}(2009)Zibetti, Charlot, \& Rix}]{Zibetti_2009}
Zibetti, S., Charlot, S., \& Rix, H.-W. 2009, \mnras, 400, 1181

\bibitem[{Zibetti \& Gallazzi(2022)}]{Zibetti_2022}
Zibetti, S. \& Gallazzi, A.~R. 2022, \mnras, 512, 1415

\bibitem[{{Zibetti} {et~al.}(2020){Zibetti}, {Gallazzi}, {Hirschmann},
  {Consolandi}, {Falc{\'o}n-Barroso}, {van de Ven}, \&
  {Lyubenova}}]{Zibetti_2019}
{Zibetti}, S., {Gallazzi}, A.~R., {Hirschmann}, M., {et~al.} 2020, \mnras, 491,
  3562

\end{thebibliography}

\appendix
\section{Effect of including additional projection angles}\label{app:angles}

Here, we use a subset of 200 randomly selected galaxies to test whether the inclusion of other projections would affect the prediction accuracy of our ML algorithm. As a test set, we set aside 50 random galaxies, each of them at one random projection angle, totalling ${\rm \sim\,2.7}$ million pixels. The remaining 150 galaxies are used to train the algorithm. First, we use only one projection angle O1, then two (O1 and O2), in each step adding an additional angle projection until all five projection angles (O1--O5) are used. The accuracy scores are located in Table\,\ref{tab:angles}.

\begin{table}[h]
    \centering
        \caption{Coefficient of determination ($R^2$) for a DNN using the four \Euclid\/\ surface brightnesses on a subset of 200 randomly selected galaxies and including different projections.}\resizebox{\columnwidth}{!}{

    \begin{tabular}{l r c c c}
\hline \hline
  & & & &   \\[-0.3cm]         
  Projection angles & Pixel count & $R^2$ ($\Sigma_{\star}$) & $R^2$ ($Z_{\star}$)  & $R^2$ ($t_{\star}$)\\
 & & &  &  \\[-0.3cm]
 \hline
 & & & &  \\[-0.3cm]
 O1 & 8\,912\,000 & 0.9319 & 0.5452 & 0.3009 \\
         O1, O2 & 17\,774\,522 & 0.9322 & 0.5484 & 0.3015 \\
         O1, O2, O3 & 26\,411\,564 & 0.9320 & 0.5899 &  0.3116 \\
         O1, O2, O3, O4 & 34\,809\,544 & 0.9323 & 0.5490 & 0.3115 \\
         O1, O2, O3, O4, O5 & 43\,214\,225 & 0.9324 & 0.5529 & 0.3116 \\

      & & & &  \\[-0.3cm]
        \hline   \hline \end{tabular}}
    \label{tab:angles}
\end{table}

\section{The effect of sample bias}\label{app:samplebias}

There is a slight bias between our training and test sets, as shown in Fig.\,\ref{fig:hist}. For this reason, we apply our trained algorithms on both the test and the training sets. The accuracy scores between these two sets, shown in Tab. \ref{tab:bias}, are within the 10\% difference.

\begin{table}[h]\label{tab:samplebias}
    \caption{Accuracy scores (as defined in Sect.\,\ref{sec:ML_metrics}) for the three target galaxy properties ($\mass$, $Z_{\star}$, and $t_{\star}$) inferred using the four \Euclid\/\ surface brightnesses for the training and test sets with the RF ML algorithm. }
    
    \smallskip
\label{tab:bias}
\smallskip
\centering
\begin{tabular}{l c c c}
\hline \hline
  & & & \\[-0.3cm]

 &Train & Test & Diff. (\%)\\
 & & &  \\[-0.3cm]
 \hline
 & & & \\[-0.3cm]
  Target & \multicolumn{3}{c}{$R^2$} \\ 
  
 & & &  \\[-0.3cm]
 \hline
 & & & \\[-0.3cm]
 $\mass {\rm \,/\, \left( M_{\odot} \, pc^{-2}\right)}$ & $0.937$& 0.927 & 1 \\
 $Z_{\star}$ & 0.606 & 0.592 & 2.4\\
 $t_{\star} {\rm \, / \, Gyr}$& 0.419 & 0.382 & 9.7 \\
  & & &  \\[-0.3cm]
 \hline
 & & & \\[-0.3cm]
   Target & \multicolumn{3}{c}{RMSE} \\ 

 & & &  \\[-0.3cm]
 \hline
 & & & \\[-0.3cm]
 $\mass {\rm \,/\, \left( M_{\odot} \, pc^{-2}\right)}$ & $0.1219$& 0.1308 & 6.8 \\
 $Z_{\star}$ & 0.0805& 0.0765 & 5.2\\
 $t_{\star} {\rm \, / \, Gyr}$& 0.1279 & 0.1285 & 9.7 \\
   & & &  \\[-0.3cm]
 \hline
 & & & \\[-0.3cm]
   Target & \multicolumn{3}{c}{$\rho$} \\ 
  
 & & &  \\[-0.3cm]
 \hline
 & & & \\[-0.3cm]
 $\mass {\rm \,/\, \left( M_{\odot} \, pc^{-2}\right)}$ & $0.968$& 0.963 & 0.5 \\
 $Z_{\star}$ & 0.778& 0.771 & 0.9\\
 $t_{\star} {\rm \, / \, Gyr}$& 0.647 & 0.622 & 4 \\

    &&&  \\[-0.3cm]
        \hline \hline
    \end{tabular}
    \tablefoot{Left to right: targets; coefficient of determination ($R^2$); root-mean-square error (${\rm RMSE}$); Pearson $\rho$ coefficient.}
\end{table}

\section{The effect of squared colours}\label{app:c2}

As this work was partly inspired by \citet{Acquaviva_2015}, we also test the effect of using squared colour could offer a higher accuracy when inferring galaxy parameters from mock \Euclid\/\ observations. We used our DNN set-up, as described in Table\,\ref{tab:dnn_hyper-parameters} in Sect.\,\ref{sec:methods_ML}.The results are shown in Table\,\ref{tab:c2_res}. The differences, if any, are minor at best, once again giving credence to our conclusion that it is possible to infer galaxy parameters using only the four \Euclid\/\ bands, as the algorithm is capable of extracting all of the information contained within them, without the need of modelling additional colours.
\begin{table}[h]
    \caption{Accuracy scores (as defined in Sect.\,\ref{sec:ML_metrics}) for the three target galaxy properties ($\mass$, $Z_{\star}$, and $t_{\star}$) inferred using the four \Euclid\/\ surface brightnesses, six colours modelled from them, and six squared colours using a DNN. }
    
    \smallskip
\label{tab:c2_res}
\smallskip
\centering
\resizebox{\columnwidth}{!}{
\begin{tabular}{l c c c c c c }

\hline \hline
  & & & &  & & \\[-0.3cm]

Target  &  $R^2$  & RMSE  & $f_{\text{out}}$ & NMAD & $\left< \Delta Y \right>$ & $\rho$ \\
 & & &  & & & \\[-0.3cm]
 \hline
 & & & & & & \\[-0.3cm]
 $\mass {\rm \,/\, \left( M_{\odot} \, pc^{-2}\right)}$ & $0.927$& $0.130$ & $0.0325$ & $0.1088$ & $0.0020$ & 0.963\\
 $Z_{\star}$ & $0.595$  & $0.076$  &$0.0007$ & $0.0719$& $0.0024$& 0.773 \\
 $t_{\star} {\rm \, / \, Gyr}$& $0.382$& $0.128$ &$0.0281$ & $0.0994$&$-0.0017$&$0.623$ \\

      & & & &  & & \\[-0.3cm]
        \hline \hline
    \end{tabular}
    }\tablefoot{ Left to right: Target; coefficient of determination ($R^2$); root-mean-square error (${\rm RMSE}$); fraction of catastrophic outliers ($f_\text{out}$); normalised median absolute deviation (${\rm NMAD}$); and bias ($\left< \Delta Y \right>$, where $Y$ represents the target galaxy property); Pearson $\rho$ coefficient.}
\end{table}
\section{Inferring colours from \Euclid\/\ surface brightnesses}\label{app:colors}

Here, we test the accuracy of inferring the six colours modelled with the four \Euclid\/\ surface brightnesses using a DNN with the same set-up as Table\,\ref{tab:dnn_hyper-parameters}. In all cases, the network converges very quickly (in under 100 epochs), and with the accuracy shown in Table\,\ref{tab:cols}, we conclude that using colours to infer galaxy parameters does not offer any additional benefit to our method.

\begin{table}[h]
    \caption{Accuracy scores (as defined in Sect.\,\ref{sec:ML_metrics}) for the six \Euclid\/\ colours inferred from the four \Euclid\/\ brightnesses using a DNN. }
    
    \smallskip
\label{tab:cols}
\smallskip
\centering
\begin{tabular}{l c c }

\hline \hline
  & & \\[-0.3cm]

Target colour  &  $R^2$  & RMSE  \\
 & &  \\[-0.3cm]
 \hline
 & & \\[-0.3cm]
 $\left( I_\text{E} - Y_\text{E} \right)$ & 0.999 &0.004 \\
 $\left( I_\text{E} - J_\text{E} \right)$ & 0.999 &0.003 \\
 $\left( I_\text{E} - H_\text{E} \right)$ & 0.998 &0.011 \\
 $\left( Y_\text{E} - J_\text{E} \right)$ & 0.999 &0.009 \\
 $\left( Y_\text{E} - H_\text{E} \right)$ & 0.999 &0.005 \\
  $\left( J_\text{E} - H_\text{E} \right)$ & 0.999 &0.005 \\
      & & \\[-0.3cm]
        \hline \hline
    \end{tabular}
    \tablefoot{Left to right: Target colour; coefficient of determination ($R^2$); root-mean-square error (${\rm RMSE}$).}
\end{table}

\end{document}